%% file: ms.tex
\documentclass[11pt,a4paper,twoside,openright]{article}
\usepackage{authblk}

\usepackage[utf8]{inputenc}
\usepackage[russian, francais, english]{babel}
\usepackage[T1]{fontenc}
\usepackage[left=3cm,right=3cm,top=3cm,bottom=3cm]{geometry}
\usepackage{amsmath}
\allowdisplaybreaks 
\usepackage{amsfonts}
\usepackage[numbers]{natbib}
\usepackage[globalcitecopy, labelstoglobalaux]{bibunits}
\usepackage{bibentry}
\nobibliography* 
\usepackage{titletoc} 
\usepackage{appendix}%
\usepackage[tabbotcap, TABBOTCAP]{subfigure}
\usepackage[hidelinks]{hyperref}
\usepackage{import}
\usepackage{amssymb}
\usepackage{graphicx}
  \usepackage[export]{adjustbox}%
\usepackage{setspace}
\usepackage{placeins}
\usepackage{changepage}
\usepackage{caption}
\usepackage{numprint}
\npstyleenglish
\npthousandsep{}
\usepackage{multicol}
\usepackage{multirow, bigdelim}
\usepackage{longtable}
\usepackage{indentfirst}
\usepackage{url}
\usepackage[outdir=fig/]{epstopdf}
\usepackage{color}
\usepackage[usenames,dvipsnames,svgnames,table]{xcolor}
\usepackage{arydshln}
\usepackage{afterpage}
\usepackage{float}
\usepackage{listings}
\usepackage{xpatch}

\usepackage{titlesec}

\titleformat{\paragraph}
{\normalfont\normalsize\bfseries}{\theparagraph}{1em}{}
\titlespacing*{\paragraph}
{0pt}{3.25ex plus 1ex minus .2ex}{1.5ex plus .2ex}

\title{\Large \textbf{A posteriori tests of subgrid-scale models in an isothermal turbulent channel flow}}
\def\runtitle{A posteriori tests of subgrid-scale models in an isothermal turbulent channel flow}

\def\runauthor{D. Dupuy, A. Toutant and F. Bataille}
\author[1]{Dorian Dupuy}
\author[1]{Adrien Toutant\thanks{Corresponding author : adrien.toutant@univ-perp.fr}}
\author[1]{Fran\c coise Bataille}
\affil[1]{PROMES CNRS, Universit\'{e} de Perpignan Via Domitia, Rambla de la thermodynamique, Tecnosud, 66100 Perpignan, France}

\date{\vspace{-4ex}\itshape\small (Published version: Physics of Fluids 31, 045105 (2019); https://doi.org/10.1063/1.5091829)\vspace{-4ex}}

\usepackage{fancyhdr}
\usepackage{scalerel,stackengine}
\stackMath
\DeclareRobustCommand\reallywidehat[1]{%
\savestack{\tmpbox}{\stretchto{%
  \scaleto{%
    \scalerel*[\widthof{\ensuremath{#1}}]{\kern-.6pt\bigwedge\kern-.6pt}%
    {\rule[-\textheight/2]{1ex}{\textheight}}
  }{\textheight}%
}{0.5ex}}%
\stackon[1pt]{\displaystyle #1}{\tmpbox}%
}
\stackMath
\DeclareRobustCommand\reallywidetilde[1]{%
\savestack{\tmpbox}{\stretchto{%
  \scaleto{%
    \scalerel*[\widthof{\ensuremath{#1}}]{\kern-1.0pt\sim\kern-1.0pt}%
    {\rule[-\textheight/2]{1ex}{\textheight}}
  }{\textheight}%
}{0.5ex}}%
\stackon[1pt]{\displaystyle #1}{\tmpbox}%
}
\newcommand{\overbar}[1]{\mkern 1.5mu\overline{\mkern-1.5mu#1}}

\DeclareMathOperator{\tr}{tr}

\let\OLDthebibliography\thebibliography
\renewcommand\thebibliography[1]{
  \OLDthebibliography{#1}
  \setlength{\parskip}{0pt}
  \setlength{\itemsep}{0pt plus 0.3ex}
}

\begin{document}

\hyphenation{aniso-thermal}

\newcommand{\der}[2]{\frac{\partial #1}{\partial #2}}
\newcommand{\dd}[2]{\frac{\partial^2 #1}{\partial #2^2}}
\newcommand{\dtot}[2]{\frac{D #1}{D #2}}
\newcommand{\ov}[1]{\overline{#1}}
\renewcommand{\f}[1]{\overbar{#1}}
\newcommand{\fa}[1]{\widetilde{#1}}
\newcommand{\FA}[1]{\reallywidetilde{#1}}
\newcommand{\ff}[1]{\widehat{#1}}
\newcommand{\FF}[1]{\reallywidehat{#1}}
\newcommand{\m}[1]{\overbar{#1}}
\newcommand{\ma}[1]{\widetilde{#1}}
\newcommand{\MA}[1]{\reallywidetilde{#1}}
\newcommand{\vv}[1]{\boldsymbol{#1}}
\newcommand{\vt}[1]{\boldsymbol{#1}}
\newcommand{\w}[1]{\widehat{#1}}
\newcommand{\W}[1]{\reallywidehat{#1}}

\newcommand{\refv}[1]{#1^{b}}
\newcommand{\adim}[1]{#1^{\circ}}

\newcommand{\upi}[0]{\pi}

\newcommand\Real{\mbox{Re}}
\newcommand\Imag{\mbox{Im}}
\newcommand\Rey{\mbox{\textit{Re}}}
\newcommand\Pran{\mbox{\textit{Pr}}}
\newcommand\Pen{\mbox{\textit{Pe}}}
\newcommand\Ai{\mbox{Ai}}
\newcommand\Bi{\mbox{Bi}}

\newcommand{\corr}[1]{{\color{red} #1}}
\newcommand{\corre}[1]{{\color{blue} #1}}

\newcommand{\e}[0]{\mathrm{e}}
\newcommand{\I}[0]{\mathrm{i}}
\newcommand{\Dee}[0]{D}
\newcommand{\ft}[1]{\widehat{#1}}
\newcommand{\FT}[1]{\reallywidehat{#1}}

\newcounter{subfigcounter}
\setcounter{subfigcounter}{1}
\DeclareRobustCommand{\lecID}{\refstepcounter{subfigcounter}}
\DeclareRobustCommand{\subfigtopleft}[1]{\lecID\def\stackalignment{l}\topinset{\text{\footnotesize(\alph{subfigcounter})}}{#1}{0.1in}{-.05in}}

\maketitle

{
\let\latexthesection\thesection %
    \renewcommand{\thesection}{\arabic{section}} 
    \renewcommand{\thefigure}{\arabic{figure}} 
    \setcounter{section}{0}
\subimport{./}{article.tex}

\let\stdthebibliography\thebibliography
\renewcommand{\thebibliography}{%
\let\chapter\section
\stdthebibliography}
  {
  \small
  \bibliography{./biblio}
  }
}

\end{document}

%% file: article.tex
\begin{abstract}
This paper studies the large-eddy simulation (LES) of
isothermal turbulent channel flows.
We investigate zero-equation algebraic models without wall function or wall model: functional models, structural models and
mixed models.
In addition to models from the literature, new models are proposed
and their relevance is examined.
Dynamic versions of each type of model are also analysed.
The performance of the subgrid-scale models is assessed
using the same finite difference numerical method and physical configuration.
The friction Reynolds number of the simulations is 180.
Three different mesh resolutions are used.
The predictions of large-eddy simulations
are compared to a direct numerical simulation filtered at the resolution
of the LES meshes.
The results are more accurate than a simulation
without model.
The predictions of functional eddy-viscosity models can be improved
using constant-parameter or dynamic tensorial methods.
\end{abstract}

\section{Introduction}

This paper addresses the large-eddy simulation (LES) of an isothermal
turbulent channel flow at low Mach number.
The large-eddy simulation of a turbulent channel flow is an important
problem as it is one of the key canonical flows to understand
wall-bounded turbulence.
A large-eddy simulation only resolves the large-scale motions of
turbulence, generally represented using a low-pass filter.
This approach is more practical than direct numerical simulation (DNS),
in which all scales of turbulence must be resolved.
However, the evolution equations of the filtered variables cannot be inferred from the
flow governing equations because the filter does not, in general,
commute with
multiplication.
We consider the large-eddy simulation of the
incompressible Navier--Stokes equations for a Newtonian fluid
using a filter ($\f{\,\,\,\cdot\,\,}$):
\begin{gather}
\der{\f{U}_{\!j}}{x_j} = 0, \\
\der{\f{U}_{\!i}}{t} = - \der{\left(\f{U}_{\!j} \f{U}_{\!i} + \tau_{ij}\right)}{x_j} - \frac{1}{\rho}\der{\f{P}}{x_i} + \nu \frac{\partial^2 \f{U}_{\!i}}{\partial x_j \partial x_i},
\end{gather}
with $\rho$ the density, $\nu$ the kinematic viscosity,
$t$ the time, $P$ the mechanical pressure,
$U_i$ the $i$-th component of velocity and
$x_i$ the Cartesian coordinate in $i$-th direction.
The momentum convection subgrid term or subgrid-scale tensor is defined as
$\tau_{ij} ={} \f{U_j U_i} - \f{U}_{\!j} \f{U}_{\!i}$.
To close the system of equations, the subgrid-scale tensor must be modelled using
an algorithm computable in a large-eddy simulation.
The addition of the subgrid-scale model should not violate the symmetry
properties of Navier--Stokes equations \citep{speziale1985galilean, oberlack2002symmetries, razafindralandy2007analysis}.
In wall-bounded flows, the subgrid-scale model should
be able to preserve the
driving mechanisms of turbulence near the walls \citep{jimenez1999physics, mathis2009large, jimenez2013near}.
The consistency of the asymptotic near-wall behaviour of
the model with the exact subgrid term is also considered very important \citep{nicoud99b, nicoud2011using}.
A review of the physical constraints that the subgrid-scale subgrid-scale model
should satisfy is given in \citet{silvis2017physical}.
Besides, for an application of these methods to another interesting problem, see
references \citep{zhou2017rayleighpart1, zhou2017rayleighpart2}.

The modelling of the subgrid-scale tensor has received a lot of attention from
the literature.
Algebraic or zero-equation models
assume that the small scales are universal
and can be expressed as functions of the resolved flow variables,
as opposed to models requiring the resolution of one or more additional
transport equations to compute the subgrid-scale model \citep{sagaut98, lu2016structural}.
The models can be classified 
into structural and functional models \citep{sagaut98}.
Functional models, also called eddy-viscosity models, make the fundamental hypothesis that the
effect of subgrid scales is analogous to viscous diffusion \citep{boussinesq1877essai},
hence strictly dissipative.
The popularity of this class of model is attributed to its robustness and
low computational complexity \citep{sagaut98, trias2015building}.
Structural models approximate the effect of the filter without assumptions on
the physical nature of the effect of the subgrid term.
A review is given by \citet{lu2016structural}.
In the literature, numerous functional and structural models have been proposed and investigated in plane channel flows
at low or moderate friction Reynolds numbers:
the Smagorinsky and dynamic Smagorinsky models \citep{iliescu2003large, vreman2004eddy, kobayashi2005subgrid, jeanmart2007investigation, toda2010subgrid, denaro2011comparative, lampitella2014large, ryu2014subgrid, ohtsuka2014toward, rieth2014comparison, trias2015building, rozema2015minimum, yu2017scale, jiang2018large}, 
the approximate deconvolution model \citep{stolz2001approximate, zhou2018structural},
the rational model \citep{iliescu2003large},
the tensorial anisotropic model \citep{abba2003analysis, denaro2011comparative, abba2015dynamic}, 
the Kobayashi model \citep{kobayashi2005subgrid}, 
the Vreman and dynamic Vreman models \citep{vreman2004eddy, ryu2014subgrid, lampitella2014large, trias2015building, jiang2016large, yu2017scale, silvis2017physical, jiang2018large}, 
the Sigma and dynamic Sigma models \citep{toda2010subgrid, rieth2014comparison, lampitella2014large, jiang2018large}, 
the QR and anisotropic minimum-dissipation (AMD) models \citep{verstappen2011does, rozema2015minimum},
the volumetric strain-stretching (VSS) model and dynamic VSS models \citep{ryu2014subgrid, yu2017scale, jiang2018large},
the anisotropy-resolving model \citep{ohtsuka2014toward},
S3PQR models \citep{trias2015building}, 
the modulated gradient model \citep{ghaisas2016dynamic}, 
the vortex-stretching model \citep{silvis2017physical},
scale-adaptive models \citep{yu2017scale} 
and mixed and dynamic mixed models \citep{morinishi2001recommended, abba2003analysis, lampitella2014large}.
Note also that large-eddy simulation can be combined with Reynolds-averaged Navier--Stokes modelling,
as in detached-eddy simulation \citep{spalart1997comments, travin2000detached, strelets2001detached, spalart2009detached},
in constrained large-eddy simulation \citep{chen2012reynolds, jiang2013constrained, larsson2016large}
or in the linear unified model \citep{gopalan2013unified, mokhtarpoor2016dynamic, mokhtarpoor2017dynamic}.
A systematic inspection of the subgrid-scale models would be useful
for the selection of the subgrid-scale model for a particular simulation
as well as for future subgrid-scale modelling developments.

This paper investigates a posteriori the modelling of the subgrid-scale tensor
in an isothermal turbulent channel flow at a friction Reynolds number of 180.
We will focus on the effect of the models on the turbulence statistics.
To assess the performance of the large-eddy simulations, the results are
compared to a direct numerical simulation filtered at the resolution of the
large-eddy simulations.
This allows the direct comparison of the results of the large-eddy simulations
and of the direct numerical simulations.
The analysis is based on the LES formalism introduced by \citet{leonard74}.
In this paradigm, the large-eddy simulation aims to provide resolved fields
whose statistics correspond to the statistics of a filtered direct numerical
simulation.
Note that the comparison with filtered direct numerical simulation is not
systematically carried out in the literature since other approaches are
possible \citep{pope2004ten}.
For practical applications, the knowledge of the filtered variables may not
be sufficient as nonfiltered variables are more relevant. This implies that a
reconstruction of the nonfiltered fields from the results of the large-eddy
simulation is required.
By comparing large-eddy simulations to filtered DNS data, we separate
the subgrid-scale modelling from the reconstruction procedure, which may
use a different model and can only address the deviatoric part of the Reynolds stresses
for traceless or partially traceless models \citep{winckelmans2002comparison}.
We address several functional and structural models from the literature
using a common numerical method and physical configuration.
In addition, tensorial eddy-viscosity models are proposed and investigated
in order to explore the relevance of the eddy-viscosity
assumption for each component of the subgrid term and
account for the anisotropy of the flow.
Mixed models combining structural and functional or tensorial models
and dynamic versions of these models are also considered.
The large-eddy simulations are carried out on three different meshes to provide an indication
of the robustness of the models to variations of the grid
resolution.

We give the subgrid-scale models investigated in section \ref{label30}. 
The channel flow configuration and the numerical method are presented in section \ref{label29}.
The results are discussed in section \ref{label31}.

\section{Subgrid-scale models}\label{label30}

Subgrid-scale models express the subgrid-scale tensor as a function of variables
resolved in the large-eddy simulation:
\begin{align}
\tau_{ij} \approx{}& \tau_{ij}^{\mathrm{mod}}(\vv{\f{U}}, \vv{\f{\Delta}}),
\end{align}%
where the function $\tau_{ij}^{\mathrm{mod}}(\vv{U},\vv{\f{\Delta}})$
depends on the model.
We investigate zero-equation algebraic models without
wall function or wall model. This includes functional models, structural models, tensorial models and
tensorial mixed models. Dynamic versions of each type of modelling are also
considered.

\subsection{Constant-parameter models}

Using functional eddy-viscosity models, the subgrid-scale tensor
is modelled by analogy with molecular diffusion,
\begin{align}
\tau_{ij}^{\mathrm{mod}}(\vv{U}, \vv{\f{\Delta}}) ={}& - 2 \nu_e^{\mathrm{mod}}(\vv{g}, \vv{\f{\Delta}}) S_{ij},
\end{align}%
with $S_{ij} = \tfrac{1}{2}\left( g_{ij} + g_{ji} \right)$ the rate of
deformation tensor and
$\vv{g}$ the velocity gradient, defined by $g_{ij} = \partial_j U_i$.
The expression of the eddy viscosity depends on the model used.
The eddy-viscosity models investigated are:
\begin{flalign}
\intertext{Smagorinsky model \citep{smagorinsky1963general}:}  \nu_e^{\mathrm{Smag.}}  (\vv{g}, \vv{\f{\Delta}}) ={}& \left( C^{\mathrm{Smag.}} \f{\Delta} \right)^2 \left|\vt{S}\right|, \\
\intertext{Wall-adapting local eddy-viscosity (WALE) model \citep{nicoud99b}:}                      \nu_e^{\mathrm{WALE}}   (\vv{g}, \vv{\f{\Delta}}) ={}& \left( C^{\mathrm{WALE}} \f{\Delta} \right)^2 \frac{\left(\mathcal{S}^d_{ij} \mathcal{S}^d_{ij}\right)^{\tfrac{3}{2}}}{\left(S_{mn} S_{mn}\right)^{\tfrac{5}{2}} + \left(\mathcal{S}^d_{mn} \mathcal{S}^d_{mn}\right)^{\tfrac{5}{4}}}, \\
\intertext{Sigma model \citep{nicoud2011using}:}               \nu_e^{\mathrm{Sigma}}  (\vv{g}, \vv{\f{\Delta}}) ={}& \left( C^{\mathrm{Sigma}} \f{\Delta} \right)^2 \frac{\sigma_3\left(\sigma_1 - \sigma_2\right)\left(\sigma_2 - \sigma_3\right)}{\sigma_1^2}, \\
\intertext{Anisotropic minimum-dissipation (AMD) model \citep{rozema2015minimum}:}               \nu_e^{\mathrm{AMD}}    (\vv{g}, \vv{\f{\Delta}}) ={}& C^{\mathrm{AMD}} \frac{\max(0, - G_{ij} S_{ij})}{g_{mn}g_{mn}}, \\
\intertext{Kobayashi model \citep{kobayashi2005subgrid}:}      \nu_e^{\mathrm{Koba.}}  (\vv{g}, \vv{\f{\Delta}}) ={}& C^{\mathrm{Koba.}} \f{\Delta}^2 \left|F_g\right|^{\frac{3}{2}} (1-F_g) \left|\vt{S}\right|, 
\end{flalign}%
where 
$\left|\vt{S}\right|=\sqrt{2 S_{ij} S_{ij}}$ is a norm of $\vt{S}$,
$\mathcal{S}^d_{ij} = \tfrac{1}{2}\left( g_{ik}g_{kj} + g_{jk}g_{ki} \right) - \tfrac{1}{3}g_{kp}g_{pk} \delta_{ij}$ the traceless symmetric part of the squared velocity gradient tensor, 
$\sigma_1 \geq \sigma_2 \geq \sigma_3$ the three singular values of $\vv{g}$,
$G_{ij} = \f{\Delta}_k^2 g_{ik} g_{jk}$ the gradient model,
$\mathrm{II}_G = \tfrac{1}{2}\left(\tr^2\left(G\right) - \tr\left(G^2\right)\right)$ its second invariant,
$R_{ij}=\beta_i g_{jj}$ the volumetric strain-stretching, with $\beta=\left(S_{23}, S_{13}, S_{12}\right)$,
and $F_g = \left(\varOmega_{ij}\varOmega_{ij} - S_{ij}S_{ij}\right)/\left(\varOmega_{mn}\varOmega_{mn} + S_{mn}S_{mn}\right)$
the coherent structure function, with $\varOmega_{ij} = \tfrac{1}{2}\left( g_{ij} - g_{ji} \right)$ the spin tensor or rate of rotation tensor.

Anisotropic eddy-viscosity models involve one length scale
per direction instead of a single length scale.
Anisotropic versions of the Smagorinsky, WALE, Sigma and Kobayashi models can
be devised.
The AMD model are already anisotropic.
We define the Anisotropic Smagorinsky model \citep{dupuy2018apriori} as,
\begin{align}
\tau_{ij}^{\mathrm{An. Smag.}}(\vv{U}, \vv{\f{\Delta}}) ={}& - 2 \nu_e^{\mathrm{Smag.}}(\vv{g^a}, \vv{\f{\Delta}}) S^a_{ij},
\end{align}
with $S^a_{ij} = \tfrac{1}{2}\left( g^a_{ij} + g^a_{ji} \right)$ the scaled
rate of deformation tensor and
$\vv{g^a}$ the scaled velocity gradient, defined by $g^a_{ij} = (\f{\Delta}_j/\f{\Delta}) \partial_j U_i$.

Using the structural gradient model \citep{leonard74}, the subgrid-scale tensor is modelled according
to a Taylor series expansion of the filter,
\begin{align}
\tau_{ij}^{\mathrm{Grad.}}(\vv{U}, \vv{\f{\Delta}}) ={}& \tfrac{1}{12} C^{\mathrm{Grad.}} G_{ij}(\vv{U}, \vv{\f{\Delta}}) = \tfrac{1}{12} C^{\mathrm{Grad.}} \f{\Delta}_k^2 g_{ik} g_{jk},
\end{align}
Using the structural scale-similarity model \citep{bardina1980improved}, the subgrid-scale tensor is modelled
following the scale-similarity assumption,
\begin{align}
\tau_{ij}^{\mathrm{Simil.}}(\vv{U}, \vv{\f{\Delta}}) ={}& C^{\mathrm{Simil.}} \left(\ft{U_j U_i} - \ft{U}_j\, \ft{U}_i\right), \label{label5}
\end{align}
where $\ft{\,\cdot\,}$ is a test filter explicitly computed in the
large-eddy simulation.
The Taylor series expansion of the filter $\ft{\,\cdot\,}$ in (\ref{label5})
leads to 
\begin{align}
\tau_{ij}^{\mathrm{Simil.}}(\vv{U}, \vv{\f{\Delta}}) = \tfrac{1}{12} C^{\mathrm{Simil.}} G_{ij}(\vv{U}, \vv{\ft{\Delta}}) = \tfrac{1}{12} C^{\mathrm{Simil.}} \ft{\Delta}_k^2 g_{ik} g_{jk}. \label{label6}
\end{align}
This corresponds to the gradient model associated with the filter lengths
$\ft{\Delta}_k^2$ of the test filter.

Tensorial eddy-viscosity models can be constructed
from any functional model.
This aims to take into account the anisotropy of the flow by
weighting of each component of the subgrid-scale model, following
the premise that the relevance of the eddy-viscosity
assumption is not the same for each component of the subgrid term.
In general, we may construct
from any algebraic model
$\tau_{ij}^{\mathrm{mod}}(\vv{U},\vv{\f{\Delta}})$,
and
second-order tensors $H^{(k)}_{ij}$
tensorial models
$\tau_{ij}^{H^{(k)}\mathrm{mod}}(\vv{U},\vv{\f{\Delta}})$
of the form
\begin{equation}
\tau_{ij}^{H^{(k)}\mathrm{mod}}(\vv{U}, \vv{\f{\Delta}}) = H^{(k)}_{ij} \tau_{ij}^{\mathrm{mod}}(\vv{U}, \vv{\f{\Delta}}),
\end{equation}
where no implicit summations over $i$ and $j$ are assumed.
We define for this purpose the tensors
$H^{(1)}_{ij} = \left[ i \neq j\right]$,
$H^{(2)}_{ij} = \left[ \chi_{ij}^{xy} \right]$,
$H^{(3)}_{ij} = \left[ \lnot \chi_{ij}^{yy} \right]$,
$H^{(4)}_{ij} = \left[ \chi_{ij}^{xy} \lor \chi_{ij}^{xz} \right]$,
$H^{(5)}_{ij} = \left[ \chi_{ij}^{xy} \lor \chi_{ij}^{yz} \right]$,
$H^{(6)}_{ij} = \left[ i=x \lor j=x \right]$ et
$H^{(7)}_{ij} = \left[ \chi_{ij}^{xx} \lor \chi_{ij}^{xy} \right]$,
where $\left[\,\cdot\,\right]$ are Iverson brackets, evaluating to 1 if the
proposition within bracket is satisfied and 0 otherwise,
$\lnot$ the logical negation (\textsc{not}),
$\land$ the logical conjunction (\textsc{and}),
$\lor$ the logical disjunction  (\textsc{or})
and with the notation
$\chi_{ij}^{ab} = \left(i=a \land j=b\right) \lor \left(i=b \land j=a\right)$.
More explicitly, we have
\begin{align}
H^{(1)}={}&
\begin{pmatrix}
0 & 1 & 1 \\ 
1 & 0 & 1 \\ 
1 & 1 & 0
\end{pmatrix}
\label{eqdefh1}
\!\!, \\
H^{(2)}_{ij}={}&
\begin{pmatrix}
0 & 1 & 0 \\ 
1 & 0 & 0 \\ 
0 & 0 & 0
\end{pmatrix}
\label{eqdefh2}
\!\!, \\
H^{(3)}_{ij}={}&
\begin{pmatrix}
1 & 1 & 1 \\ 
1 & 0 & 1 \\
1 & 1 & 1
\end{pmatrix}
\label{eqdefh3}
\!\!, \\
H^{(4)}_{ij}={}&
\begin{pmatrix}
0 & 1 & 1 \\ 
1 & 0 & 0 \\ 
1 & 0 & 0
\end{pmatrix}
\label{label2}
\!\!, \\
H^{(5)}_{ij}={}&
\begin{pmatrix}
0 & 1 & 0 \\ 
1 & 0 & 1 \\
0 & 1 & 0
\end{pmatrix}
\label{label3}
\!\!, \\
H^{(6)}_{ij}={}&
\begin{pmatrix}
1 & 1 & 1 \\ 
1 & 0 & 0 \\ 
1 & 0 & 0
\end{pmatrix}
\label{label4}
\!\!, \\
H^{(7)}_{ij}={}&
\begin{pmatrix}
1 & 1 & 0 \\ 
1 & 0 & 0 \\
0 & 0 & 0
\end{pmatrix}
\!\!.
\label{eqdefh7}
\end{align}%

Functional and structural models may also be combined to form mixed models.
To be more general, we consider 
tensorial mixed models, which combine the two models with a different weighting for each
component.
This may be used to combine structural and functional models for each component
or to model each component with either a functional or a structural model. 
Tensorial mixed models are constructed from two algebraic models
$\tau_{ij}^{\mathrm{one}}(\vv{U},\vv{\f{\Delta}})$ and
$\tau_{ij}^{\mathrm{two}}(\vv{U},\vv{\f{\Delta}})$,
and two constant second-order tensors
$H^{(k)}$ and $H^{(l)}$,
\begin{equation}
\begin{aligned}
\tau_{ij}^{(1-H^{(k)})\mathrm{one} + H^{(l)}\mathrm{two}}(\vv{\f{U}}, \vv{\f{\Delta}}) ={}& (1-H^{(k)}_{ij}) \tau_{ij}^{\mathrm{one}}(\vv{\f{U}}, \vv{\f{\Delta}}) \\
&+ H^{(l)}_{ij} \tau_{ij}^{\mathrm{two}}(\vv{\f{U}}, \vv{\f{\Delta}}).
\end{aligned}
\end{equation}
where no implicit summations over $i$ and $j$ are assumed.

Unless stated otherwise, we implicitly use the model parameters
$C^{\mathrm{Smag.}} = 0.10$,
$C^{\mathrm{WALE}} = 0.55$,
$C^{\mathrm{Sigma}}=1.5$,
$C^{\mathrm{AMD}}=0.3$ and
$C^{\mathrm{Koba.}}=0.045$.
We compute the filter length scale using $\f{\Delta}=(\f{\Delta}_x\f{\Delta}_y\f{\Delta}_z)^{1/3}$ \citep{deardorff1970numerical}.
The reader may refer to \citet{trias2017new} for a review of alternative definitions.

\subsection{Dynamic models}

For any constant-parameter algebraic subgrid-scale model, dynamic models may be
constructed using the approach introduced by \citet{germano91}.
A new
model $\tau_{ij}^{\mathrm{dyn,mod}}(\vv{U},\vv{\f{\Delta}})$, referred to as
the dynamic version of the model, can be constructed
from any
algebraic model $\tau_{ij}^{\mathrm{mod}}(\vv{U},\vv{\f{\Delta}})$,
\begin{equation}
\tau_{ij}^{\mathrm{dyn,mod}}(\vv{\f{U}}, \vv{\f{\Delta}}) = C^{\mathrm{dyn}} \tau_{ij}^{\mathrm{mod}}(\vv{\f{U}}, \vv{\f{\Delta}}).
\end{equation}
Following the approach of \citet{lilly1992proposed}, the parameter $C^{\mathrm{dyn}}$ is computed
to minimise the variance of the residual
$E_{ij}(\vv{\f{U}}, \vv{\f{\Delta}}) = L_{ij}(\vv{\f{U}}) - C^{\mathrm{dyn}} m_{ij}(\vv{\f{U}}, \vv{\f{\Delta}})$.
This leads to
\begin{equation}
C^{\mathrm{dyn}} = \frac{\left\langle m_{ij}(\vv{\f{U}}, \vv{\f{\Delta}}) L_{ij}(\vv{\f{U}}) \right\rangle}{\left\langle m_{mn}(\vv{\f{U}}, \vv{\f{\Delta}}) m_{mn}(\vv{\f{U}}, \vv{\f{\Delta}}) \right\rangle}. \label{eqdinlillyme}
\end{equation}
$L_{ij}(\vv{\f{U}}) = \ft{\f{U}_{\!j} \f{U}_{\!i}} - \ft{\f{U}}_j\, \ft{\f{U}}_i$,
with
$m_{ij}(\vv{\f{U}}, \vv{\f{\Delta}}) = \tau_{ij}^{\mathrm{mod}}(\vv{\ft{\f{U}}}, \vv{\ft{\f{\Delta}}}) - \FT{\tau_{ij}^{\mathrm{mod}}(\vv{\f{U}}, \vv{\f{\Delta}})}$
and where ($\ft{\,\cdot\,}$) is a test filter.
The value of $\ft{\f{\Delta}}$ is best approximated as
$\ft{\f{\Delta}} = (\f{\Delta}_i^2+\ft{\Delta}_i^2)^{1/2}$
for Gaussian and box filters \citep{germano1992turbulence, vreman1994formulation}.
Tensorial dynamic methods can extend the dynamic procedure to the construction
of models of the form
\begin{equation}
\tau_{ij}^{\mathrm{ten,dyn,mod}}(\vv{\f{U}}, \vv{\f{\Delta}}) = C_{ij}^{\mathrm{dyn}} \tau_{ij}^{\mathrm{mod}}(\vv{\f{U}}, \vv{\f{\Delta}}),
\end{equation}
where no implicit summations over $i$ and $j$ are assumed.
As in the (scalar) dynamic method, the tensorial parameter of the model is
computed dynamically to minimise for all $i$ and $j$ the variance of the
residual \citep{abba2003analysis}.
This leads to
\begin{equation}
C_{ij}^{\mathrm{dyn}} = \frac{\left\langle m_{ij}(\vv{\f{U}}, \vv{\f{\Delta}}) L_{ij}(\vv{\f{U}}) \right\rangle}{\left\langle m_{ij}(\vv{\f{U}}, \vv{\f{\Delta}}) m_{ij}(\vv{\f{U}}, \vv{\f{\Delta}}) \right\rangle},
\end{equation}
where no implicit summations over $i$ and $j$ are assumed.

In addition, dynamic mixed models can be constructed using analogous
procedures.
The dynamic mixed model
$\tau_{ij}^{\mathrm{dyn,one,two}}(\vv{U},\vv{\f{\Delta}})$
may be expressed
from two algebraic models
$\tau_{ij}^{\mathrm{one}}(\vv{U},\vv{\f{\Delta}})$ and
$\tau_{ij}^{\mathrm{two}}(\vv{U},\vv{\f{\Delta}})$
as
\begin{equation}
\tau_{ij}^{\mathrm{dyn,one,two}}(\vv{\f{U}}, \vv{\f{\Delta}}) = C^{\mathrm{one}} \tau_{ij}^{\mathrm{one}}(\vv{\f{U}}, \vv{\f{\Delta}}) + C^{\mathrm{two}} \tau_{ij}^{\mathrm{two}}(\vv{\f{U}}, \vv{\f{\Delta}}).
\end{equation}
Several methods have been suggested to compute the parameters
$C^{\mathrm{one}}$ and $C^{\mathrm{two}}$:
\begin{itemize}
\item Two-parameter dynamic mixed method: The parameters of the two models are computed dynamically to minimise the
variance of the residual
$E_{ij} = L_{ij} - C^{\mathrm{one}} m_{ij}^{\mathrm{one}} - C^{\mathrm{two}} m_{ij}^{\mathrm{two}}$ \citep{salvetti1995priori, salvetti1997large, horiuti1997new, sarghini1999scale}.
This leads to
\begin{equation}
C^{\mathrm{two}} = \frac{\left\langle m_{ij}^{\mathrm{one}} m_{ij}^{\mathrm{one}} \right\rangle\left\langle L_{kl} m_{kl}^{\mathrm{two}} \right\rangle - \left\langle m_{ij}^{\mathrm{one}} m_{ij}^{\mathrm{two}}\right\rangle\left\langle m_{kl}^{\mathrm{one}} L_{kl} \right\rangle}{\left\langle m_{mn}^{\mathrm{one}} m_{mn}^{\mathrm{one}} \right\rangle\left\langle m_{pq}^{\mathrm{two}} m_{pq}^{\mathrm{two}} \right\rangle - \left\langle m_{mn}^{\mathrm{one}} m_{mn}^{\mathrm{two}} \right\rangle\left\langle m_{pq}^{\mathrm{one}} m_{pq}^{\mathrm{two}} \right\rangle}. \label{defdtw}
\end{equation}
with
$m_{ij}^{\mathrm{one}}(\vv{\f{U}}, \vv{\f{\Delta}}) = \tau_{ij}^{\mathrm{one}}(\vv{\ft{\f{U}}}, \vv{\ft{\f{\Delta}}}) - \FT{\tau_{ij}^{\mathrm{one}}(\vv{\f{U}}, \vv{\f{\Delta}})}$ and
$m_{ij}^{\mathrm{two}}(\vv{\f{U}}, \vv{\f{\Delta}}) = \tau_{ij}^{\mathrm{two}}(\vv{\ft{\f{U}}}, \vv{\ft{\f{\Delta}}}) - \FT{\tau_{ij}^{\mathrm{two}}(\vv{\f{U}}, \vv{\f{\Delta}})}$.
The parameter $C^{\mathrm{one}}$ may be computed from the permutation of the
exponents ``one'' and ``two'' in the above expression.

\item One-parameter dynamic mixed method: The parameter of one of the two models is arbitrarily set,
for instance $C^{\mathrm{one}}$, then the parameter of the
other model is computed dynamically to minimise the variance of the residual
\citep{zang1993dynamic, vreman1994formulation}.
This leads to
\begin{equation}
C^{\mathrm{two}} = \frac{\left\langle m_{ij}^{\mathrm{two}} \left(L_{ij} - C^{\mathrm{one}} m_{ij}^{\mathrm{one}}\right)\right\rangle}{\left\langle m_{mn}^{\mathrm{two}} m_{mn}^{\mathrm{two}} \right\rangle}. \label{label1}
\end{equation}
The parameter of the first model $C^{\mathrm{one}}$ may be set to a constant.
Alternatively, it may be computed with the classical dynamic method,
that is without taking into consideration the second model.  This has been
suggested in order to improve the
two-parameter dynamic procedure \citep{anderson1999effects, morinishi2001recommended}.
\end{itemize}
A generalisation of dynamic mixed models to an arbitrary number of parameters
is given by \citet{sagaut2000general}.

The dynamic procedure may be extended to the construction of a model using
tensorial parameters $C_{ij}^{\mathrm{one}}$ and $C_{ij}^{\mathrm{two}}$,
\begin{equation}
\tau_{ij}^{\mathrm{dyn,one,two}}(\vv{\f{U}}, \vv{\f{\Delta}}) = C_{ij}^{\mathrm{one}} \tau_{ij}^{\mathrm{one}}(\vv{\f{U}}, \vv{\f{\Delta}}) + C_{ij}^{\mathrm{two}} \tau_{ij}^{\mathrm{two}}(\vv{\f{U}}, \vv{\f{\Delta}}),
\end{equation}
where no implicit summations over $i$ and $j$ are assumed.
The dynamic methods (\ref{defdtw}) and (\ref{label1}) can be
extended to tensorial parameters:
\begin{itemize}
\item Tensorial two-parameter dynamic mixed method: As in the (scalar) two-parameter dynamic mixed method, the parameters of the
two models are computed dynamically to minimise for all $i$ and $j$ the
variance of the residual.
This leads to
\begin{equation}
C_{ij}^{\mathrm{two}} = \frac{\left\langle m_{ij}^{\mathrm{one}} m_{ij}^{\mathrm{one}} \right\rangle\left\langle L_{ij} m_{ij}^{\mathrm{two}} \right\rangle - \left\langle m_{ij}^{\mathrm{one}} m_{ij}^{\mathrm{two}}\right\rangle\left\langle m_{ij}^{\mathrm{one}} L_{ij} \right\rangle}{\left\langle m_{ij}^{\mathrm{one}} m_{ij}^{\mathrm{one}} \right\rangle\left\langle m_{ij}^{\mathrm{two}} m_{ij}^{\mathrm{two}} \right\rangle - \left\langle m_{ij}^{\mathrm{one}} m_{ij}^{\mathrm{two}} \right\rangle\left\langle m_{ij}^{\mathrm{one}} m_{ij}^{\mathrm{two}} \right\rangle},
\end{equation}
where no implicit summations over $i$ and $j$ are assumed.
The parameter $C^{\mathrm{one}}$ may be computed from the permutation of the
exponents ``one'' and ``two'' in the above expression.
\item Tensorial one-parameter dynamic mixed method: As in the (scalar) one-parameter dynamic mixed method, the parameters of one
of the two models are arbitrarily set, the parameters of the other model being
computed dynamically to minimise for all $i$ and $j$ the variance of the
residual.
This leads to
\begin{equation}
C_{ij}^{\mathrm{two}} = \frac{\left\langle m_{ij}^{\mathrm{two}} \left(L_{ij} - C^{\mathrm{one}} m_{ij}^{\mathrm{one}}\right)\right\rangle}{\left\langle m_{ij}^{\mathrm{two}} m_{ij}^{\mathrm{two}} \right\rangle},
\end{equation}
where no implicit summations over $i$ and $j$ are assumed.
The parameter of the first model $C^{\mathrm{one}}$ either be set to a
constant or computed using the classical tensorial
dynamic method.
\end{itemize}
For each dynamic procedure, the average $\left\langle \,\cdot\, \right\rangle$
can be computed as a plane average, that is over the homogeneous directions,
or as a global average \citep{park2006dynamic, you2007dynamic, lee2010dynamic, toda2010dynamic, singh2013dynamic}, that is over the volume of the channel.
The parameter of plane-average dynamic procedures is a function of time and
the wall-normal coordinate. The parameter of global-average dynamic
procedures is a function of time.

\section{Numerical study configuration}\label{label29}

\begin{table*}
\centerline{\begin{tabular}{llllll}
$Re_{\tau}$ & Name                 &\multicolumn{1}{c}{ Number of grid points                }&\multicolumn{1}{c}{ Dimension of the domain                          }&\multicolumn{1}{c}{ Cell sizes in wall units}&\multicolumn{1}{c}{ Mesh dilatation parameter}     \\
            &                          &\multicolumn{1}{c}{ $N_x\times N_y\times N_z$            }&\multicolumn{1}{c}{ $L_x\times L_y\times L_z\phantom{(/3)}$        }&\multicolumn{1}{c}{ $\phantom{0}\Delta_x^+\hfill;\hfill\Delta_y^+(0)$\hfill--\hfill$\Delta_y^+(h)\hfill;\hfill\Delta_z^+$}&\multicolumn{1}{c}{$a$}\\[.5em]
\hline\\[-.5em]
180 & 48B   &\multicolumn{1}{c}{ $48\times50\times48$    }&\multicolumn{1}{c}{ $4\pi h\times2 h\times2\pi h\phantom{(/3)}$ }   &\multicolumn{1}{c}{ $\phantom{0.}\numprint{68}\hfill;\hfill\phantom{0}\numprint{0.50}$\hfill--\hfill$\numprint{25}\phantom{.}\hfill;\hfill\numprint{34}\phantom{.}$ }&\multicolumn{1}{c}{$0.981$} \\
180 & 36C   &\multicolumn{1}{c}{ $36\times40\times36$    }&\multicolumn{1}{c}{ $4\pi h\times2 h\times2\pi h\phantom{(/3)}$ }   &\multicolumn{1}{c}{ $\phantom{0.}\numprint{91}\hfill;\hfill\phantom{00}\numprint{2.0}$\hfill--\hfill$\numprint{22}\phantom{.}\hfill;\hfill\numprint{45}\phantom{.}$ }&\multicolumn{1}{c}{$0.910$} \\
180 & 24C   &\multicolumn{1}{c}{ $24\times28\times24$    }&\multicolumn{1}{c}{ $4\pi h\times2 h\times2\pi h\phantom{(/3)}$ }   &\multicolumn{1}{c}{ $\phantom{.}\numprint{136}\hfill;\hfill\phantom{00}\numprint{2.0}$\hfill--\hfill$\numprint{35}\phantom{.}\hfill;\hfill\numprint{68}\phantom{.}$ }&\multicolumn{1}{c}{$0.949$}\\[.5em]
180 & DNS   &\multicolumn{1}{c}{ $384\times266\times384$ }&\multicolumn{1}{c}{ $4\pi h\times2 h\times2\pi h\phantom{(/3)}$ }   &\multicolumn{1}{c}{ $\phantom{0}\numprint{5.8}\hfill;\hfill\numprint{0.085}$\hfill--\hfill$\numprint{2.9}\hfill;\hfill\numprint{2.9}$ }&\multicolumn{1}{c}{$0.971$}
\end{tabular}}
\caption[Computational domain and grid spacing.]{Computational domain and grid spacing of the DNS mesh and the three LES meshes.
The cell sizes in wall units are computed using the friction velocity of the direct numerical simulation.
\label{label11}}
\end{table*}

\subsection{Channel flow configuration}\label{label28}

We investigate the large-eddy simulation of a fully developed three-dimensional
turbulent channel flow.
The channel is periodic in the streamwise ($x$) and
spanwise ($z$) directions and enclosed by two plane walls in the wall-normal
direction ($y$).
The flow is isothermal and incompressible.
The mean friction
Reynolds number is $Re_{\tau} = 180$.
The domain size is $4\pi h \times 2h \times 2 \pi h$.
To analyse the results, we use a direct numerical simulation of the same
channel presented in \citet{dupuy2018turbulence}, and validated against the
reference data of \citet{moser1999a, bolotnov2010, vreman2014comparison} and \citet{lee2015direct}.

\subsection{Numerical settings}

The channel flow presented in section \ref{label28} is simulated using
three meshes, referred to as ``48B'', ``36C'' and
``24C''.
The meshes are rectilinear. The grid spacing is uniform in the homogeneous directions ($x$
and $z$) and follows a hyperbolic
tangent law in the wall-normal coordinate direction ($y$),
\begin{equation}
y_k = L_y \left( 1 + \frac{1}{a} \tanh\left[ \left(\frac{k-1}{N_y-1} - 1\right)\tanh^{-1}(a)\right] \right),
\end{equation}
with $a$ the mesh dilatation parameter and $N_y$ the number of grid points
in the wall-normal direction.
The domain size and grid spacing of the simulations are given in table \ref{label11}.
We use a finite difference method in a staggered grid system \citep{morinishi1998fully, nicoud2000conservative}
with a fourth-order centred momentum convection scheme,
a second-order centred diffusion scheme
and a third-order Runge--Kutta time scheme \citep{williamson1980low}.
If present, the eddy viscosity is computed and discretised at the center of
control volumes in the same way as molecular viscosity and
the operators $\partial_j (\mu g_{ij})$ and $\partial_j (\mu g_{ji})$
are discretised directly using second-order centred schemes.
The simulations are performed using the TrioCFD software \citep{calvin2002object}.
This software has been used in many numerical simulations of fluid flows \citep{toutant2013, aulery2016, dupuy2019effect}.

The direct numerical simulations use the same numerical method as the
large-eddy simulations and have the same domain size.

\subsection{Filtering process}

In order to allow the direct comparison of the results of the large-eddy
simulations and of the direct numerical simulations, we filter the
instantaneous DNS data at the resolution
of the LES meshes.
We use a top-hat filter to perform this filtering.
To carry out the box filter, we first interpolate the DNS data using a cubic
spline then compute the filter from the interpolated data, as in \citep{dupuy2018study, dupuy2018apriori}.
The cubic spline interpolation allows the computation of the filter with an
arbitrary filter length and without mesh restrictions.
The spline interpolation adds an additional filtering to the box filter.
However, this additional filter is small compared to the box filter with the
DNS mesh used and can be neglected.

Filtering is also required to compute the test filter involved in some
subgrid-scale models.
These filters are computed using other methods because the spline interpolation
is too computationally expensive to be used in a large-eddy simulation.
The test filter of dynamic methods, referred to as ``filter A'' is computed as
an average over three cells in the three directions.
This approximates a top-hat filter whose width is thrice as large as the LES
mesh.
The test filter of the scale-similarity model has been computed using
the filter A and another filter.
The second filter, referred to as ``filter T'', uses the Taylor
series expansion of the box filter
using the local cell size as the filter width.

\section{Results and discussion}\label{label31}

The large-eddy simulations are used to study the relevance of
models for the subgrid-scale tensor.
Before proceeding to the comparison of the subgrid-scale models, we briefly
discuss the simulation of the channel without subgrid-scale model.
To analyse the results, we use two types of scaling.
With the wall scaling ($^+$), all quantities are scaled using a combination of
the friction velocity $U_{\tau}$ and the kinematic viscosity $\nu$,
$y^+ = y U_{\tau}/\nu$ and
$\vv{U}^+ = \vv{U}/U_{\tau}$.
With the scaling ($^\circ$), all quantities are scaled using a combination of
the channel half-height and the kinematic viscosity $\nu$,
$y^\circ = y/h$ and
$\vv{U}^\circ = \vv{U} h/\nu$.

\subsection{Simulation without subgrid-scale models}\label{secgeneresulles}

\begin{figure*}
\setcounter{subfigcounter}{0}
\centerline{
\subfigtopleft{\includegraphics[width=0.44\textwidth, trim={0 5 5 5}, clip]{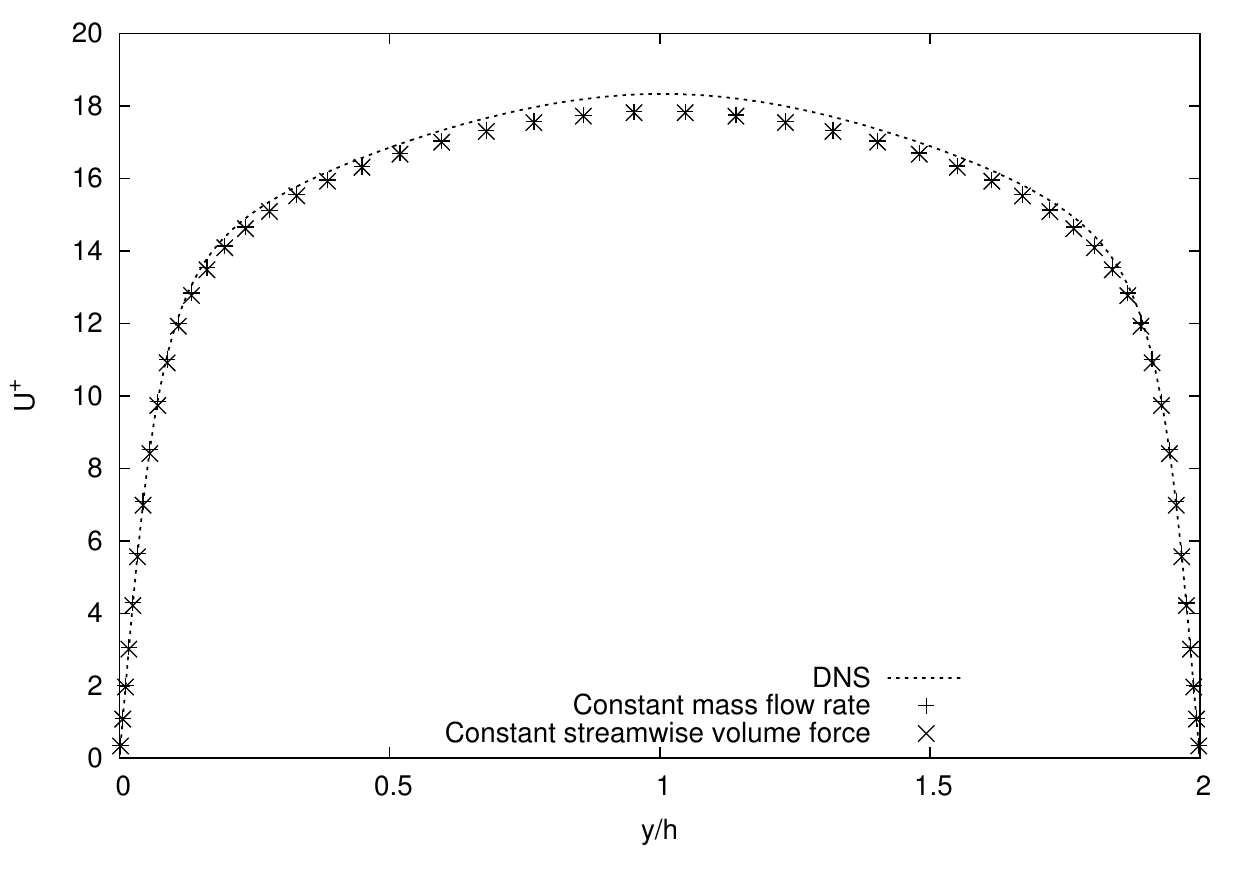}}
\subfigtopleft{\includegraphics[width=0.44\textwidth, trim={0 5 5 5}, clip]{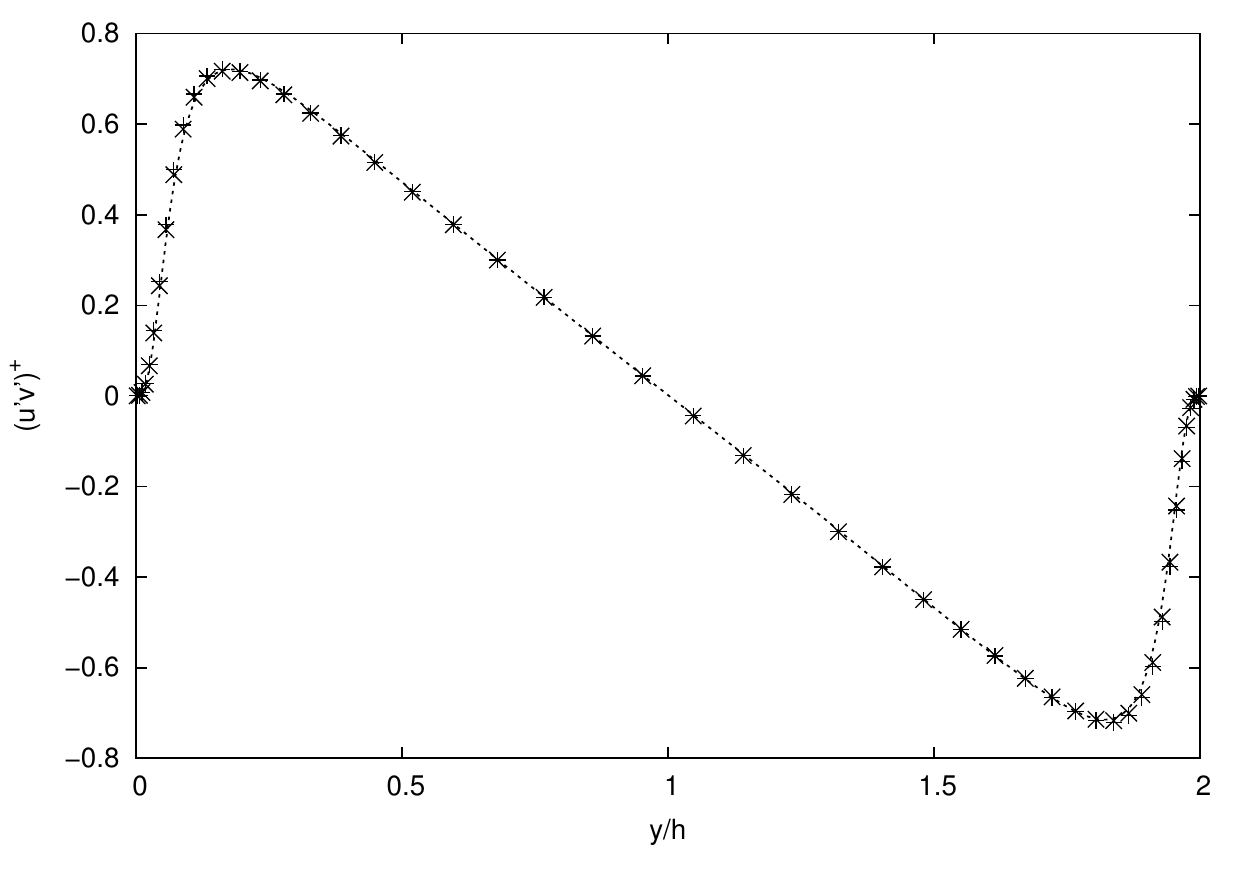}}
}
\caption[Comparison of simulations with no subgrid-scale model with constant mass flow rate and constant streamwise volume force.]{
Comparison of simulations
with no subgrid-scale model
with constant mass flow rate and constant streamwise volume force
for the profiles of the mean streamwise velocity $\left\langle U_x \right\rangle$ (left) and the covariance of streamwise and wall-normal velocity $\left\langle u_{\smash[b]{x}}' u_{\smash[b]{y}}' \right\rangle$ (right)
with the mesh 48B.
\label{label15}}
\end{figure*}

Simulations without subgrid-scale model are carried out
with the meshes 24C, 36C and 48B.
These simulations aim to provide reference data without any subgrid-scale modelling.
Although no explicit subgrid-scale models are used, the approach differs from
implicit large-eddy simulation, which is usually based on the use
of the dissipative properties of specific numerical schemes to model the
small-scale motions \citep{sagaut98}, as our numerical method is
intended to cause minimal dissipation.
The mass flow rate of the simulations is imposed using a control loop to
adjust the streamwise volume force $f$.
The targeted mass flow rate is the same as in the direct numerical
simulations.
Accordingly, the simulations have the same mass flow rate than the
direct numerical simulations but predict a different wall shear stress.
With the mesh 48B, the error on the friction velocity is 2\%.
Imposing a constant streamwise volume force would maintain the wall shear stress at the
same level as the direct numerical simulations, but results in an error
of 2\% on the mass flow rate.
The results of simulations with constant mass flow rate and constant
streamwise volume force are compared
in figure \ref{label15}.
The scaling of the profiles takes into account the differences of mass flow
rate. Nevertheless, the two approaches are not completely equivalent because
the Reynolds number differences between the two methods may induce low
Reynolds number effects.

\begin{figure*}
\setcounter{subfigcounter}{0}
\centerline{
\subfigtopleft{\includegraphics[width=0.44\textwidth, trim={0 5 5 5}, clip]{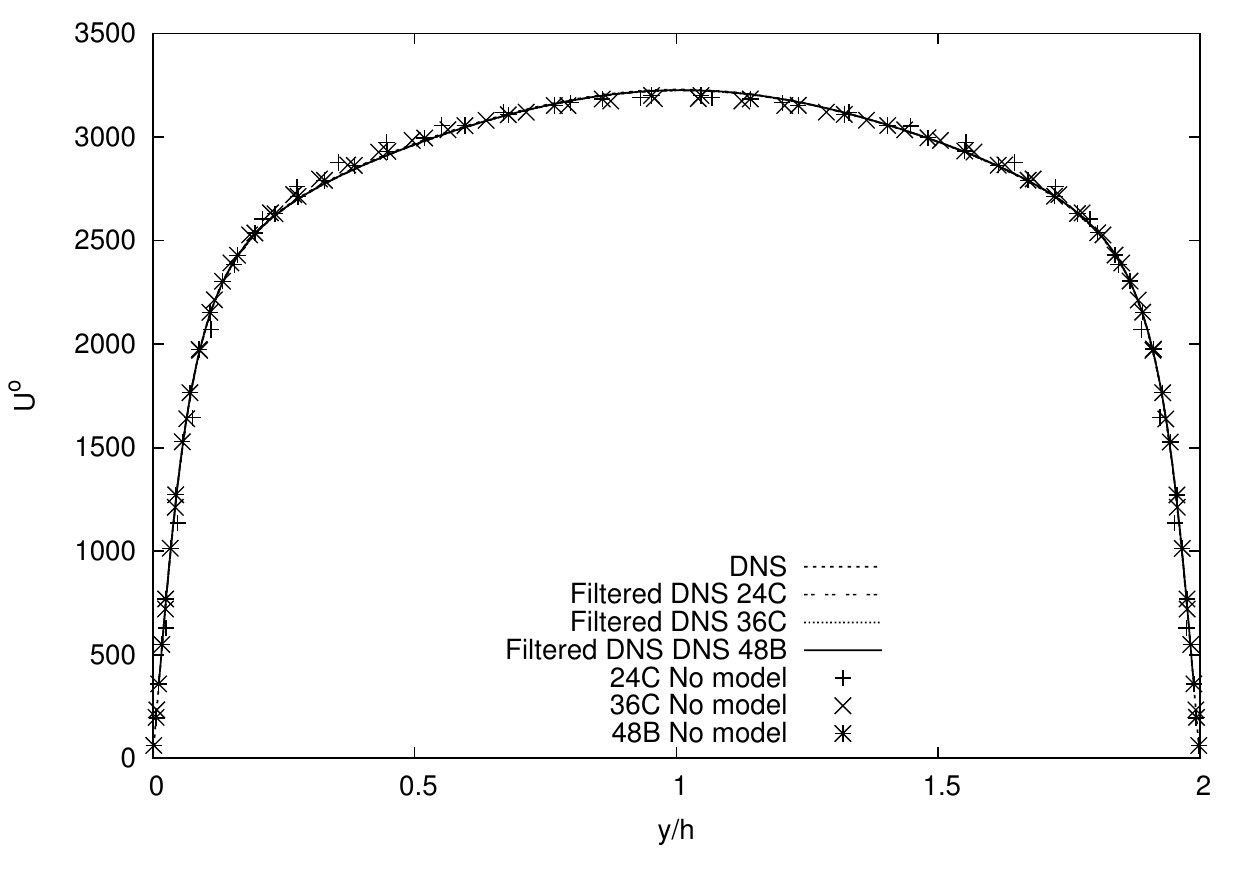}}
\subfigtopleft{\includegraphics[width=0.44\textwidth, trim={0 5 5 5}, clip]{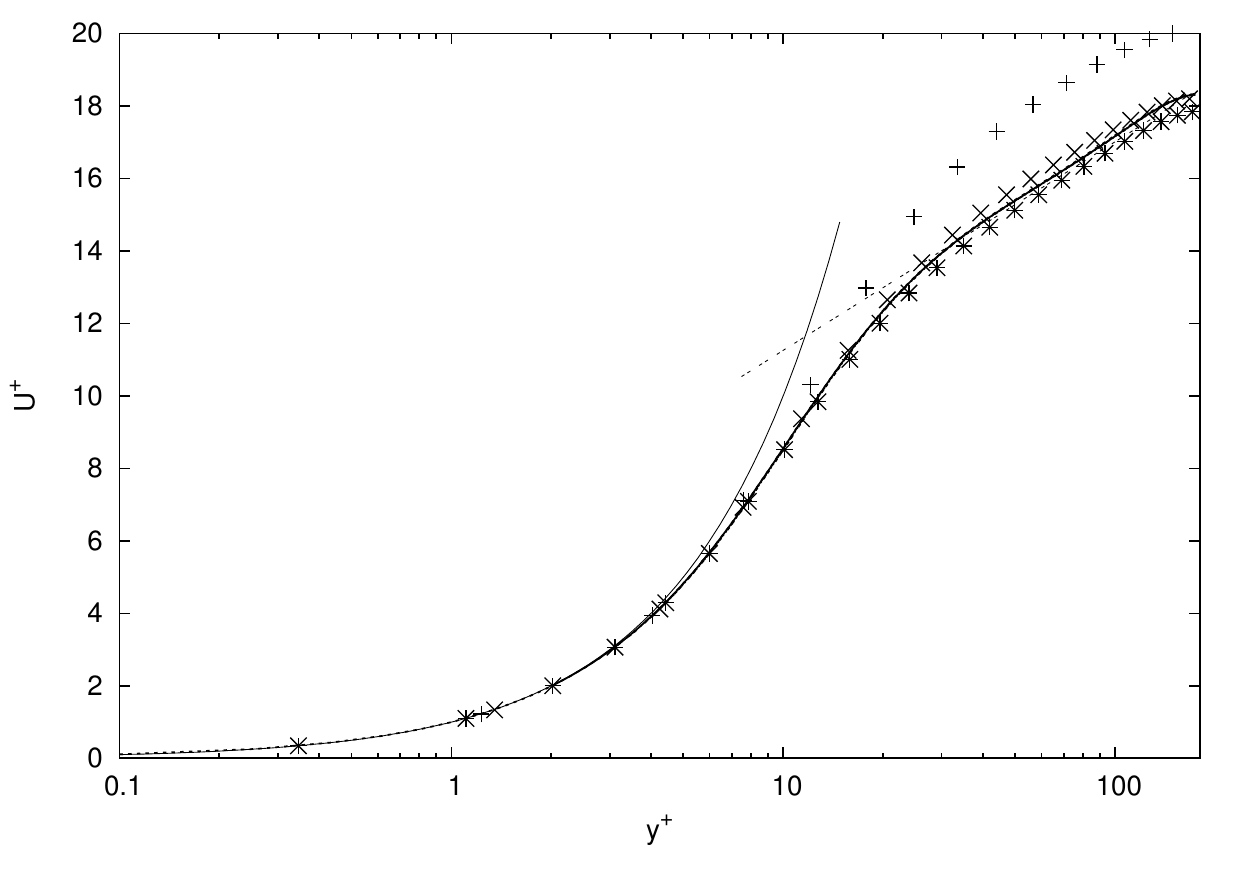}}
}\centerline{
\subfigtopleft{\includegraphics[width=0.44\textwidth, trim={0 5 5 5}, clip]{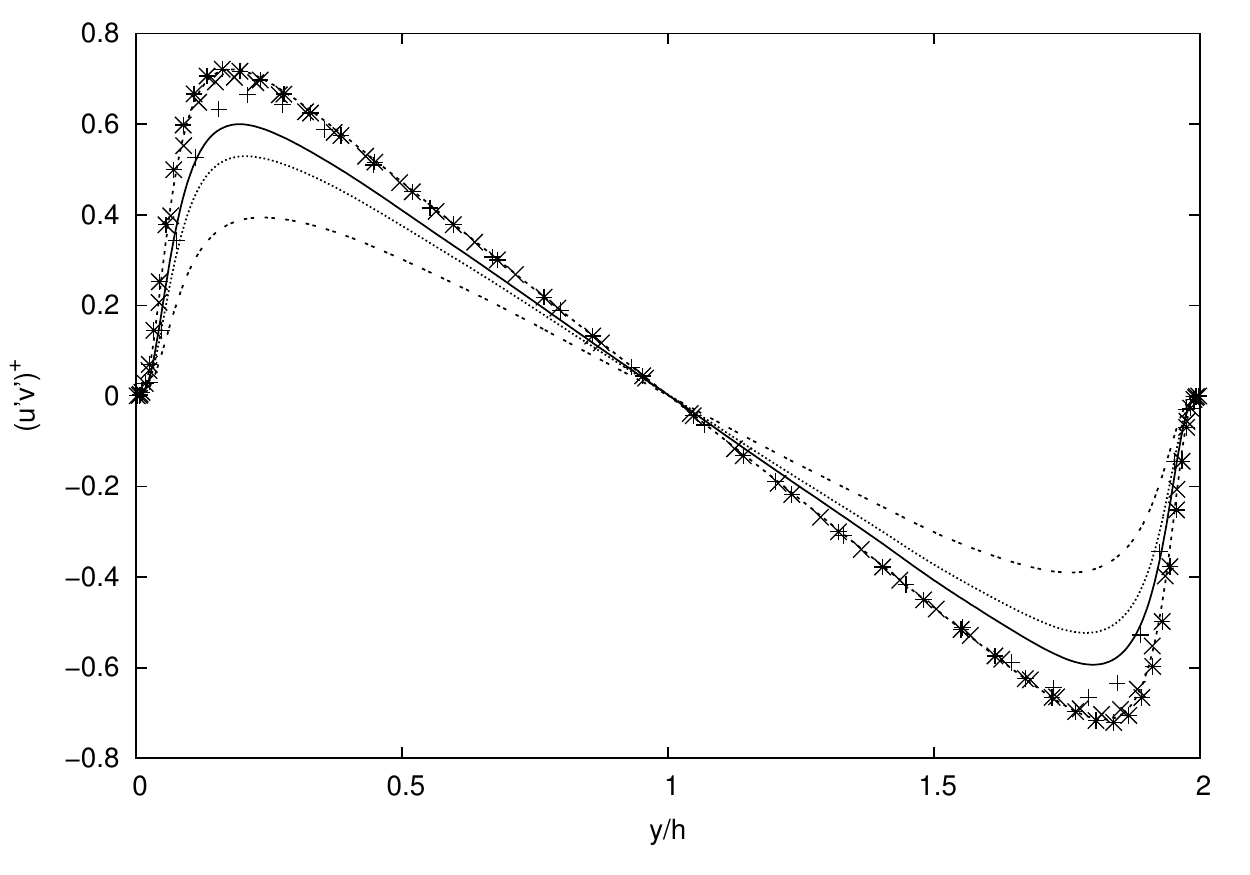}}
\subfigtopleft{\includegraphics[width=0.44\textwidth, trim={0 5 5 5}, clip]{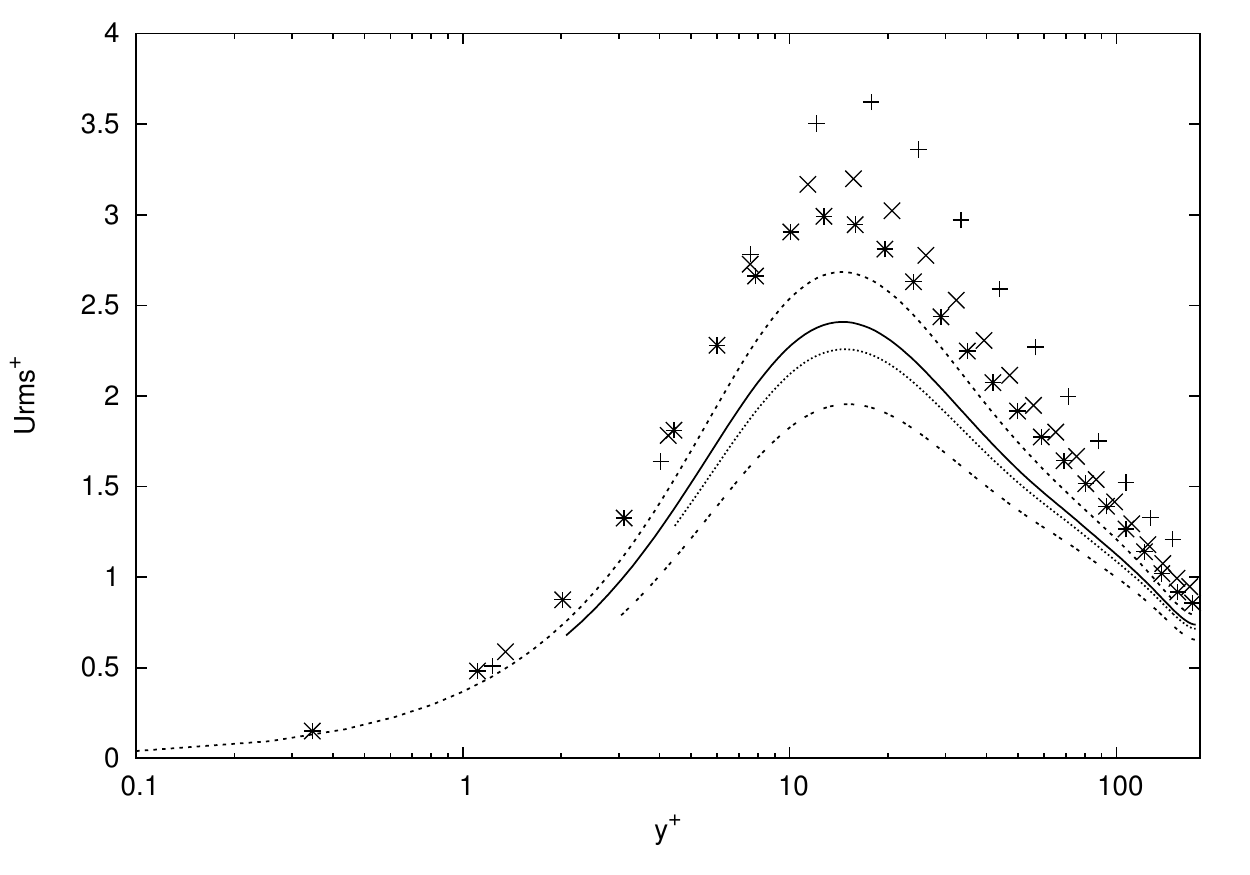}}
}\centerline{
\subfigtopleft{\includegraphics[width=0.44\textwidth, trim={0 5 5 5}, clip]{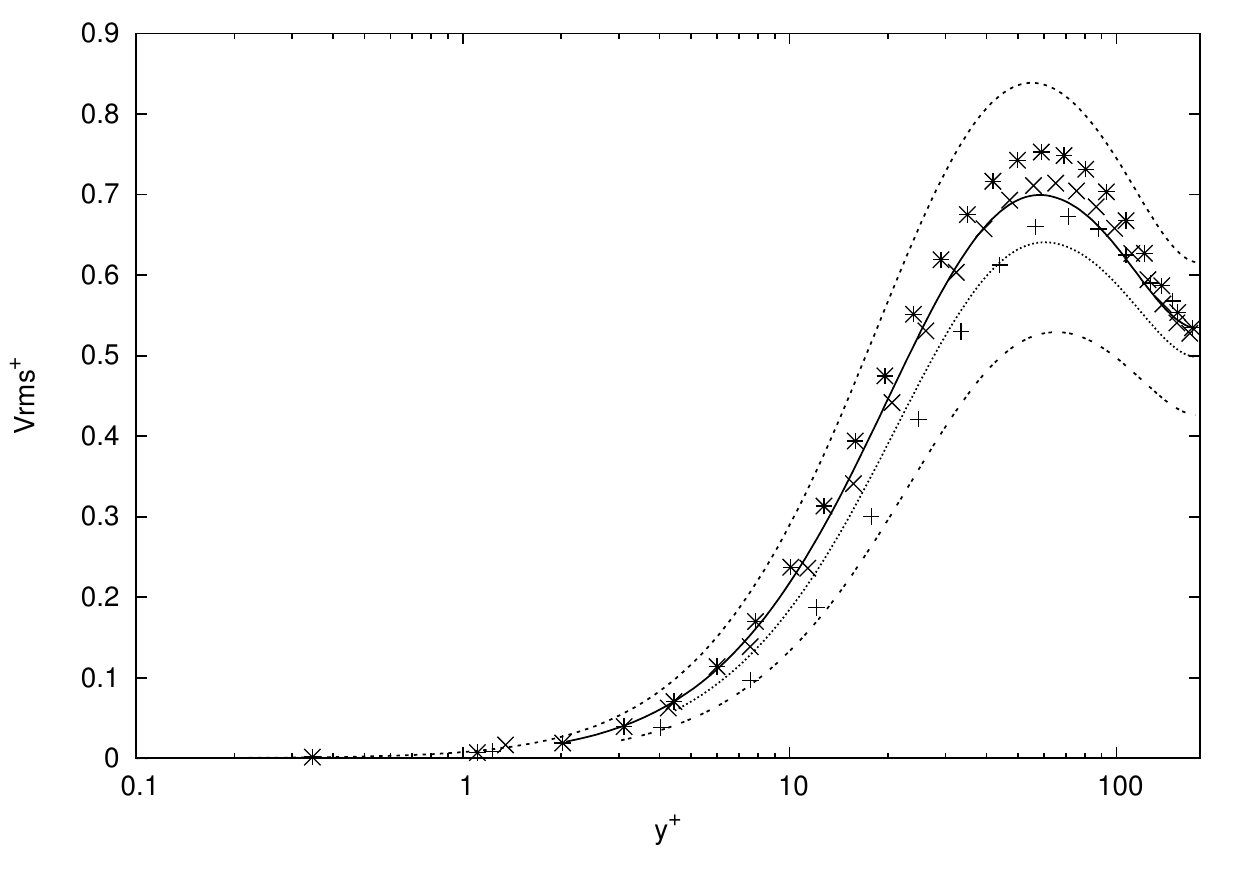}}
\subfigtopleft{\includegraphics[width=0.44\textwidth, trim={0 5 5 5}, clip]{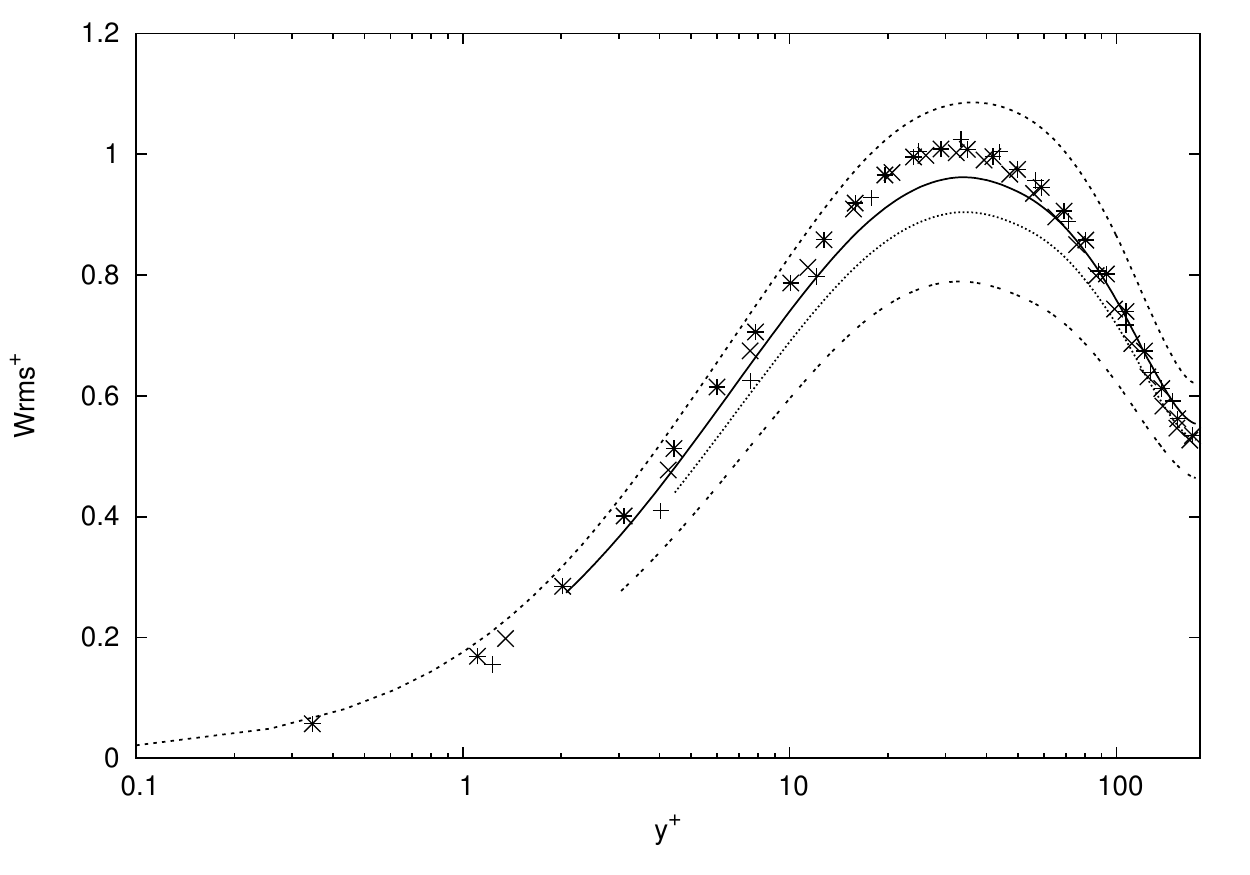}}
}
\caption[Comparison of simulations with no subgrid-scale model with the meshes 24C, 36C and 48B.]{
Comparison of simulations
with no subgrid-scale model
with the meshes 24C, 36C and 48B
for the profiles of the mean streamwise velocity $\left\langle U_x \right\rangle$ (a, b), the covariance of streamwise and wall-normal velocity $\left\langle u_{\smash[b]{x}}' u_{\smash[b]{y}}' \right\rangle$ (c), the standard deviation of streamwise velocity $\smash[t]{\sqrt{\left\langle u_{\smash[b]{x}}'^2 \right\rangle}}$ (d), wall-normal velocity $\smash[t]{\sqrt{\left\langle u_{\smash[b]{y}}'^2 \right\rangle}}$ (e) and spanwise velocity $\smash[t]{\sqrt{\left\langle u_{\smash[b]{z}}'^2 \right\rangle}}$ (f).
\label{label12}}
\end{figure*}

The results of the simulations without subgrid-scale model
with the meshes 24C, 36C and 48B are compared
in figure \ref{label12}.
The profiles of the turbulence statistics
are compared to direct numerical simulations filtered at the resolution of the
simulation meshes.
The mean nonfiltered and filtered streamwise
velocity are almost identical, because the filtered field has sufficient spectral content \citep{winckelmans2002comparison}.
Nevertheless, the simulation with the mesh 24C underestimates
significantly the friction velocity, and thus the mean streamwise velocity near the wall.
The mean streamwise velocity is
without scaling satisfactory at the center of the channel for all
simulations (figure \ref{label12}).

The filtering of the DNS data decreases significantly
the maximum value of the covariance of streamwise and wall-normal velocity
and the standard deviation of velocity components.
The decrease is larger for a larger filter width.
The decrease ranges from around 10\%
with the mesh 48B to around 30\% with the mesh 24C.
However, the simulations without model lead with the three meshes to a
similar covariance of streamwise and wall-normal velocity and standard
deviation of spanwise velocity, while the standard deviation of streamwise
velocity increases with mesh derefinement (figure \ref{label12}).
The interpretation of these results should take into account the effect of the
classical scaling, as the underestimation of the wall shear stress in the
coarser simulations offsets a slight decrease of the covariance of streamwise
and wall-normal velocity and the standard deviation of spanwise velocity
without scaling.

The error on the friction velocity is
9\% with the mesh 24C,
6\% with the mesh 36C and
2\% with the mesh 48B.
The relative accuracy of the wall shear stress with the mesh 48B is partly
due to its non-monotonous convergence of the prediction with mesh refinement.
As identified by \citet{meyers2007plane}, the non-monotonous convergence
allows the existence of a grid-resolution line where the error on the wall
shear stress is zero.
The simulation of the channel with a finer $72\times68\times72$ mesh
leads to an error of 4\% for the wall shear stress.
This is less accurate than with the mesh 48B,
confirming that the mesh 48B is close to Meyers' no error line.
Due to this non-monotonous convergence of the wall shear stress and the
turbulence statistics, it is important to verify the robustness of the
subgrid-scale models to a range of grid
resolutions.

Thus, while the mean streamwise velocity is fairly well represented by the
simulations without model, more complex turbulence statistics, such
as the Reynolds stresses, are less accurate.
In the following, we will study the simulation
of the channel with subgrid-scale models,
that is its large-eddy simulation, and examine whether the addition of
a subgrid-scale model can improve these results.
To study the modelling of the subgrid-scale tensor, we carry out
large-eddy simulations with several functional
models, structural models, tensorial models and tensorial mixed models.

\subsection{Functional modelling}

\begin{figure*}
\setcounter{subfigcounter}{0}
\centerline{
\subfigtopleft{\includegraphics[width=0.44\textwidth, trim={0 5 5 5}, clip]{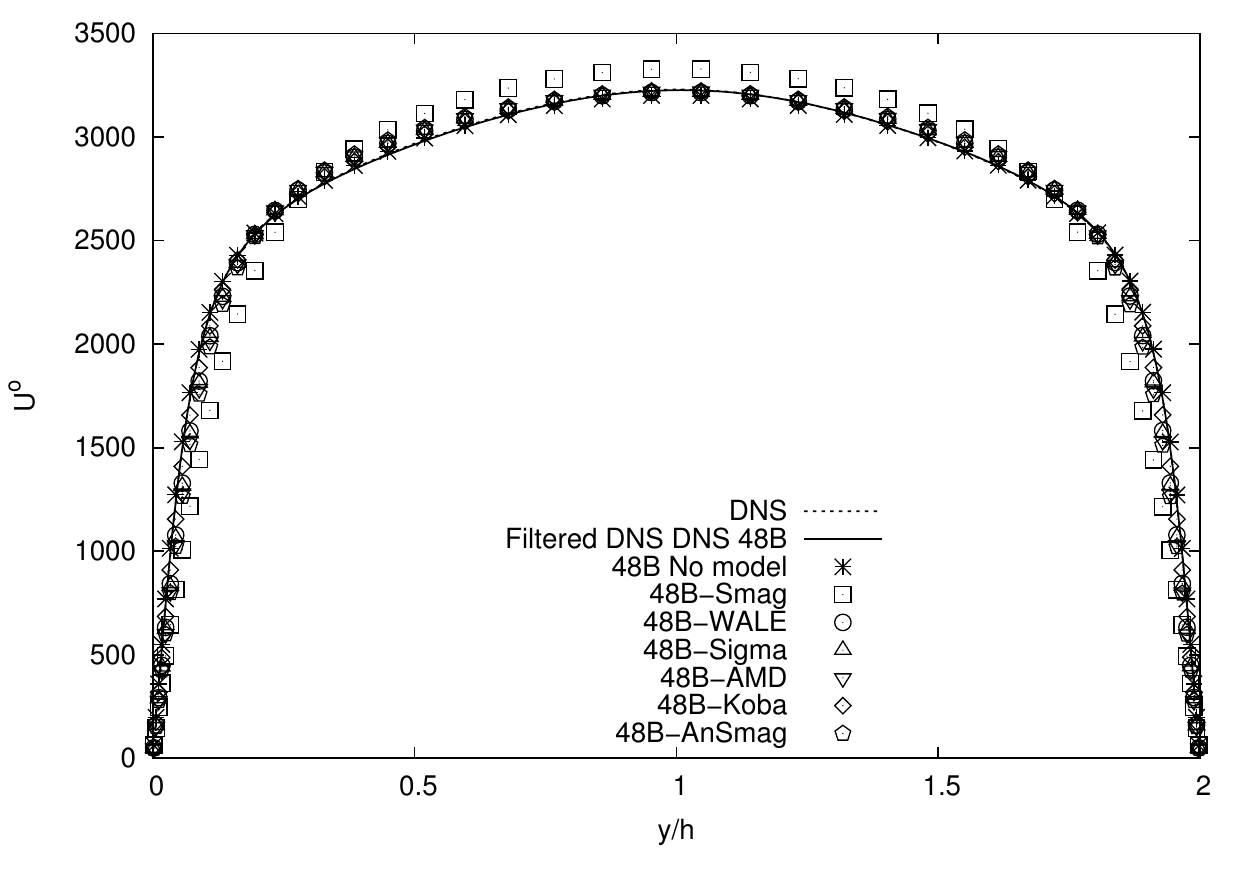}}
\subfigtopleft{\includegraphics[width=0.44\textwidth, trim={0 5 5 5}, clip]{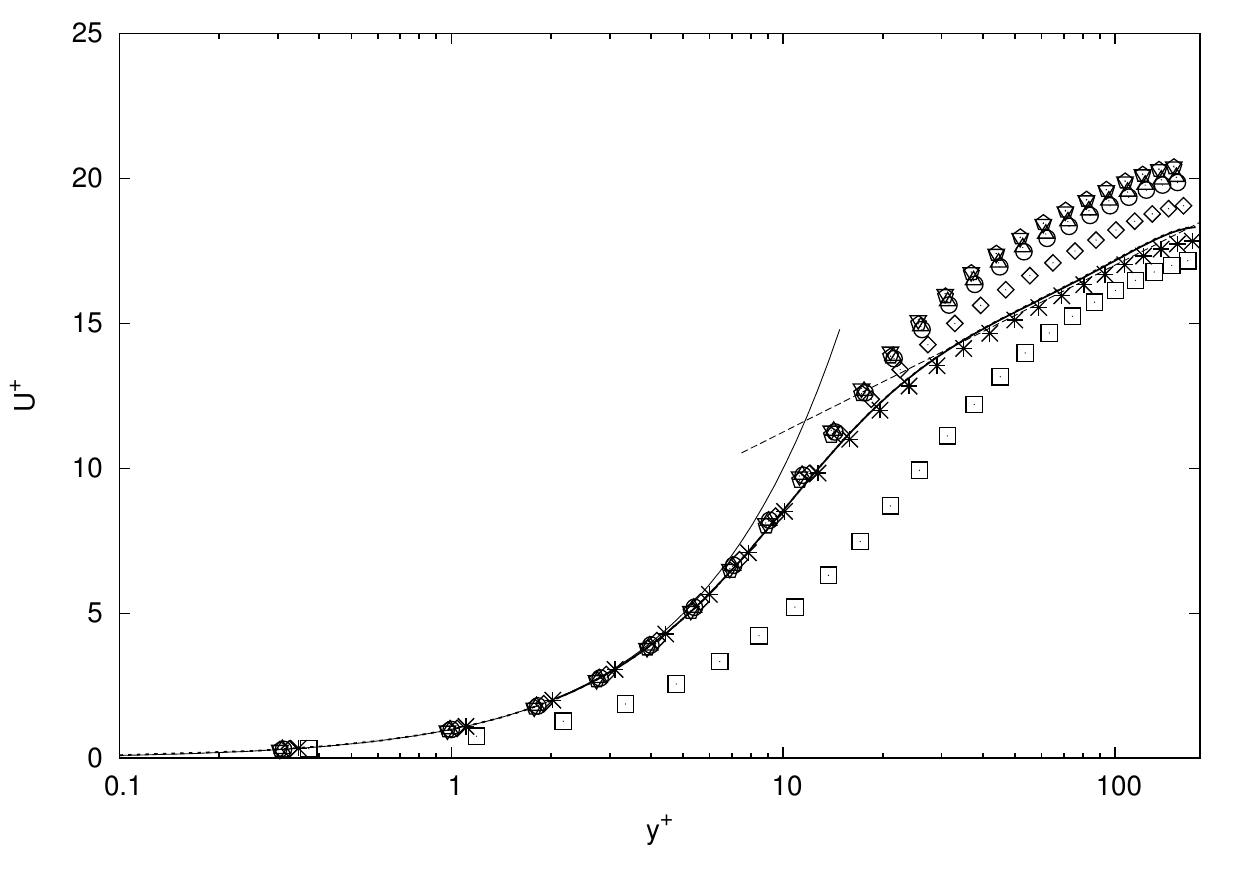}}
}\centerline{
\subfigtopleft{\includegraphics[width=0.44\textwidth, trim={0 5 5 5}, clip]{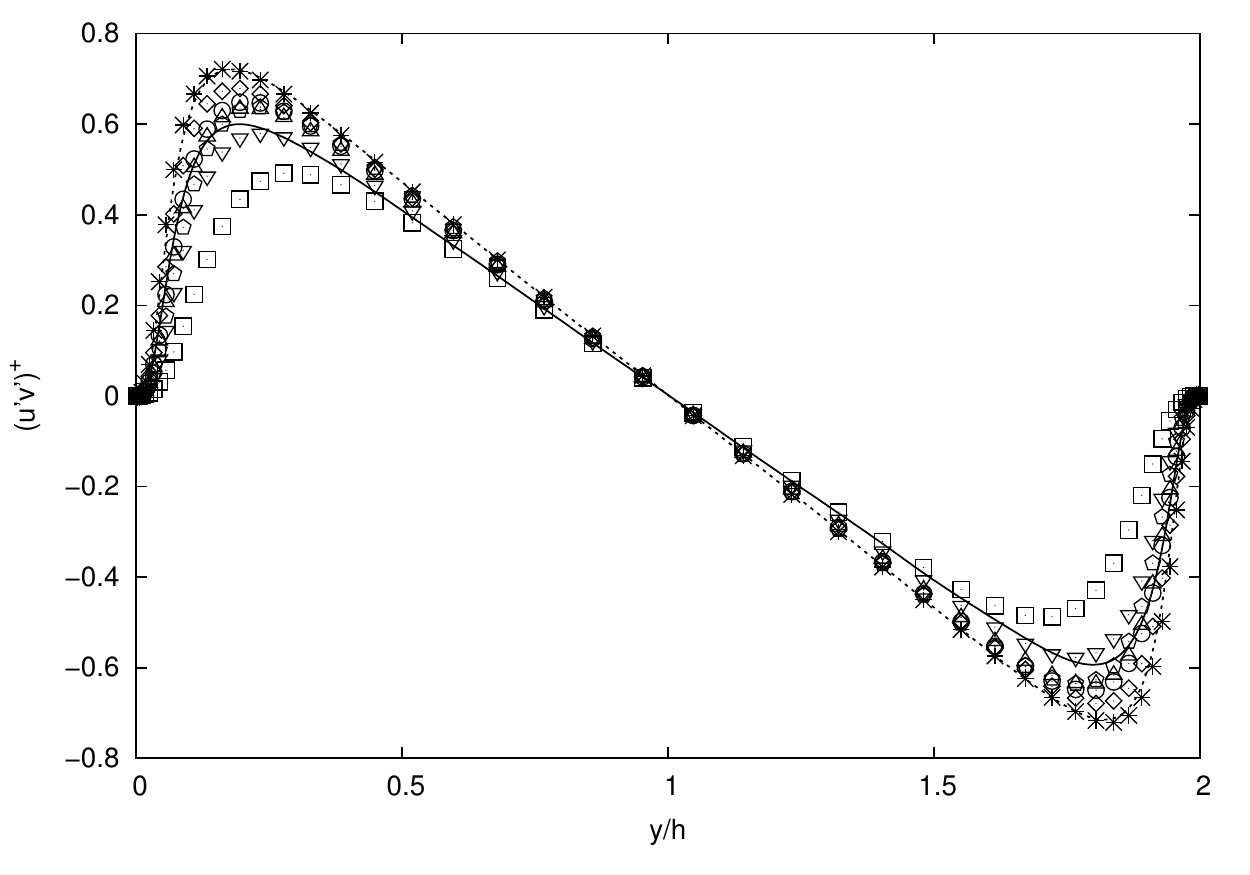}}
\subfigtopleft{\includegraphics[width=0.44\textwidth, trim={0 5 5 5}, clip]{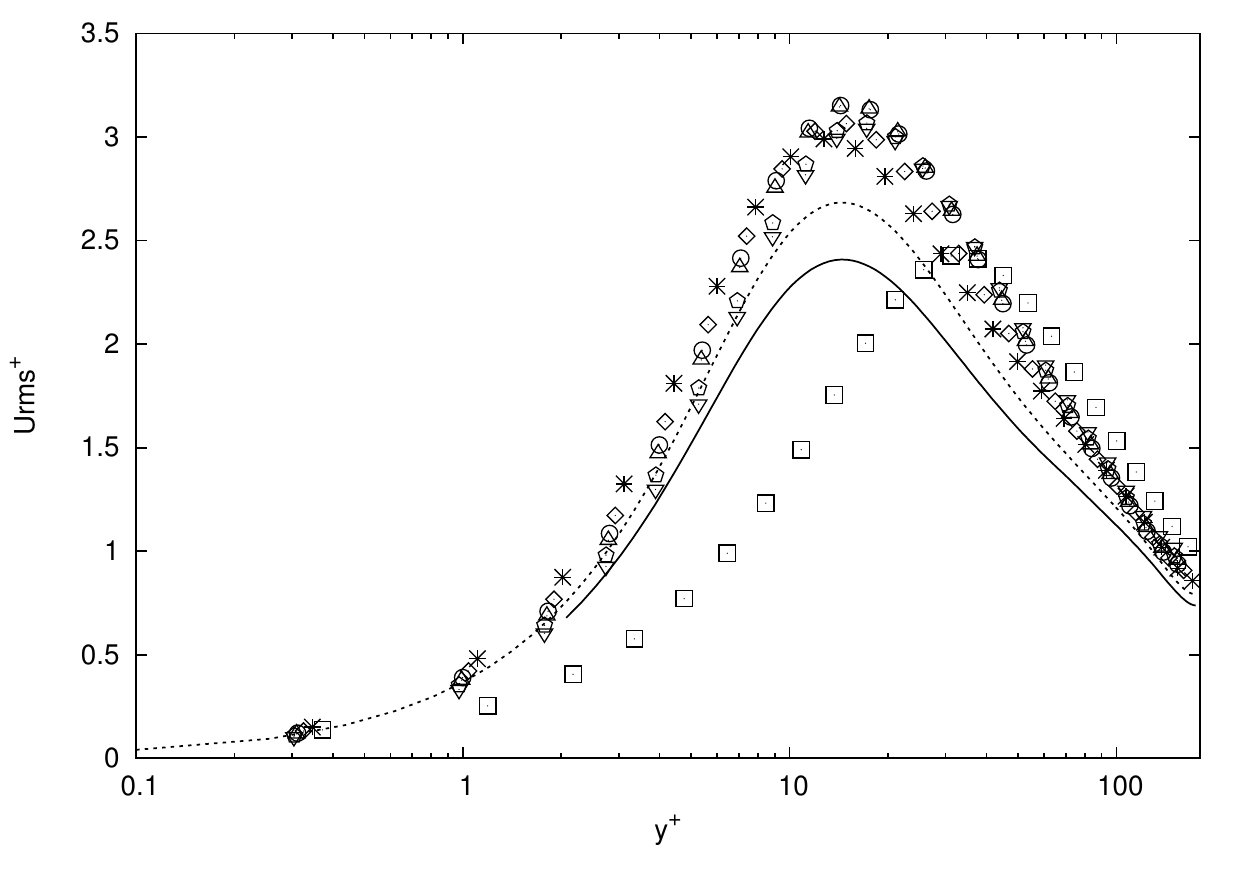}}
}\centerline{
\subfigtopleft{\includegraphics[width=0.44\textwidth, trim={0 5 5 5}, clip]{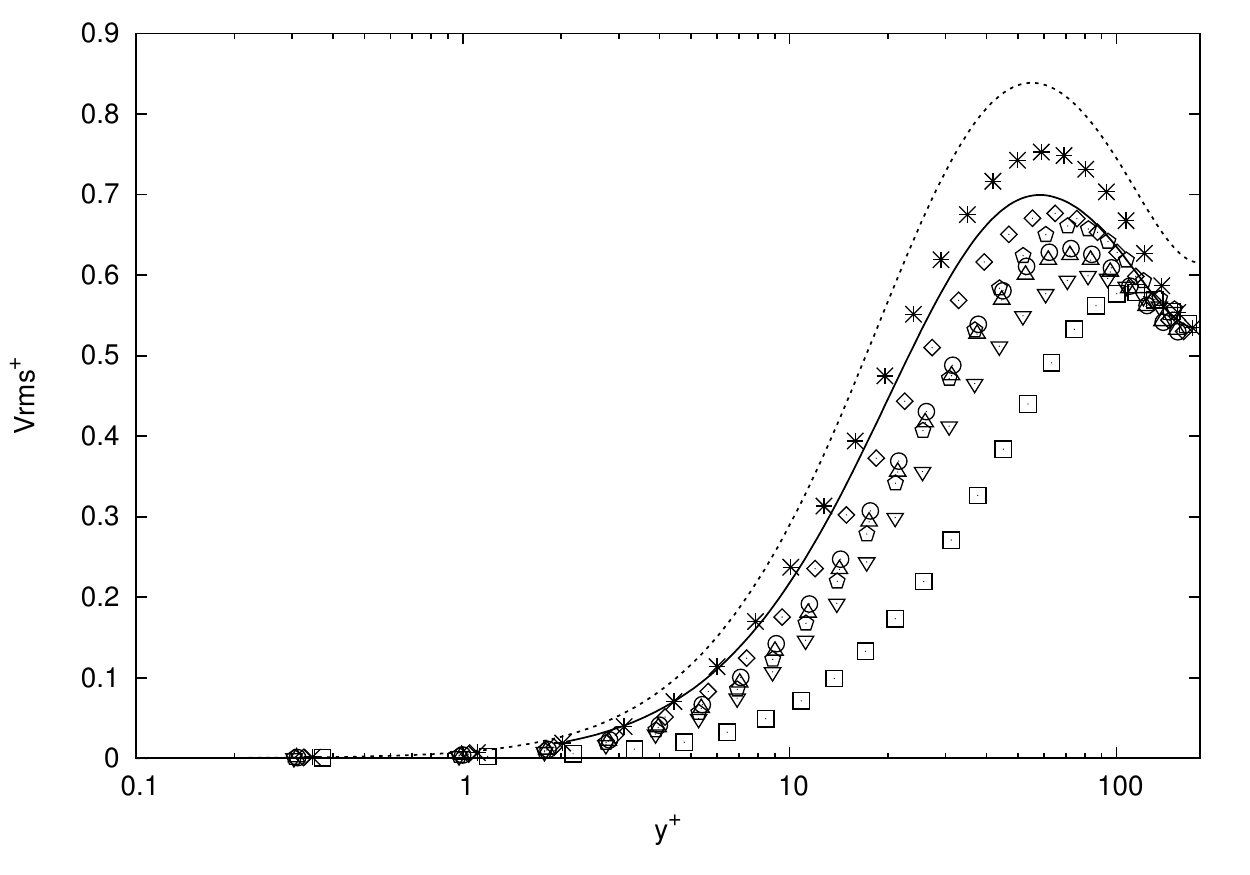}}
\subfigtopleft{\includegraphics[width=0.44\textwidth, trim={0 5 5 5}, clip]{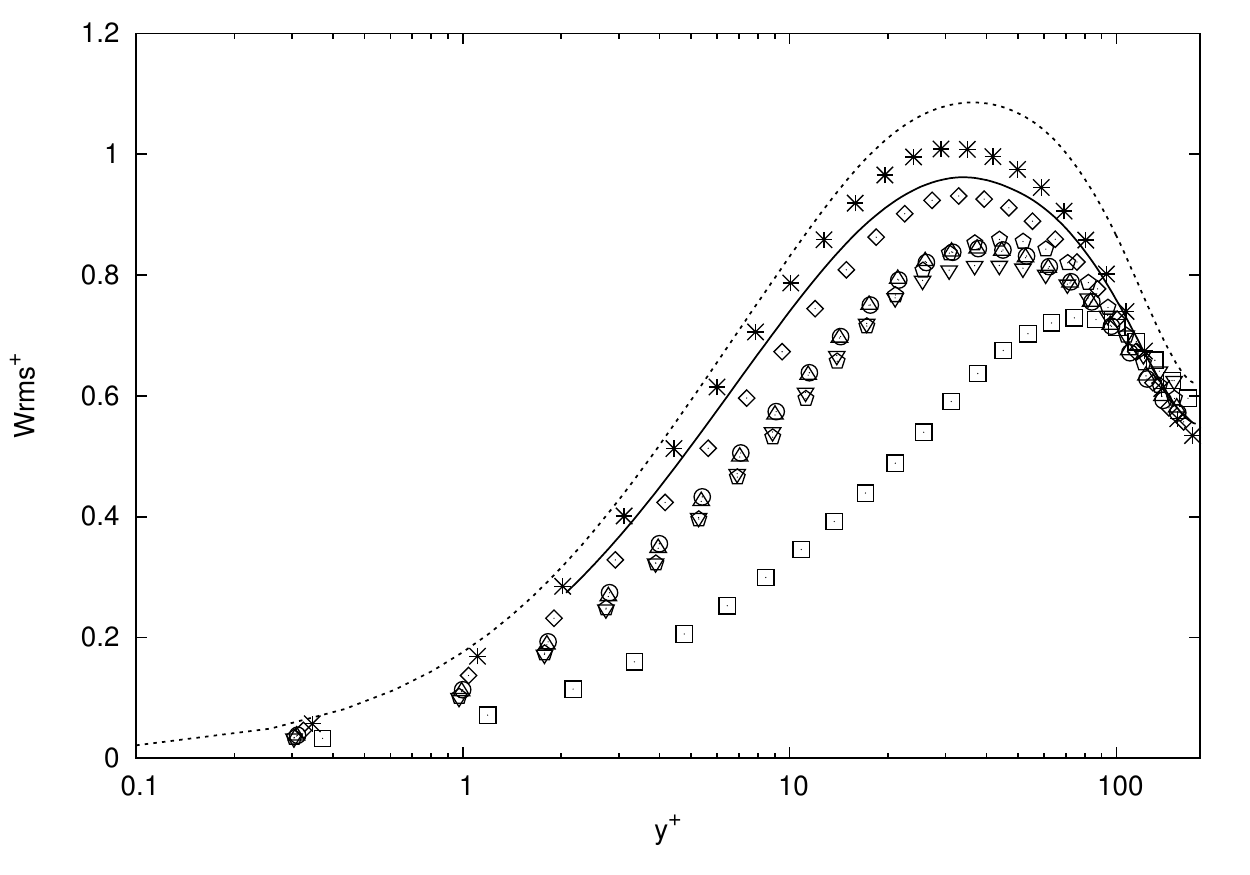}}
}
\caption[Comparison of large-eddy simulations with the Smagorinsky, WALE, Sigma, AMD, Kobayashi and Anisotropic Smagorinsky models.]{
Comparison of large-eddy simulations
with the Smagorinsky, WALE, Sigma, AMD, Kobayashi and Anisotropic Smagorinsky models
for the profiles of the mean streamwise velocity $\left\langle U_x \right\rangle$ (a, b), the covariance of streamwise and wall-normal velocity $\left\langle u_{\smash[b]{x}}' u_{\smash[b]{y}}' \right\rangle$ (c), the standard deviation of streamwise velocity $\smash[t]{\sqrt{\left\langle u_{\smash[b]{x}}'^2 \right\rangle}}$ (d), wall-normal velocity $\sqrt{\left\langle u_{\smash[b]{y}}'^2 \right\rangle}$ (e) and spanwise velocity $\sqrt{\left\langle u_{\smash[b]{z}}'^2 \right\rangle}$ (f)
with the mesh 48B.
\label{label22}}
\end{figure*}

\begin{figure*}
\setcounter{subfigcounter}{0}
\centerline{
\subfigtopleft{\includegraphics[width=0.44\textwidth, trim={0 5 5 5}, clip]{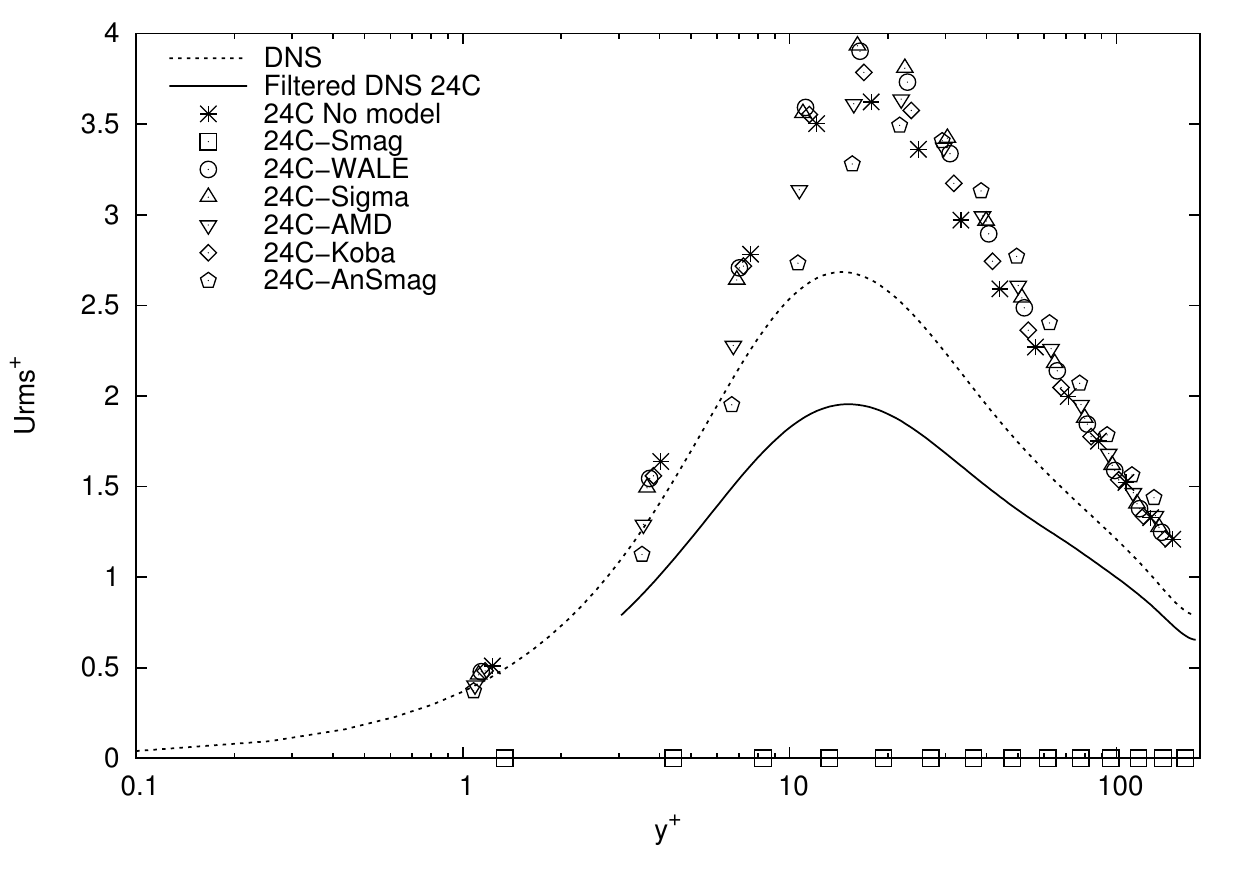}}
\subfigtopleft{\includegraphics[width=0.44\textwidth, trim={0 5 5 5}, clip]{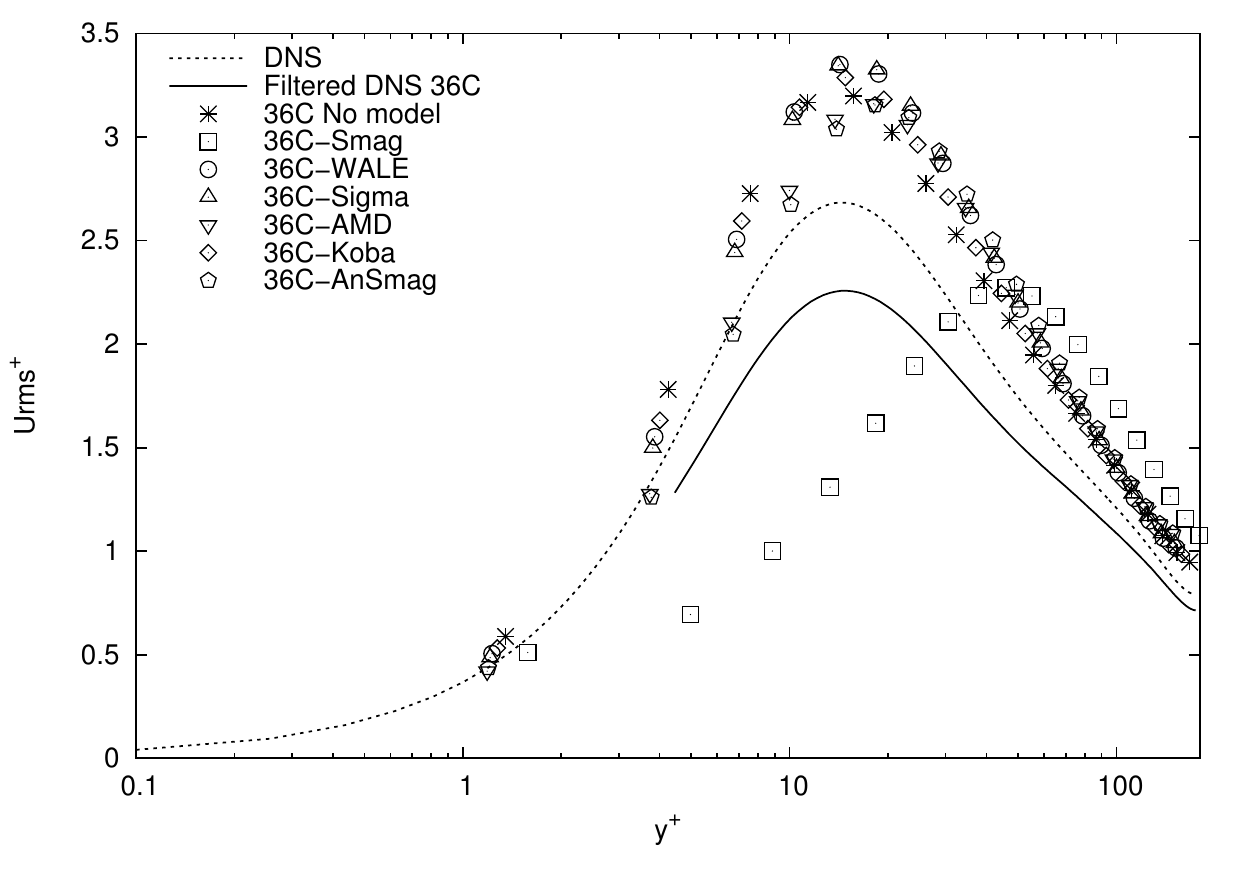}}
}\centerline{
\subfigtopleft{\includegraphics[width=0.44\textwidth, trim={0 5 5 5}, clip]{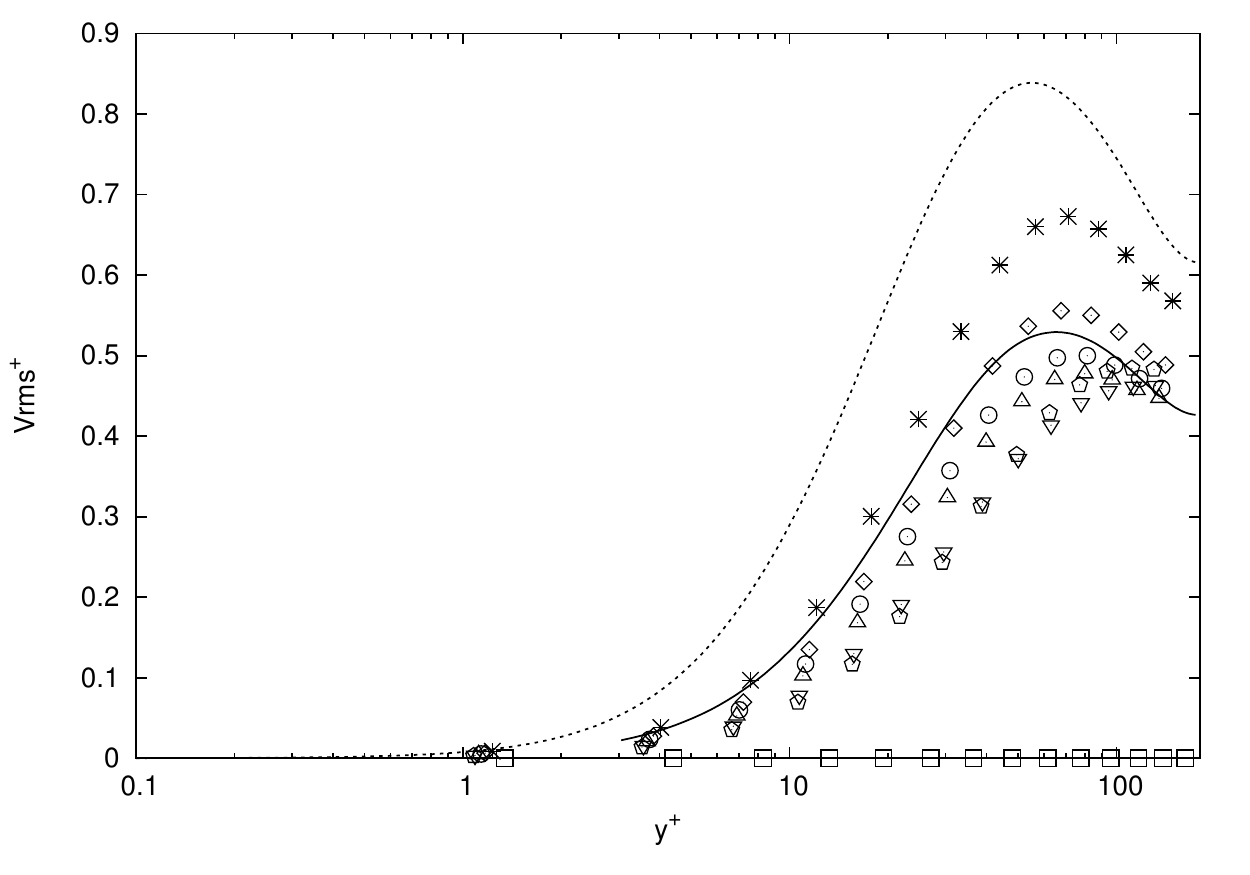}}
\subfigtopleft{\includegraphics[width=0.44\textwidth, trim={0 5 5 5}, clip]{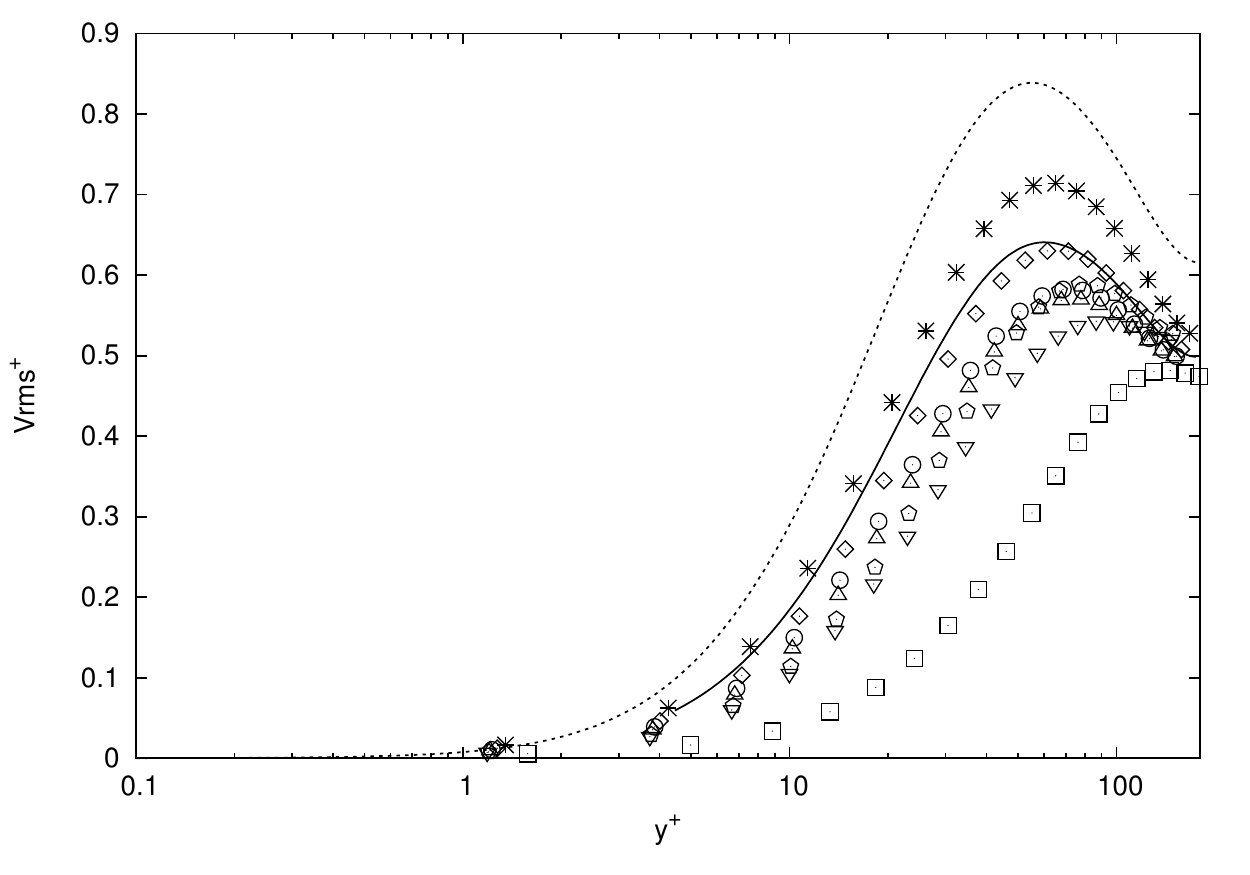}}
}\centerline{
\subfigtopleft{\includegraphics[width=0.44\textwidth, trim={0 5 5 5}, clip]{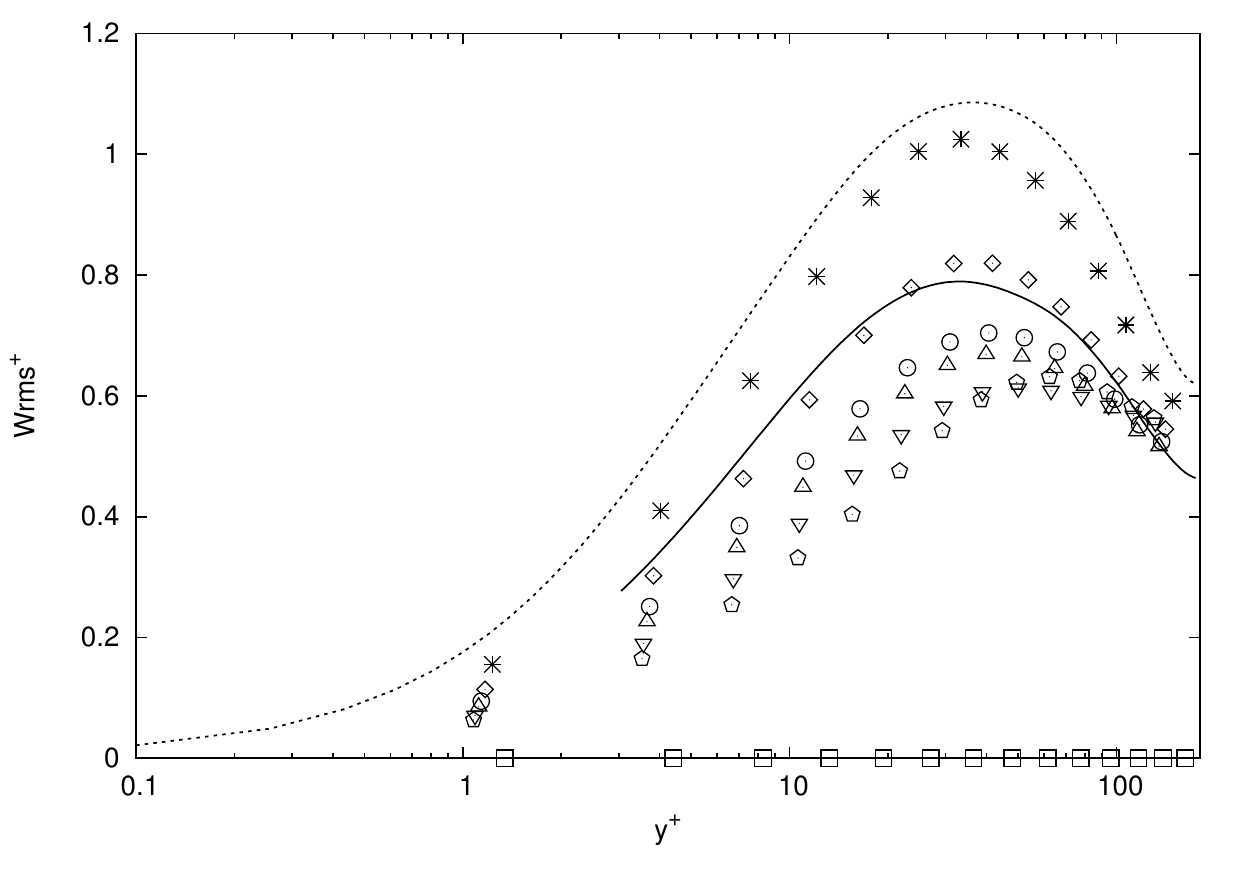}}
\subfigtopleft{\includegraphics[width=0.44\textwidth, trim={0 5 5 5}, clip]{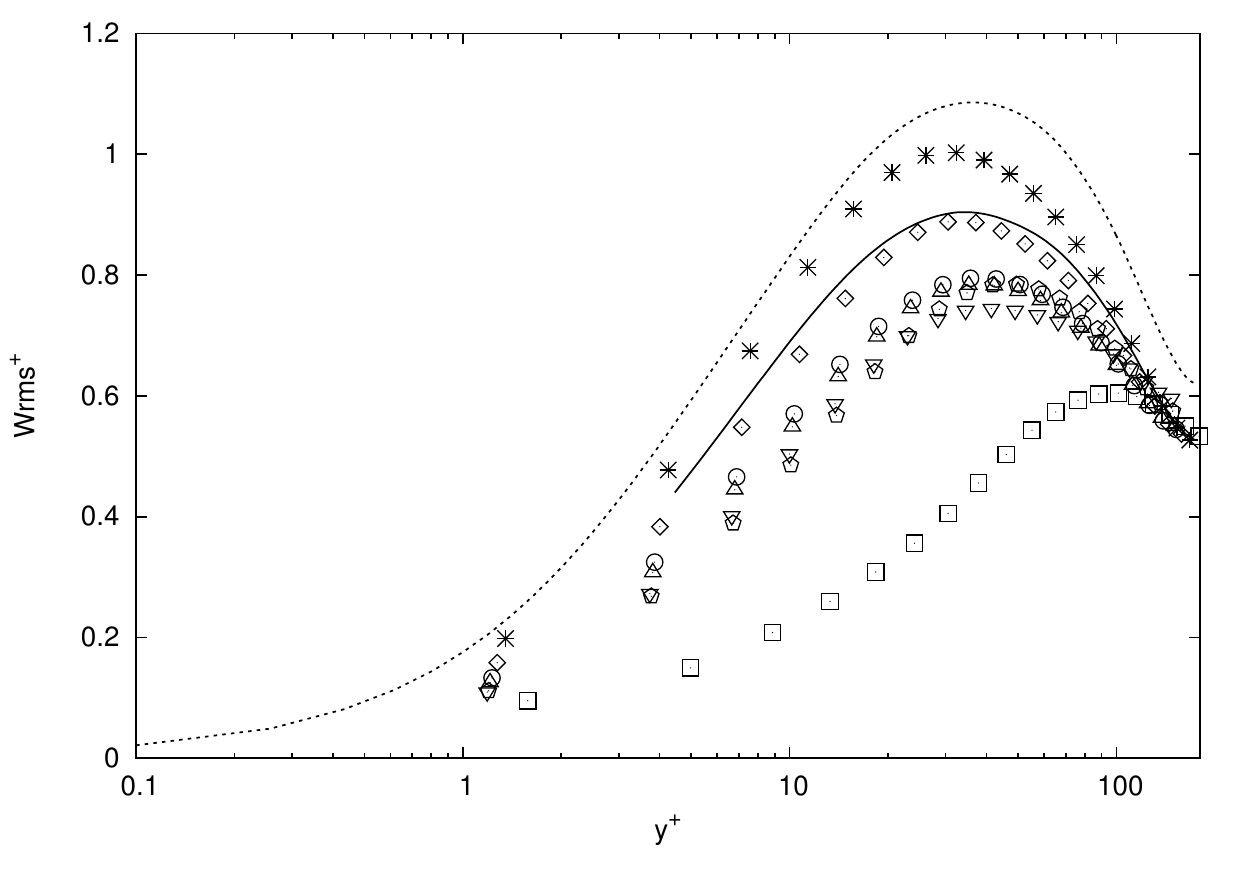}}
}
\caption[Comparison of large-eddy simulations with the Smagorinsky, WALE, Sigma, AMD, Kobayashi and Anisotropic Smagorinsky models with the meshes 24C and 36C.]{
Comparison of large-eddy simulations
with the Smagorinsky, WALE, Sigma, AMD, Kobayashi and Anisotropic Smagorinsky models
with the meshes 24C (left) and 36C (right)
for the profiles of the standard deviation of streamwise velocity $\smash[t]{\sqrt{\left\langle u_{\smash[b]{x}}'^2 \right\rangle}}$ (a, b), wall-normal velocity $\sqrt{\left\langle u_{\smash[b]{y}}'^2 \right\rangle}$ (c, d) and spanwise velocity $\sqrt{\left\langle u_{\smash[b]{z}}'^2 \right\rangle}$ (e, f).
\label{label16}}
\end{figure*}

\begin{figure*}
\setcounter{subfigcounter}{0}
\centerline{
\subfigtopleft{\includegraphics[width=0.44\textwidth, trim={0 5 5 5}, clip]{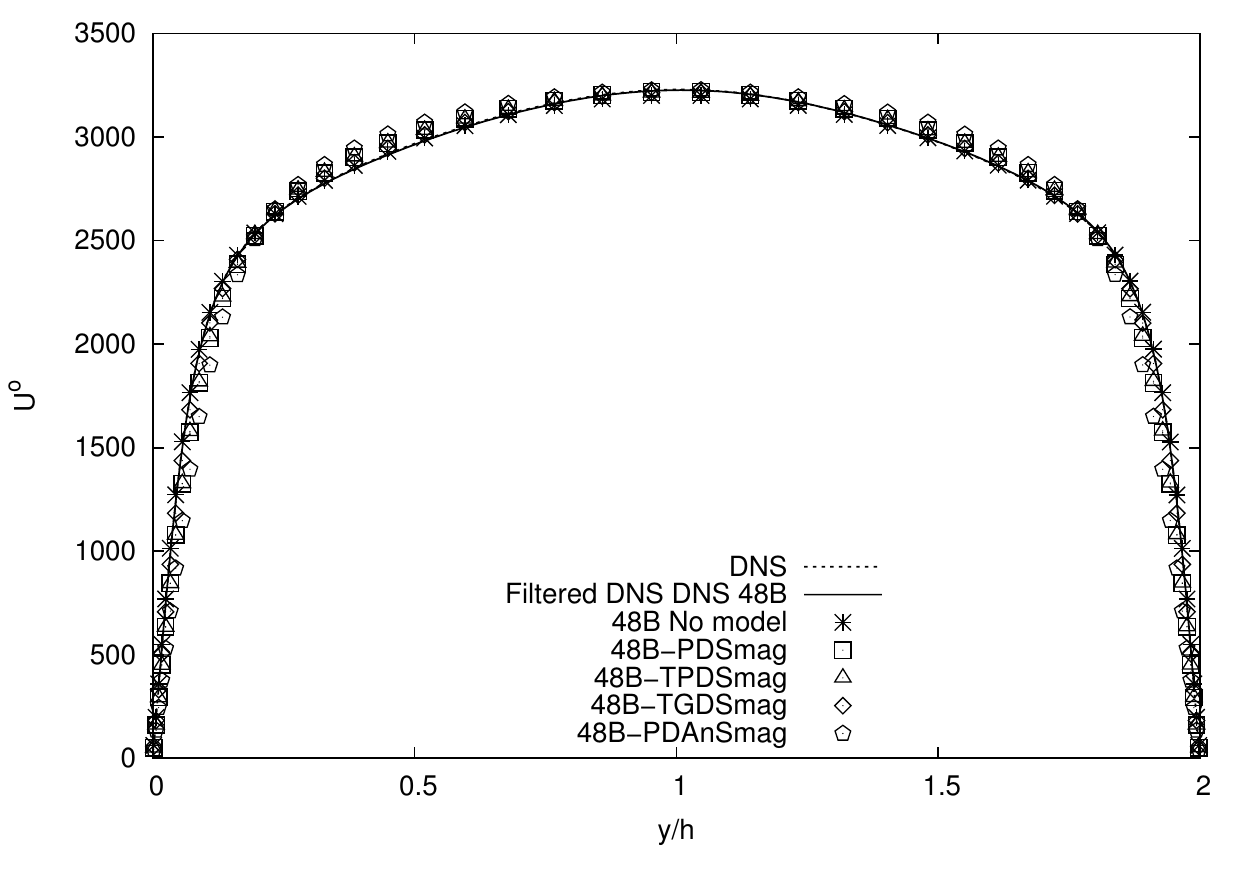}}
\subfigtopleft{\includegraphics[width=0.44\textwidth, trim={0 5 5 5}, clip]{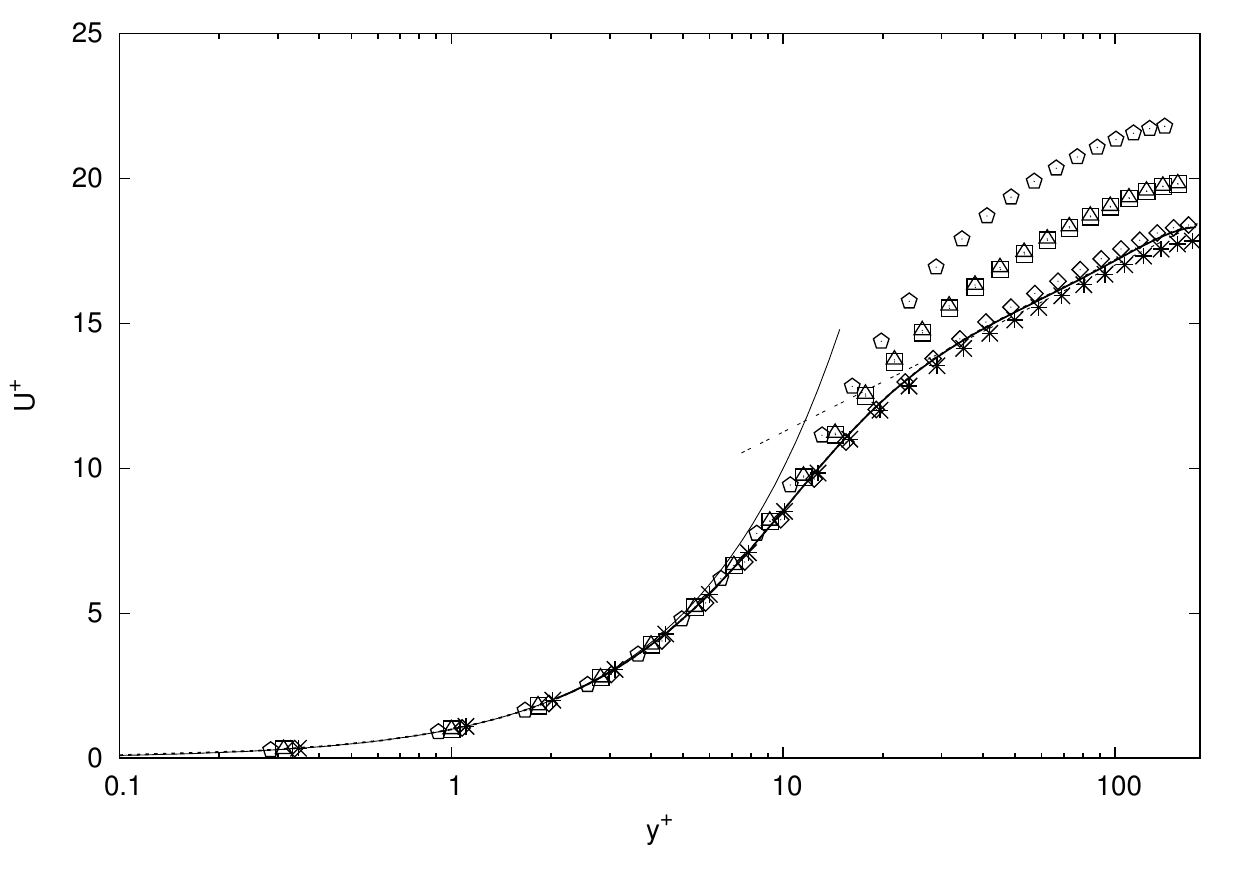}}
}\centerline{
\subfigtopleft{\includegraphics[width=0.44\textwidth, trim={0 5 5 5}, clip]{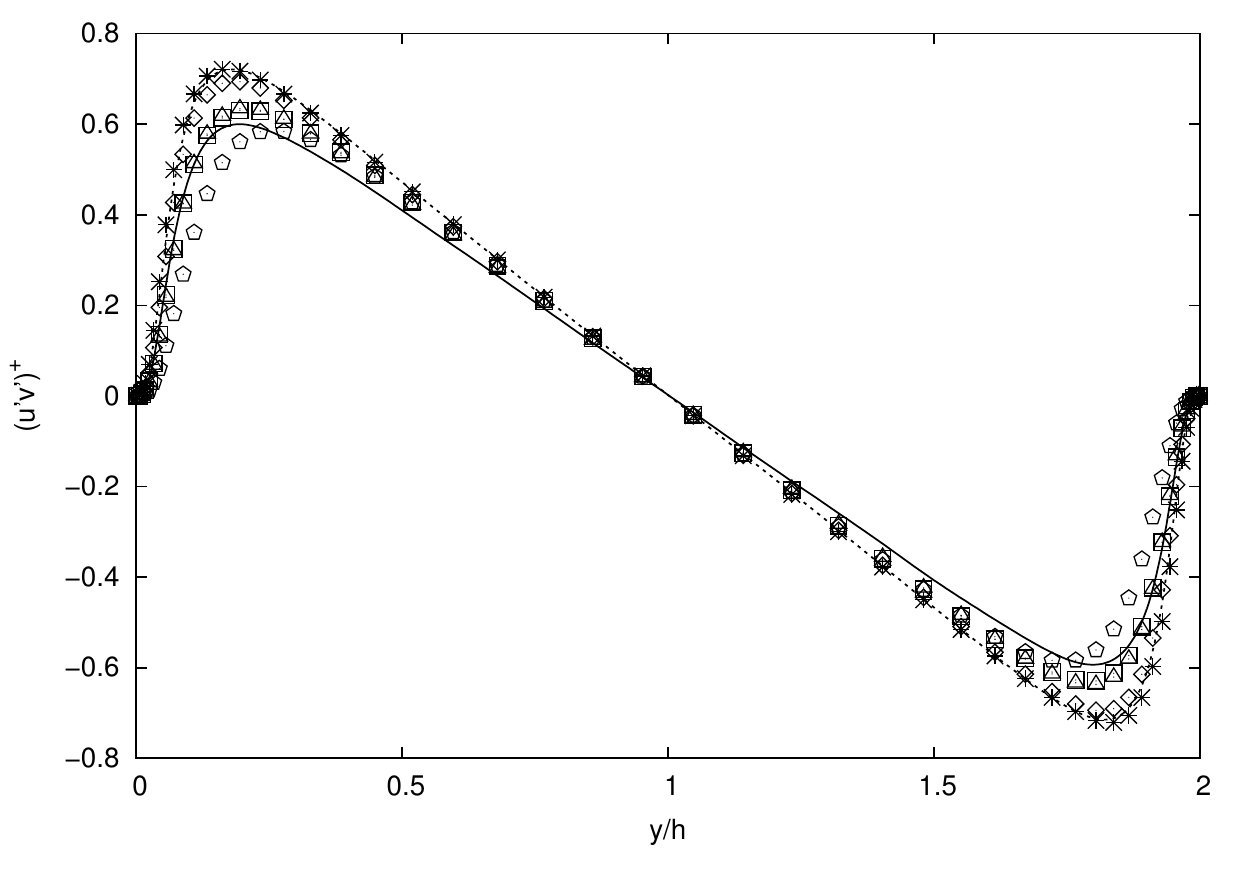}}
\subfigtopleft{\includegraphics[width=0.44\textwidth, trim={0 5 5 5}, clip]{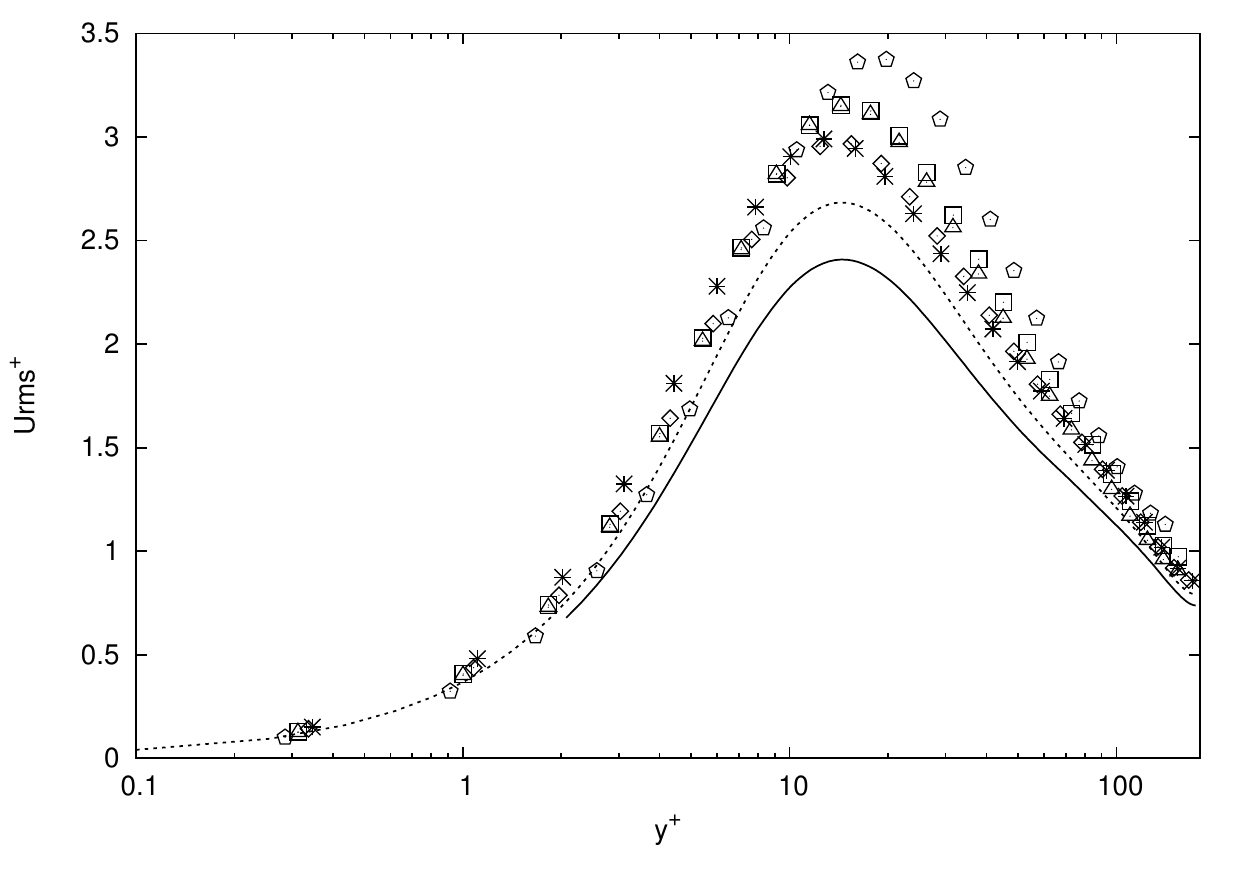}}
}\centerline{
\subfigtopleft{\includegraphics[width=0.44\textwidth, trim={0 5 5 5}, clip]{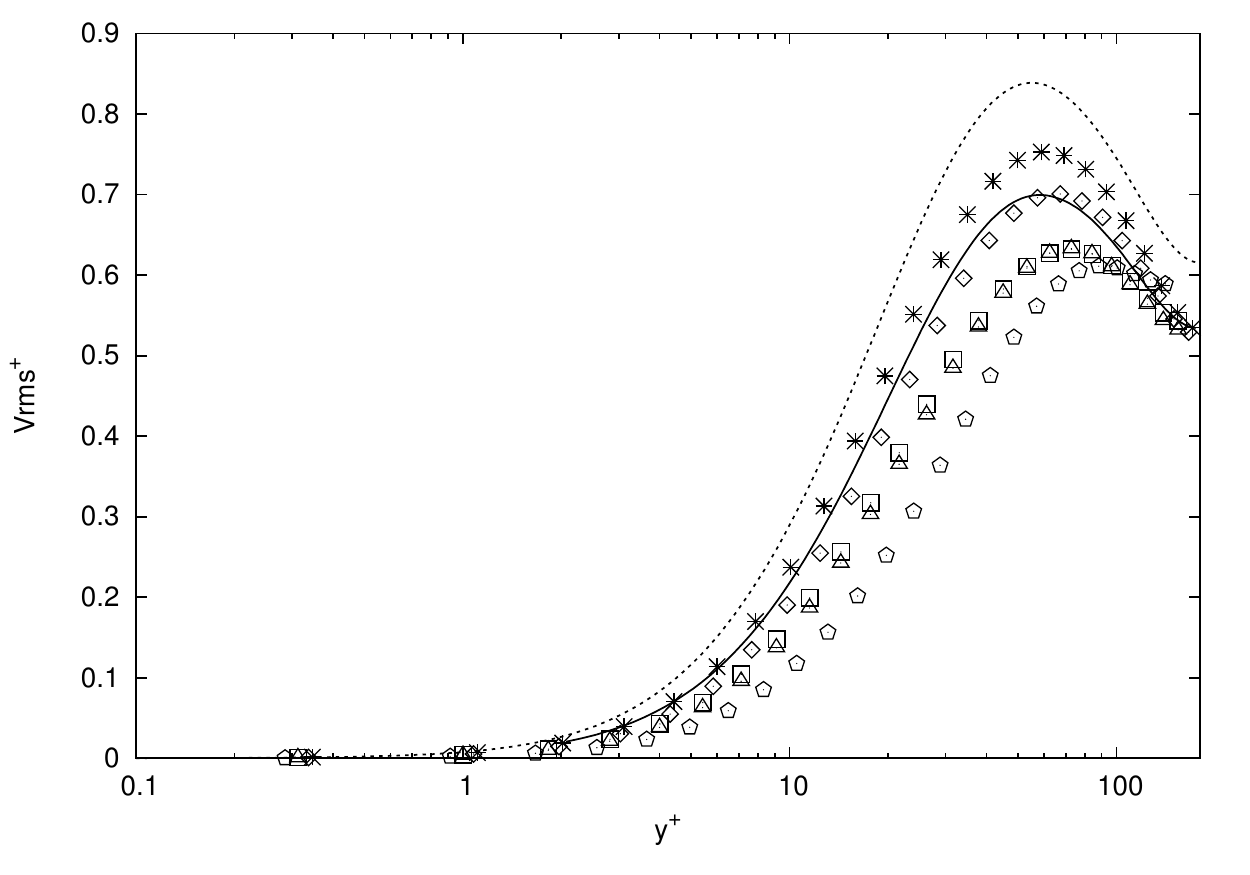}}
\subfigtopleft{\includegraphics[width=0.44\textwidth, trim={0 5 5 5}, clip]{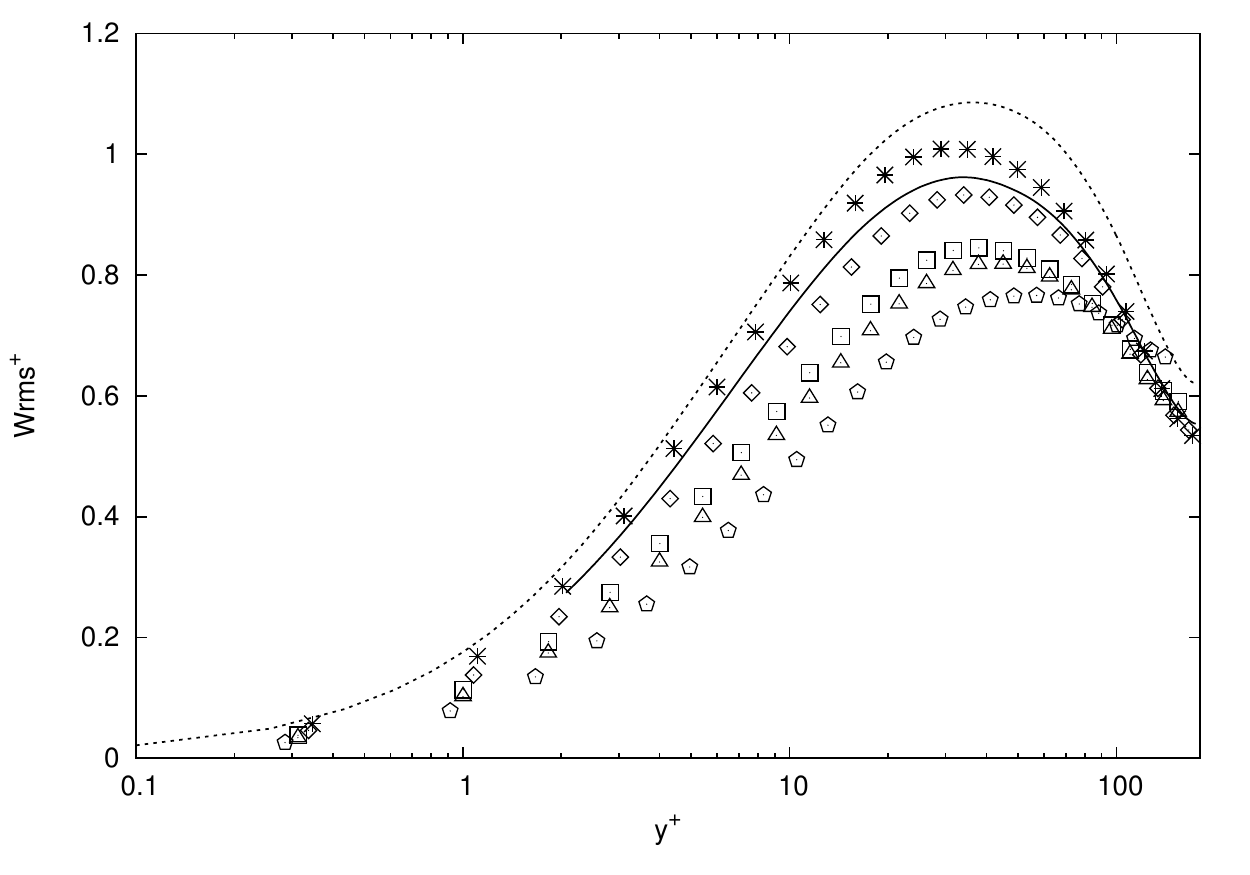}}
}
\caption[Comparison of large-eddy simulations with the plane-average, tensorial plane-average and tensorial global-average dynamic Smagorinsky model and the plane-average dynamic Anisotropic Smagorinsky model.]{
Comparison of large-eddy simulations
with the plane-average, tensorial plane-average and tensorial global-average dynamic Smagorinsky model and the plane-average dynamic Anisotropic Smagorinsky model
for the profiles of the mean streamwise velocity $\left\langle U_x \right\rangle$ (a, b), the covariance of streamwise and wall-normal velocity $\left\langle u_{\smash[b]{x}}' u_{\smash[b]{y}}' \right\rangle$ (c), the standard deviation of streamwise velocity $\smash[t]{\sqrt{\left\langle u_{\smash[b]{x}}'^2 \right\rangle}}$ (d), wall-normal velocity $\sqrt{\left\langle u_{\smash[b]{y}}'^2 \right\rangle}$ (e) and spanwise velocity $\sqrt{\left\langle u_{\smash[b]{z}}'^2 \right\rangle}$ (f)
with the mesh 48B.
\label{label24}}
\end{figure*}

\begin{figure*}
\setcounter{subfigcounter}{0}
\centerline{
\subfigtopleft{\includegraphics[width=0.44\textwidth, trim={0 5 5 5}, clip]{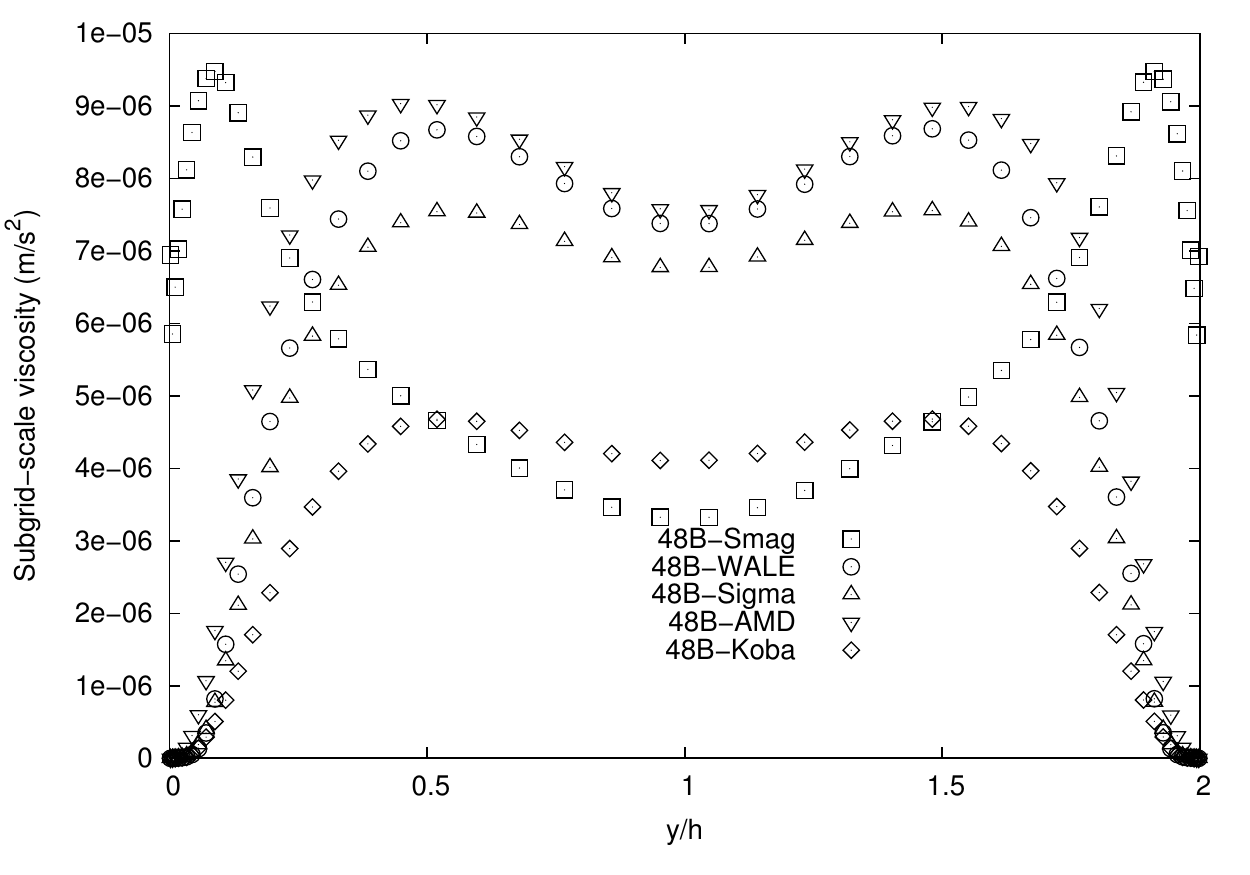}}
\subfigtopleft{\includegraphics[width=0.44\textwidth, trim={0 5 5 5}, clip]{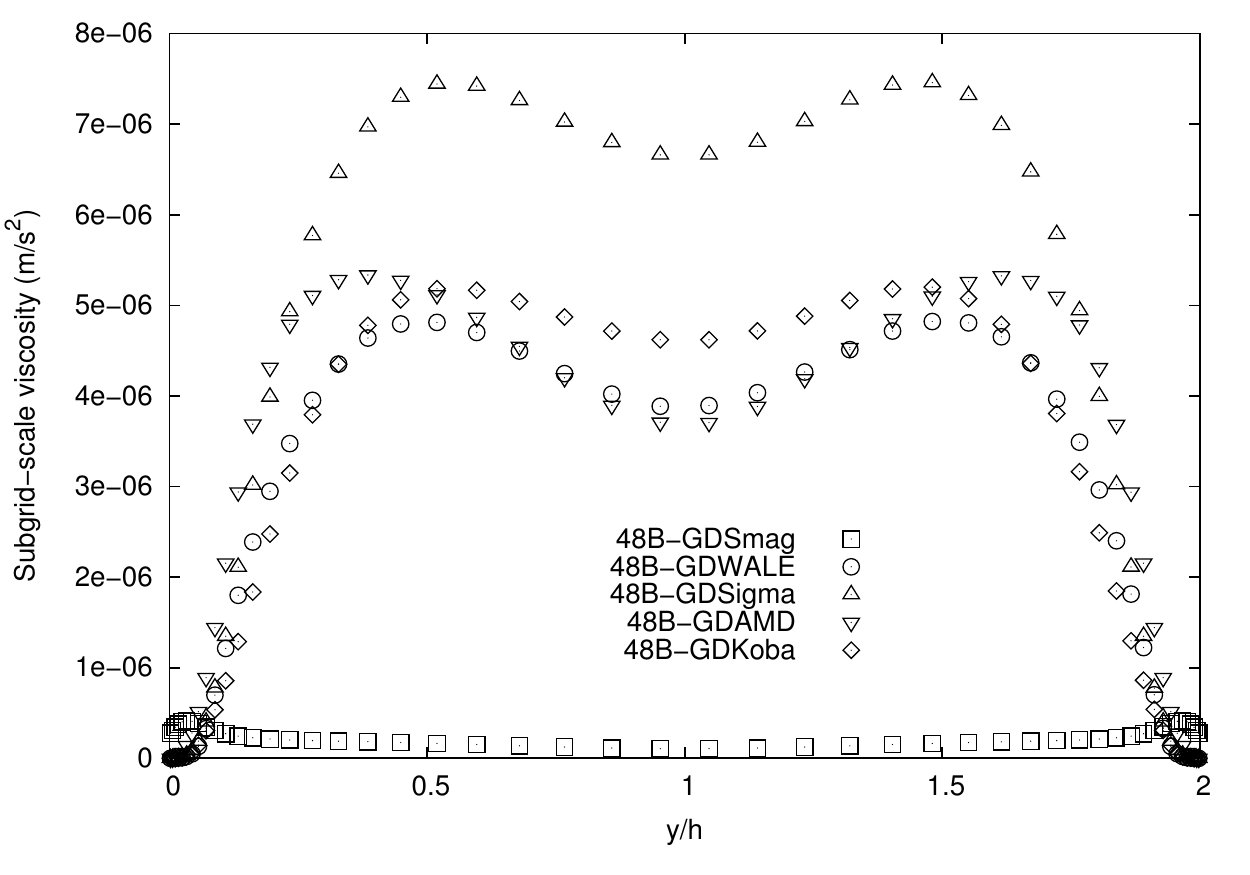}}
}
\caption[Comparison of simulations with the constant-parameters and global-average dynamic Smagorinsky, WALE, Sigma, AMD, Kobayashi and Anisotropic Smagorinsky models for the profiles of the subgrid-scale viscosity.]{
Comparison of simulations
with the constant-parameters and global-average dynamic Smagorinsky, WALE, Sigma, AMD, Kobayashi and Anisotropic Smagorinsky models
for the profiles of the subgrid-scale viscosity
with the mesh 48B.
\label{label23}}
\end{figure*}

In this section, we investigate the functional modelling of the subgrid-scale tensor.
The functional models investigated are the Smagorinsky, WALE, Sigma, AMD,
Kobayashi and Anisotropic Smagorinsky models, as well as dynamic versions
of these models.
The results of large-eddy simulations with these models are compared
in figure \ref{label22}
with the mesh 48B.
As consistently found in the literature \citep[see e.g.][]{vreman2004eddy},
the Smagorinsky model does not perform well in shear flow and considerably
deteriorates the profiles of the turbulence statistics.
The Anisotropic Smagorinsky model improves significantly the predictions
compared to the Smagorinsky model, providing similar results to the WALE,
Sigma and AMD models.
The WALE, Sigma, AMD, Kobayashi and Anisotropic Smagorinsky models
underestimate the wall shear stress, thus do not lead to
a good representation of the scaled mean streamwise velocity.
The additional dissipation provided by the model is able to decrease the
maximum value of the standard deviation of wall-normal and
spanwise velocity, but the standard deviation of streamwise
velocity is increased further away from the filtered DNS profile.
The no-model simulation yields a better prediction of the friction Reynolds
number, the mean streamwise velocity and the standard deviation of velocity
components than the large-eddy simulations with functional models.
The points discussed above are also valid for the meshes 24C and 36C.
The larger filter widths amplify the reduction
of the standard deviation of wall-normal and spanwise velocity
following approximately the same behaviour as the filtered direct numerical
simulation (figure \ref{label16}).
On the other hand, the standard deviation of streamwise velocity
is even with the 24C mesh not reduced compared to the no-model simulation,
further enhancing the discrepancy with the filtered direct numerical
simulation.
The predictions of the large-eddy simulations depend on the
amplitude of the subgrid-scale viscosity.
In our simulations, a lower subgrid-scale
viscosity is obtained with the Kobayashi model (figure \ref{label23}(a)).
This leads to more accurate results with the meshes and numerical method of this
study.

Dynamic models provide a less arbitrary comparison of functional models
in the sense that it is not complicated by the choice of the model parameter.
We study plane-average, global-average, tensorial plane-average and
tensorial global-average dynamic methods.
The main purpose of the plane-average dynamic method is the local adaptation of
the model parameter, which may compensate an unsatisfactory asymptotic near-wall
behaviour of the model \citep{silvis2017physical}.
This is particularly well-suited to the Smagorinsky model.
The plane-average dynamic Smagorinsky model (figure \ref{label24})
gives similar results to the non-dynamic WALE and Sigma models.
With the plane-average dynamic procedure, the Anisotropic Smagorinsky model
deteriorates the predictions of the Smagorinsky model.
Large-eddy simulations with the plane-average dynamic
WALE, Sigma, AMD and Kobayashi models are not stable.
This is consistent with the observation by \citet{toda2010subgrid}
that the plane-average dynamic method might degrade subgrid-scale models
with a proper asymptotic near-wall behaviour and lead to numerical instabilities.

\begin{table}
\centerline{\begin{tabular}{lccc}
                             & \multicolumn{3}{l}{ \rlap{\hspace{-5em}          Average of the dynamic parameter (standard deviation), } } \\
                             & \multicolumn{3}{c}{           $\left\langle C^{\mathrm{mod}} \right\rangle$ ($\sqrt{\left\langle (C^{\mathrm{mod}})^2 \right\rangle - \left\langle C^{\mathrm{mod}} \right\rangle^2}$)}\\
                             & Mesh 24C                                & Mesh 36C                                 & Mesh 48 B \\[.5em]
\hline\\[-.5em]
Smag.                        & $\numprint{0.009}$ ($\numprint{0.001}$) & $\numprint{0.016}$ ($\numprint{0.001}$)  &   $\numprint{0.033}$ ($\numprint{0.001}$) \\
WALE                         & $\numprint{1.337}$ ($\numprint{0.554}$) & $\numprint{0.591}$ ($\numprint{0.073}$)  &   $\numprint{0.494}$ ($\numprint{0.038}$) \\
Sigma                        & $\numprint{1.723}$ ($\numprint{0.263}$) & $\numprint{1.182}$ ($\numprint{0.083}$)  &   $\numprint{0.982}$ ($\numprint{0.046}$) \\
AMD                          & ---                                     & ---                                      &   $\numprint{0.455}$ ($\numprint{0.019}$) \\
Kobayashi                    & ---                                     & ---                                      &   $\numprint{1.151}$ ($\numprint{0.088}$) \\
An. Smag.                    & $\numprint{0.694}$ ($\numprint{0.080}$) & $\numprint{1.008}$ ($\numprint{0.075}$)  &   $\numprint{1.544}$ ($\numprint{0.101}$)\\[.5em]
\hline\\[-.5em]                                                                                               
Gradient                     & ---                                     & ---                                      &   $\numprint{2.593}$ ($\numprint{0.053}$) \\
\end{tabular}}
\caption[Dynamic parameter of large-eddy simulations with global-average dynamic models.]{Average and normalised standard deviation of the dynamic parameter
of the large-eddy simulations
with the global-average dynamic Smagorinsky, WALE, Sigma, AMD, Kobayashi and Anisotropic Smagorinsky models
with the meshes 24C, 36C and 48B.
\label{label7}}
\end{table}

\begin{table*}
\centerline{\begin{tabular}{lcccccc}
                             & \multicolumn{6}{c}{           Average of the dynamic parameter (standard deviation), } \\
                             & \multicolumn{6}{c}{           $\left\langle C^{\mathrm{mod}} \right\rangle$ ($\sqrt{\left\langle (C^{\mathrm{mod}})^2 \right\rangle - \left\langle C^{\mathrm{mod}} \right\rangle^2}$)} \\
                             & $xx$                                               & $xy$                                     & $xz$                                    & $yy$                                              & $zy$                                               & $zz$\\[.5em]
\hline\\[-.5em]                                                                                                                                                                                                                                                                 
Smag.                        & $          \numprint{ 6.755}$ ($\numprint{0.434}$) & $\numprint{0.028} $ ($\numprint{0.001}$) & $\numprint{0.073}$ ($\numprint{0.049}$) & $          \numprint{ 0.269}$ ($\numprint{0.025}$)& $\!\!\!\!\!\numprint{-0.011}$ ($\numprint{0.005}$) & $\!\!\!\!\!\numprint{-0.266}$ ($\numprint{0.036}$)   \\       
WALE                         & $\!\!\!\!\!\numprint{-0.732}$ ($\numprint{0.159}$) & $\numprint{1.463} $ ($\numprint{0.093}$) & $\numprint{0.432}$ ($\numprint{0.041}$) & $          \numprint{ 0.225}$ ($\numprint{0.066}$)& $\!\!\!\!\!\numprint{-0.035}$ ($\numprint{0.023}$) & $          \numprint{ 0.418}$ ($\numprint{0.056}$)   \\       
Sigma                        & $          \numprint{ 1.664}$ ($\numprint{0.267}$) & $\numprint{1.662} $ ($\numprint{0.077}$) & $\numprint{0.617}$ ($\numprint{0.069}$) & $          \numprint{ 0.155}$ ($\numprint{0.058}$)& $          \numprint{ 0.002}$ ($\numprint{0.030}$) & $          \numprint{ 0.160}$ ($\numprint{0.050}$)   \\       
AMD                          & $          \numprint{ 0.624}$ ($\numprint{0.097}$) & $\numprint{0.750} $ ($\numprint{0.035}$) & $\numprint{0.245}$ ($\numprint{0.038}$) & $          \numprint{ 0.001}$ ($\numprint{0.034}$)& $          \numprint{ 0.035}$ ($\numprint{0.022}$) & $\!\!\!\!\!\numprint{-0.064}$ ($\numprint{0.026}$)   \\       
Kobayashi                    & $          \numprint{ 2.546}$ ($\numprint{0.274}$) & $\numprint{2.719} $ ($\numprint{0.175}$) & $\numprint{0.785}$ ($\numprint{0.133}$) & $\!\!\!\!\!\numprint{-0.145}$ ($\numprint{0.081}$)& $          \numprint{ 0.048}$ ($\numprint{0.046}$) & $\!\!\!\!\!\numprint{-0.030}$ ($\numprint{0.067}$)   \\       
An. Smag.                    & $          \numprint{ 5.435}$ ($\numprint{0.480}$) & $\numprint{1.618} $ ($\numprint{0.148}$) & $\numprint{0.275}$ ($\numprint{0.094}$) & $          \numprint{ 1.531}$ ($\numprint{0.259}$)& $\!\!\!\!\!\numprint{-0.236}$ ($\numprint{0.086}$) & $\!\!\!\!\!\numprint{-0.323}$ ($\numprint{0.078}$)\\[.5em] 
\hline\\[-.5em]
Gradient                     & $          \numprint{ 2.587}$ ($\numprint{0.061}$) & $\numprint{2.404} $ ($\numprint{0.039}$) & $\numprint{1.379}$ ($\numprint{0.039}$) & $          \numprint{ 2.928}$ ($\numprint{0.067}$)& $          \numprint{ 1.566}$ ($\numprint{0.041}$) & $          \numprint{ 1.949}$ ($\numprint{0.031}$)   \\       
\end{tabular}}
\caption[Dynamic parameters of large-eddy simulations with the tensorial global-average dynamic models.]{Average and normalised standard deviation of the dynamic parameters
of the large-eddy simulations
with the tensorial global-average dynamic Smagorinsky, WALE, Sigma, AMD, Kobayashi and Anisotropic Smagorinsky models
with the mesh 48B.
\label{label8}}
\end{table*}

\begin{figure*}
\setcounter{subfigcounter}{0}
\centerline{
\subfigtopleft{\includegraphics[width=0.44\textwidth, trim={0 5 5 5}, clip]{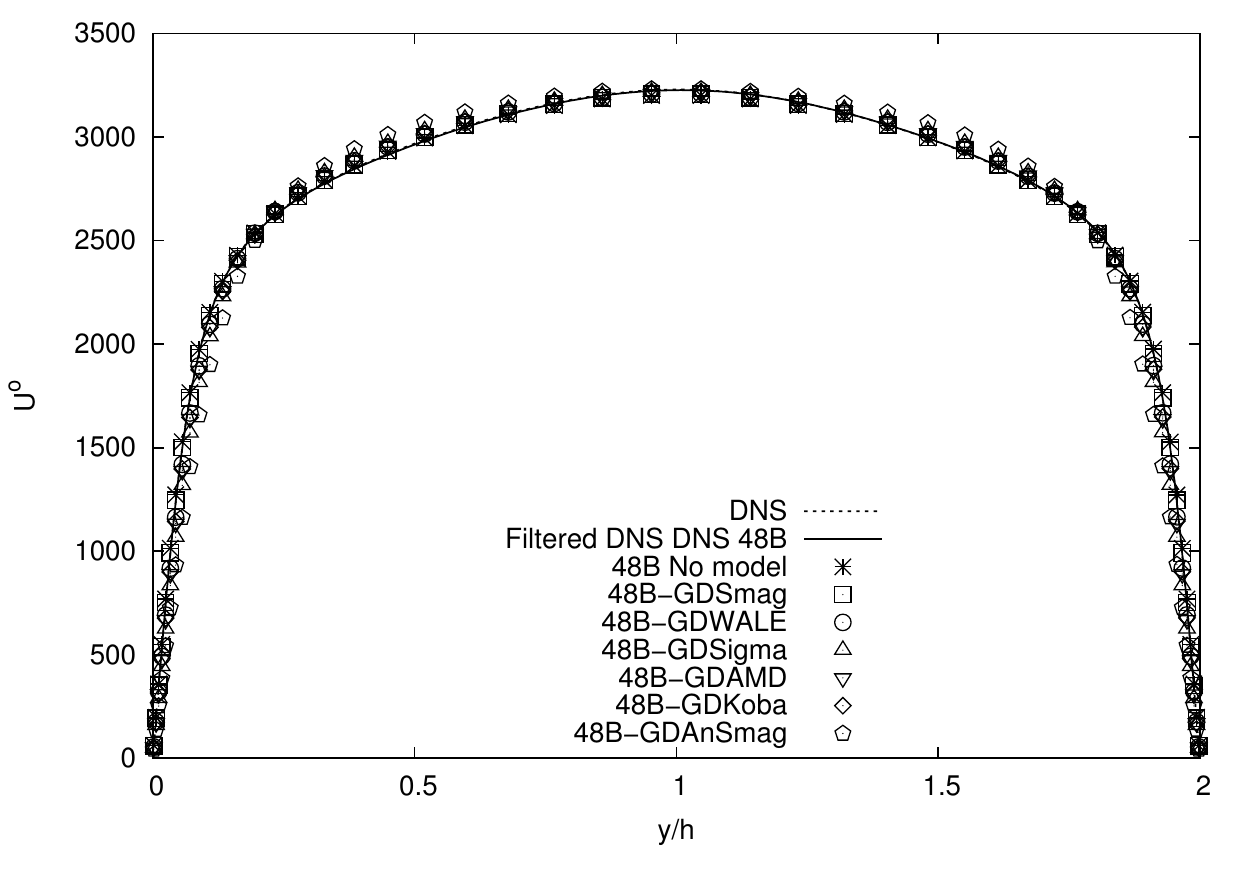}}
\subfigtopleft{\includegraphics[width=0.44\textwidth, trim={0 5 5 5}, clip]{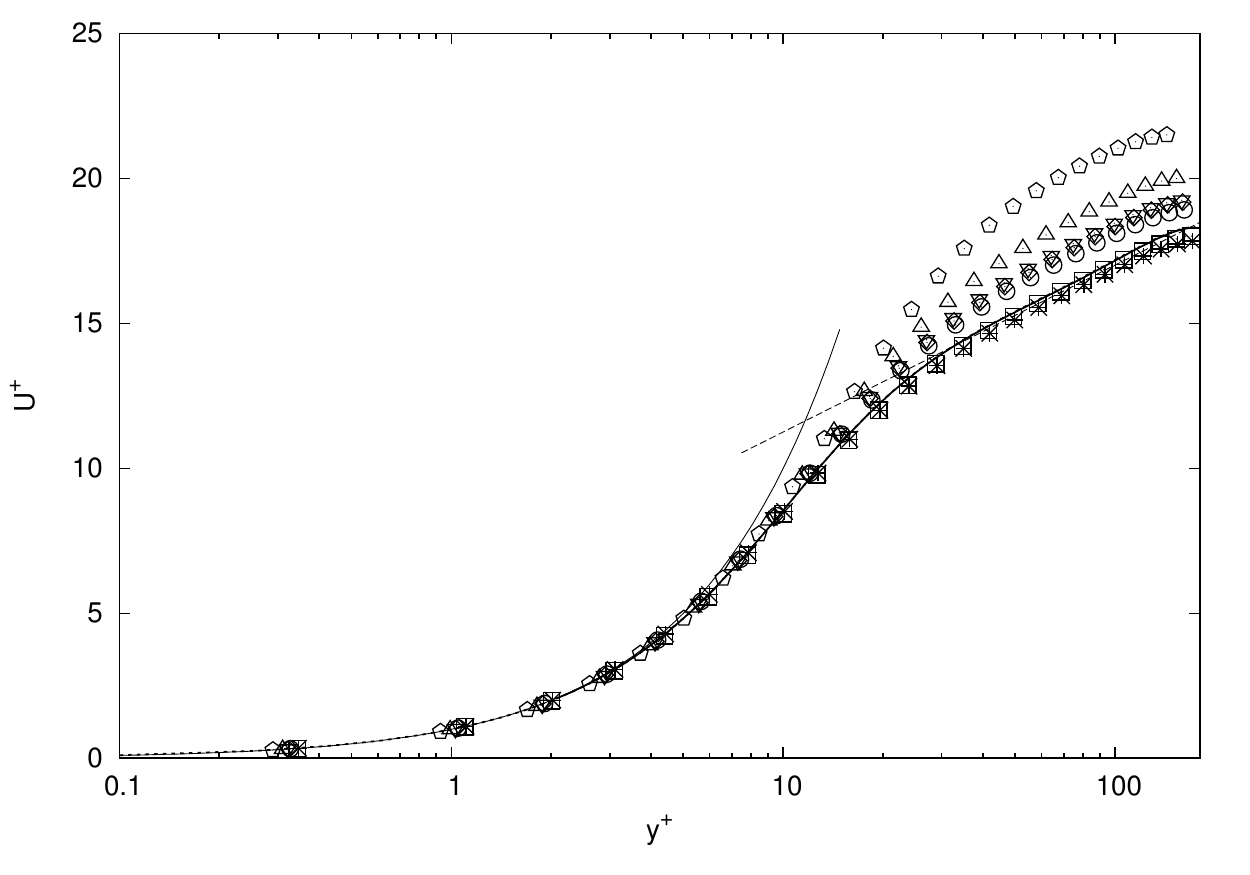}}
}\centerline{
\subfigtopleft{\includegraphics[width=0.44\textwidth, trim={0 5 5 5}, clip]{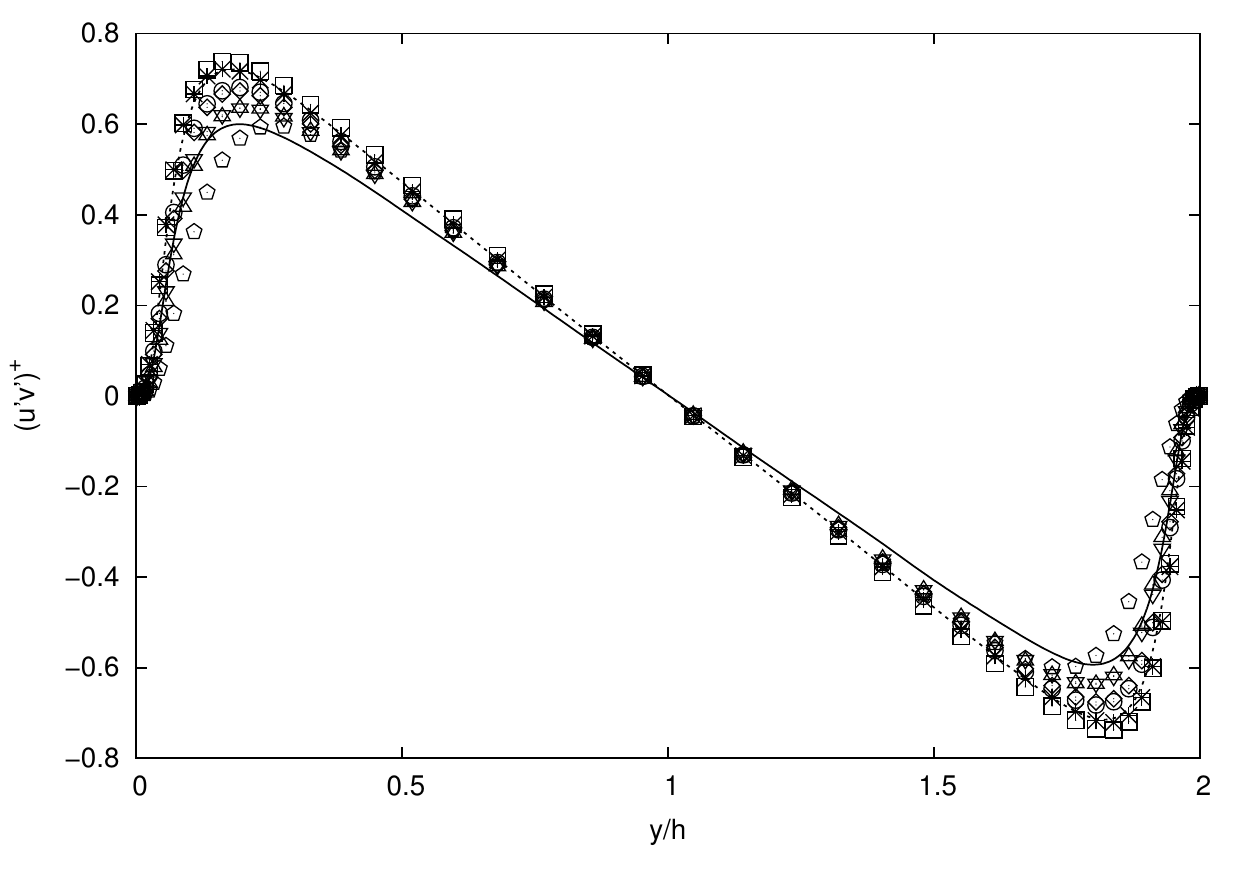}}
\subfigtopleft{\includegraphics[width=0.44\textwidth, trim={0 5 5 5}, clip]{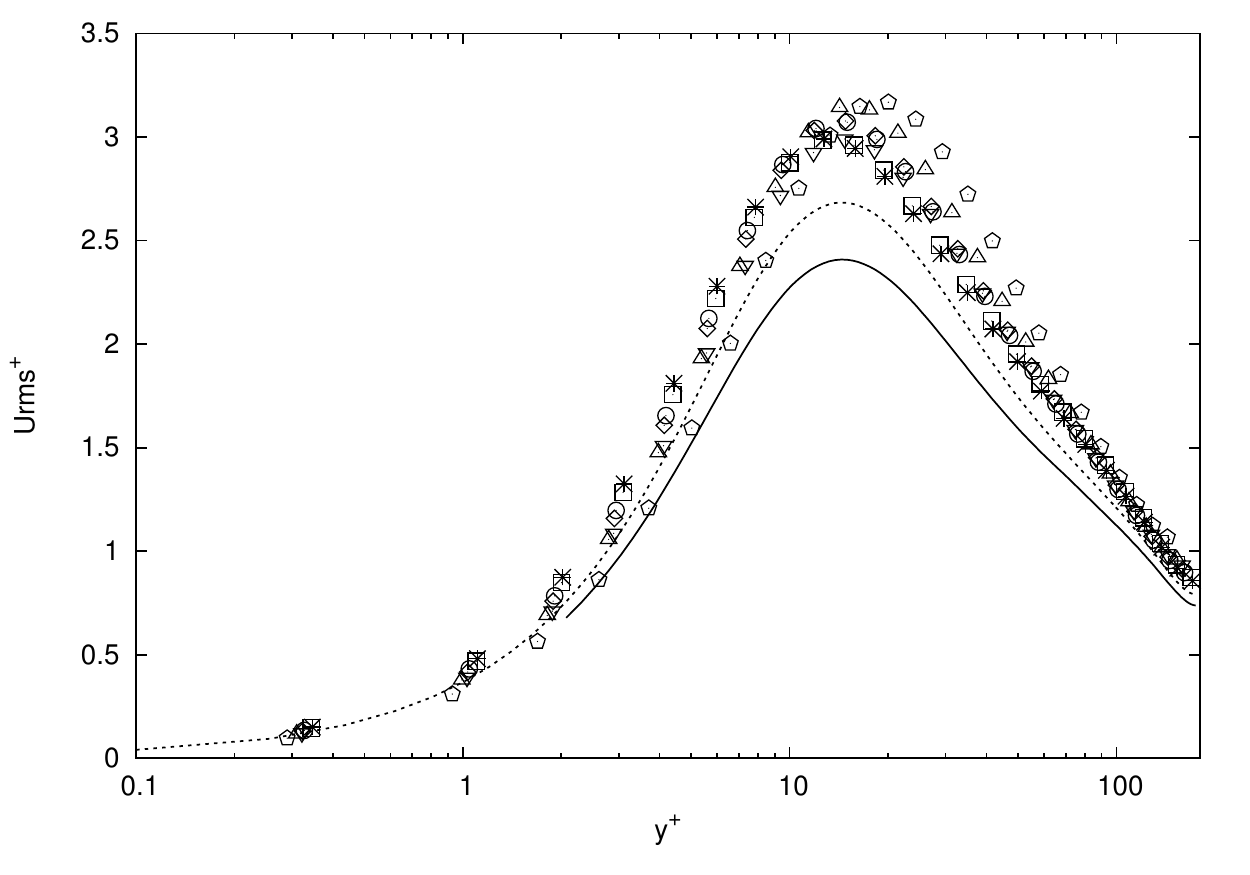}}
}\centerline{
\subfigtopleft{\includegraphics[width=0.44\textwidth, trim={0 5 5 5}, clip]{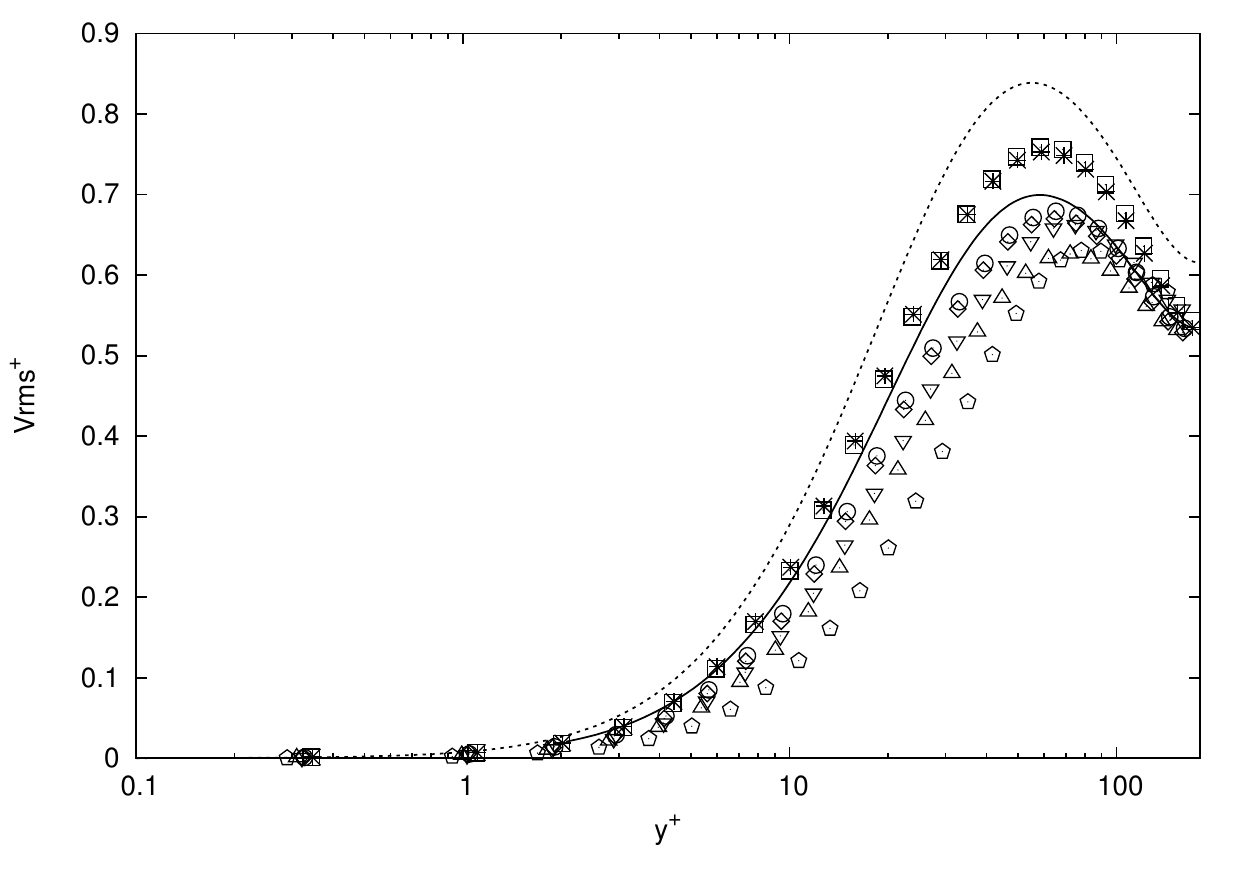}}
\subfigtopleft{\includegraphics[width=0.44\textwidth, trim={0 5 5 5}, clip]{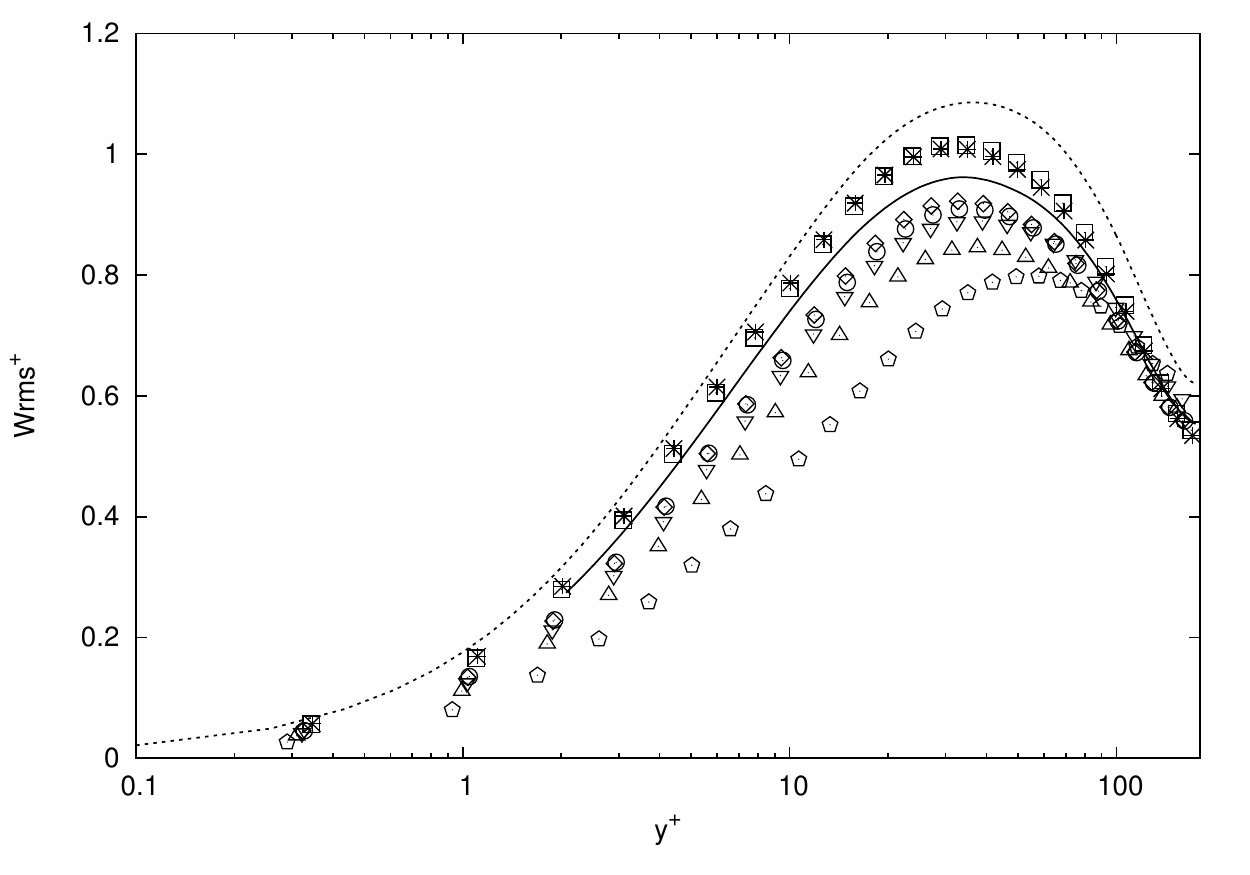}}
}
\caption[Comparison of large-eddy simulations with the global-average dynamic Smagorinsky, WALE, Sigma, AMD, Kobayashi and Anisotropic Smagorinsky models.]{
Comparison of large-eddy simulations
with the global-average dynamic Smagorinsky, WALE, Sigma, AMD, Kobayashi and Anisotropic Smagorinsky models
for the profiles of the mean streamwise velocity $\left\langle U_x \right\rangle$ (a, b), the covariance of streamwise and wall-normal velocity $\left\langle u_{\smash[b]{x}}' u_{\smash[b]{y}}' \right\rangle$ (c), the standard deviation of streamwise velocity $\smash[t]{\sqrt{\left\langle u_{\smash[b]{x}}'^2 \right\rangle}}$ (d), wall-normal velocity $\sqrt{\left\langle u_{\smash[b]{y}}'^2 \right\rangle}$ (e) and spanwise velocity $\sqrt{\left\langle u_{\smash[b]{z}}'^2 \right\rangle}$ (f)
with the mesh 48B.
\label{label25}}
\end{figure*}

\begin{figure*}
\setcounter{subfigcounter}{0}
\centerline{
\subfigtopleft{\includegraphics[width=0.44\textwidth, trim={0 5 5 5}, clip]{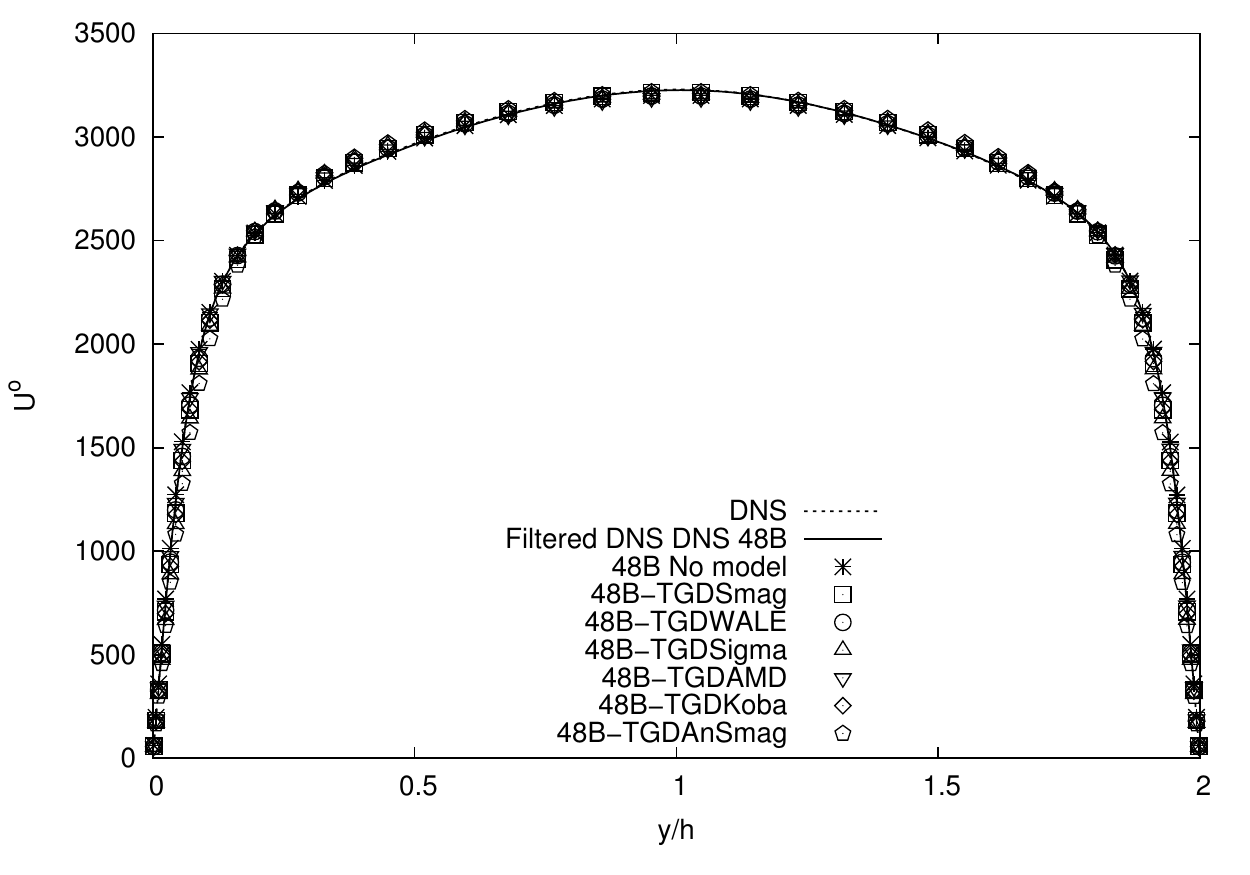}}
\subfigtopleft{\includegraphics[width=0.44\textwidth, trim={0 5 5 5}, clip]{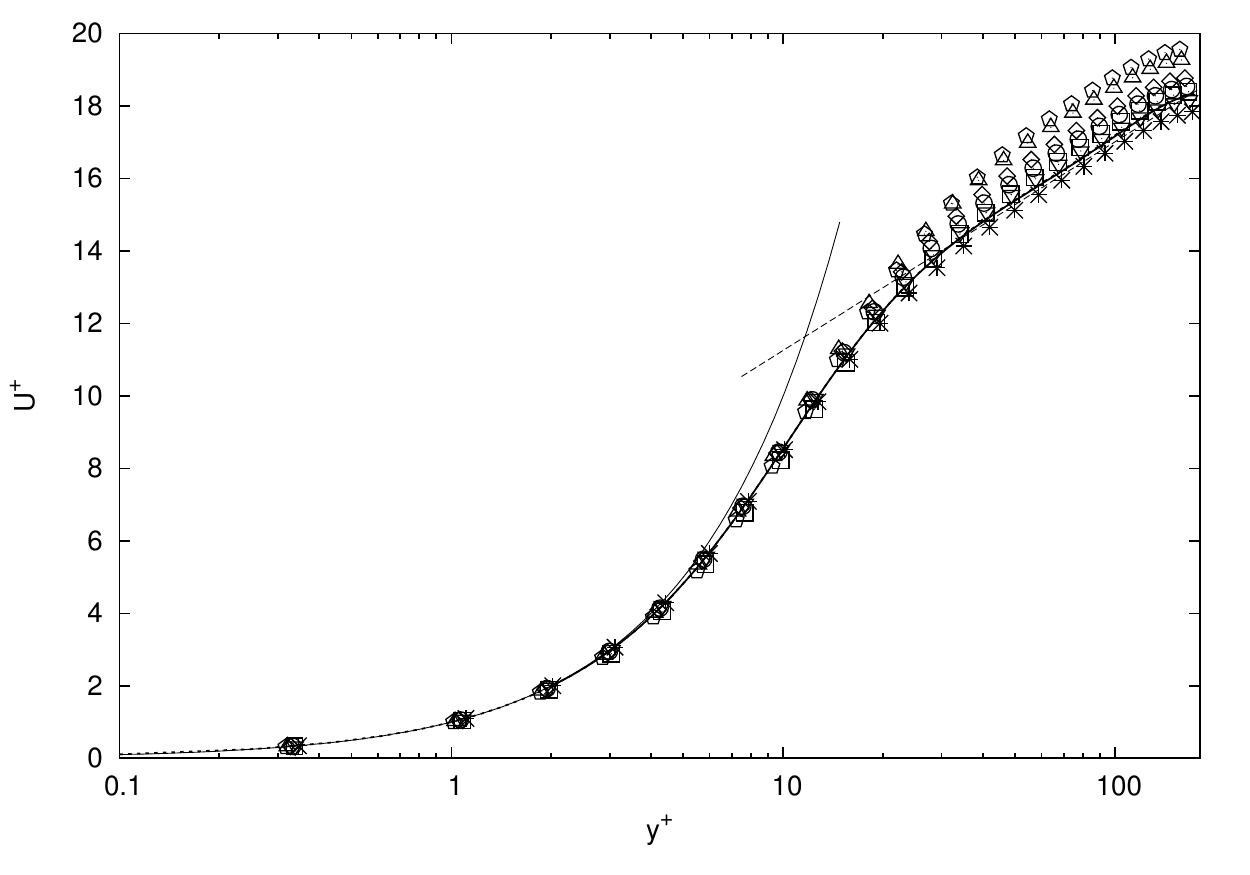}}
}\centerline{
\subfigtopleft{\includegraphics[width=0.44\textwidth, trim={0 5 5 5}, clip]{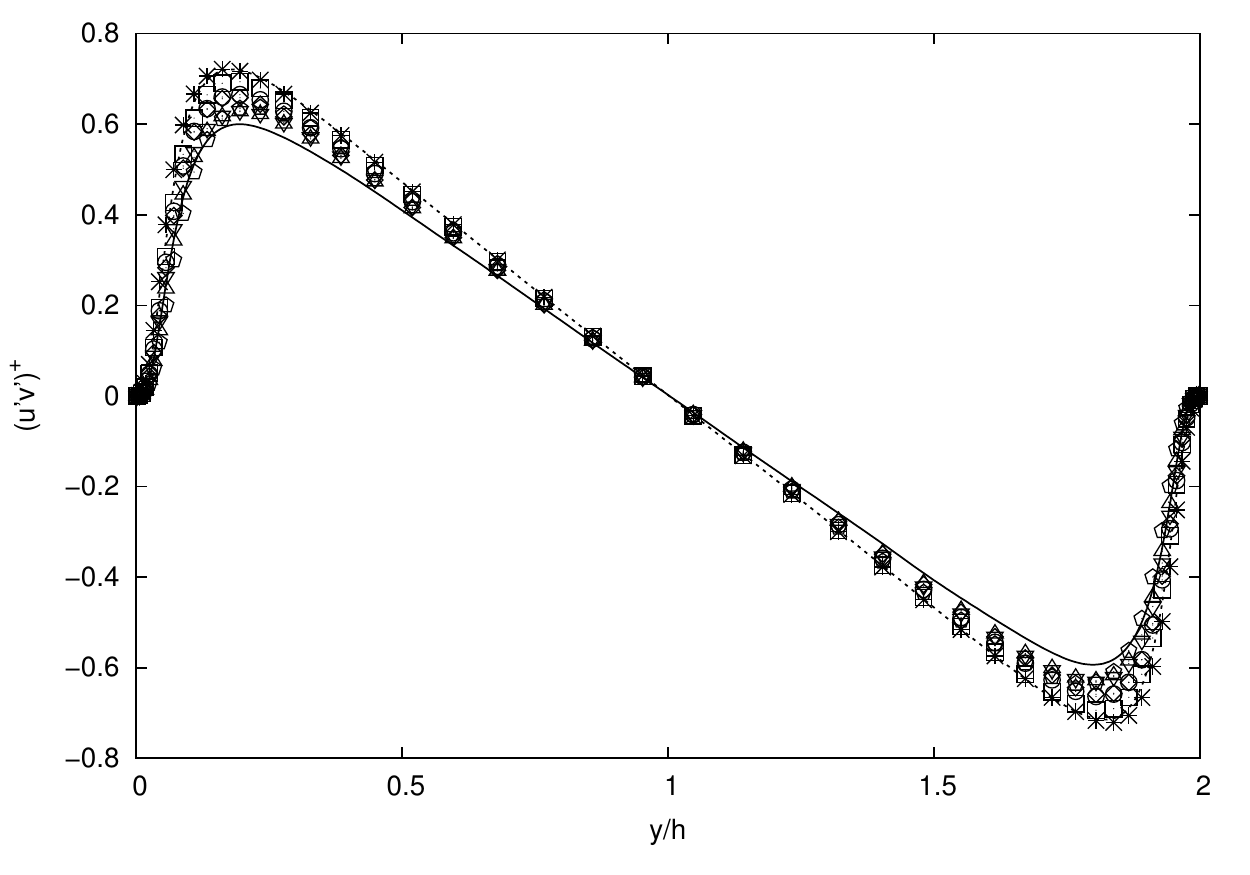}}
\subfigtopleft{\includegraphics[width=0.44\textwidth, trim={0 5 5 5}, clip]{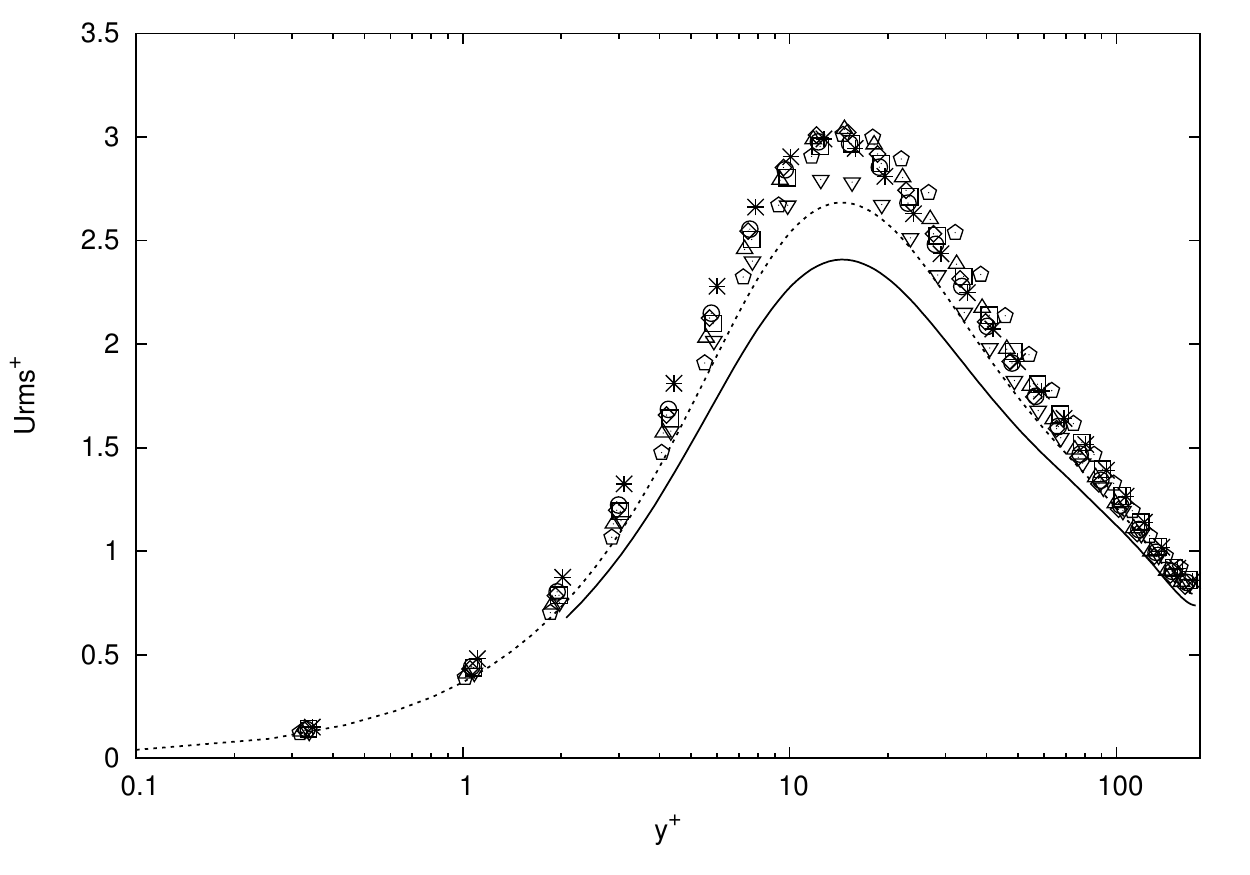}}
}\centerline{
\subfigtopleft{\includegraphics[width=0.44\textwidth, trim={0 5 5 5}, clip]{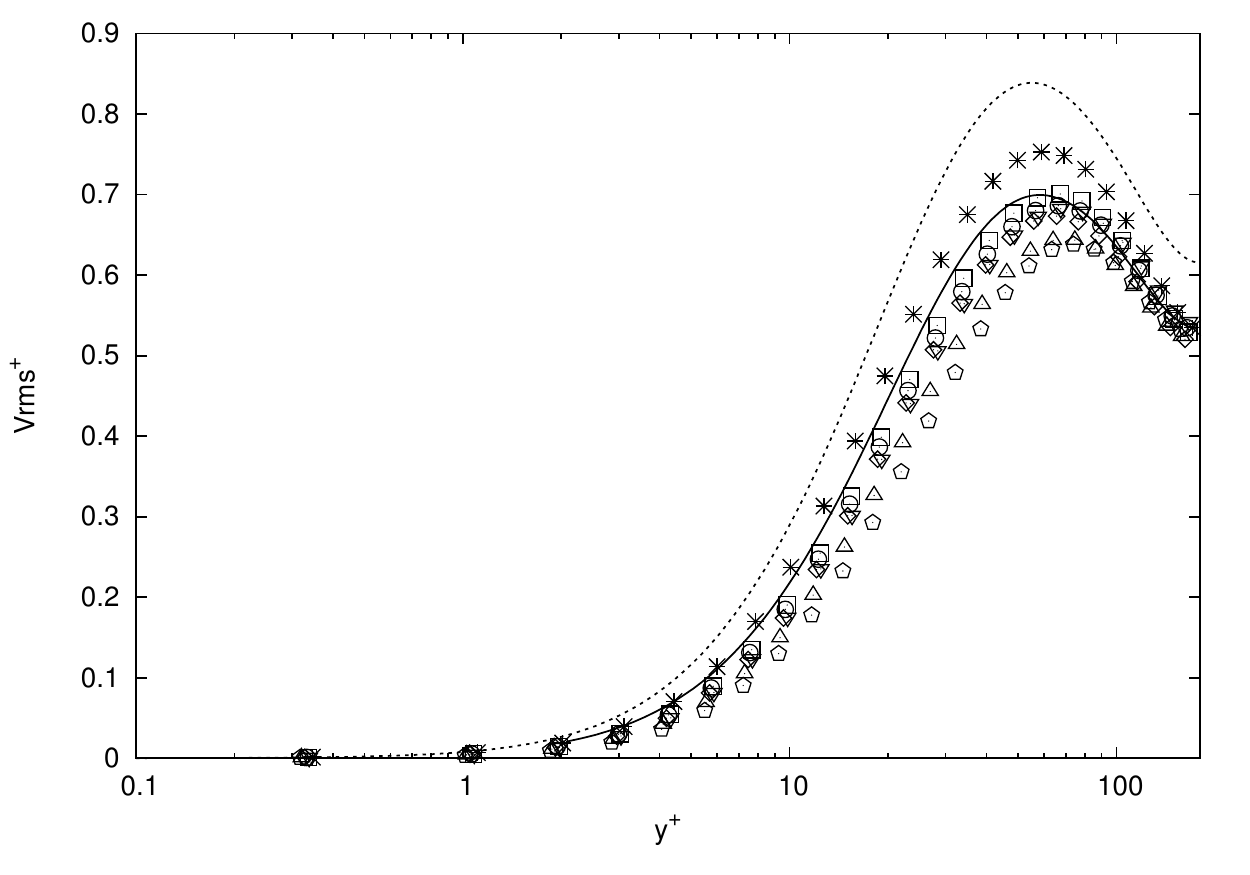}}
\subfigtopleft{\includegraphics[width=0.44\textwidth, trim={0 5 5 5}, clip]{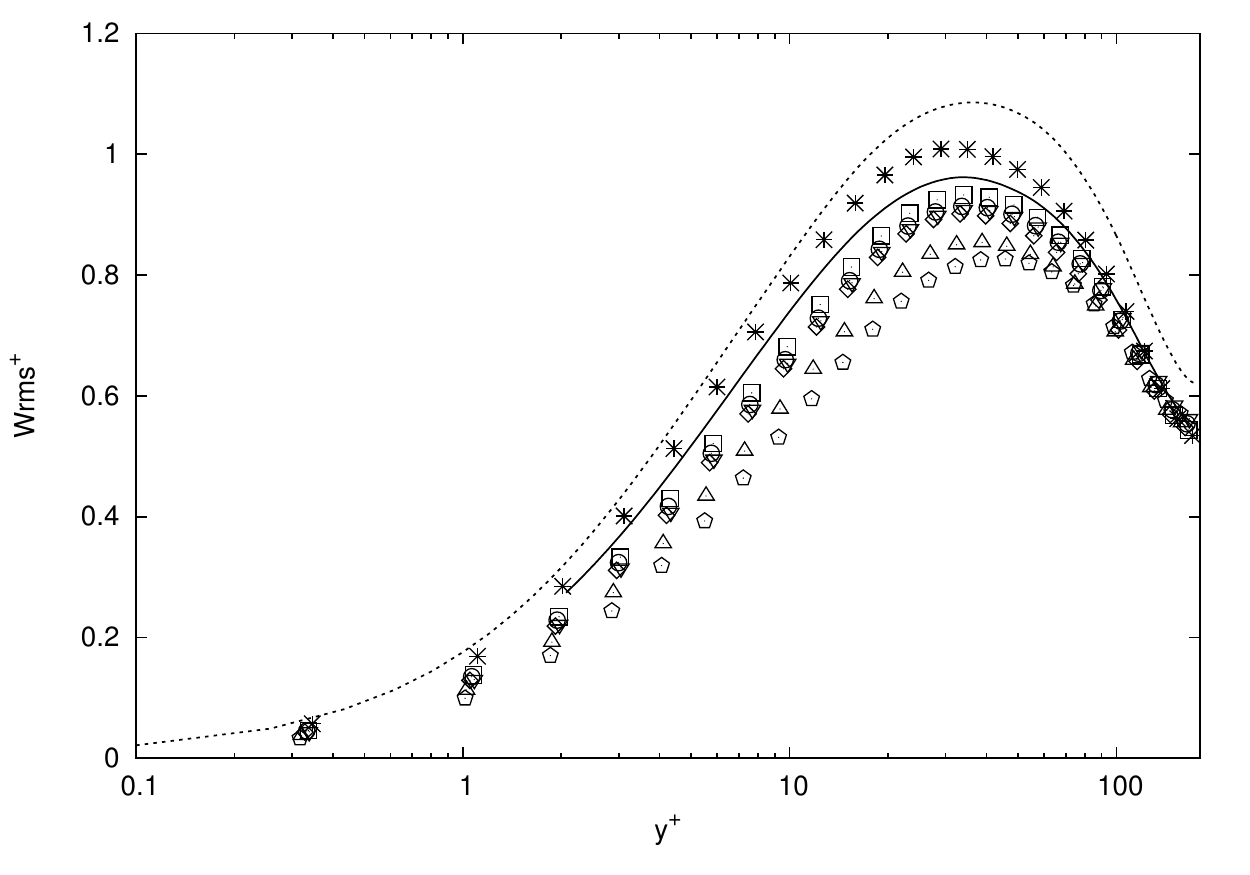}}
}
\caption[Comparison of large-eddy simulations with the tensorial global-average dynamic Smagorinsky, WALE, Sigma, AMD, Kobayashi and Anisotropic Smagorinsky models.]{
Comparison of large-eddy simulations
with the tensorial global-average dynamic Smagorinsky, WALE, Sigma, AMD, Kobayashi and Anisotropic Smagorinsky models
for the profiles of the mean streamwise velocity $\left\langle U_x \right\rangle$ (a, b), the covariance of streamwise and wall-normal velocity $\left\langle u_{\smash[b]{x}}' u_{\smash[b]{y}}' \right\rangle$ (c), the standard deviation of streamwise velocity $\smash[t]{\sqrt{\left\langle u_{\smash[b]{x}}'^2 \right\rangle}}$ (d), wall-normal velocity $\sqrt{\left\langle u_{\smash[b]{y}}'^2 \right\rangle}$ (e) and spanwise velocity $\sqrt{\left\langle u_{\smash[b]{z}}'^2 \right\rangle}$ (f)
with the mesh 48B.
\label{label26}}
\end{figure*}

The global-average dynamic method multiplies the subgrid-scale models by a time-dependent
function without modifying the local behaviour of the model. 
The average and standard deviation of the dynamic parameters are
given in table \ref{label7}.
The global-average dynamic procedure increases the subgrid-scale viscosity of the Kobayashi and Anisotropic Smagorinsky models
but reduces the subgrid-scale viscosity of the WALE and AMD models, except with the mesh 24C.
The Smagorinsky model is made negligible to prevent its detrimental near-wall
influence  (figure \ref{label23}(b)).
The global-average dynamic WALE, AMD and Kobayashi models lead to a good prediction of
the standard deviation of wall-normal and spanwise velocity, but
the standard deviation of streamwise velocity is not improved
compared to the no-model simulation (figure \ref{label25}).
The Sigma and Anisotropic Smagorinsky models do not provide good results with
the global-average dynamic procedure.

The tensorial global-average dynamic method alters the relative contribution
of each component of the subgrid-scale models.
Excluding the Anisotropic Smagorinky model, the tensorial global-average
dynamic procedure decreases heavily the relative amplitude of the ``$yy$'', ``$yz$'' and ``$zz$''
components, moderately decreases the ``$xz$'' component and amplifies the ``$xy$'' (table \ref{label8}).
The effect of the tensorial global-average dynamic procedure on the ``$xx$''
component is strongly dependent of the model.
Depending on the model and the component, negative average parameters are
obtained. In other words, the models are not purely dissipative.
The Sigma model is the only functional model investigated with only
positive parameters. The ``$xy$'' and ``$xz$'' are positive for all models
while the ``$zz$'' component is negative for most models.
The tensorial global-average dynamic Smagorinsky model
decreases the standard deviation of wall-normal and spanwise velocity without
increasing the standard deviation of streamwise velocity (figure \ref{label24}).
Similar results are obtained with the tensorial global-average dynamic WALE,
AMD and Kobayashi models (figure \ref{label26}).
This is an improvement compared to the global-average dynamic
procedure.
The tensorial global-average dynamic AMD model is able to decrease the maximum
value of the standard deviation of streamwise velocity,
improving the results compared to the no-model simulation.
It is the only investigated functional model with this property.

\subsection{Structural modelling}

\begin{figure*}
\setcounter{subfigcounter}{0}
\centerline{
\subfigtopleft{\includegraphics[width=0.44\textwidth, trim={0 5 5 5}, clip]{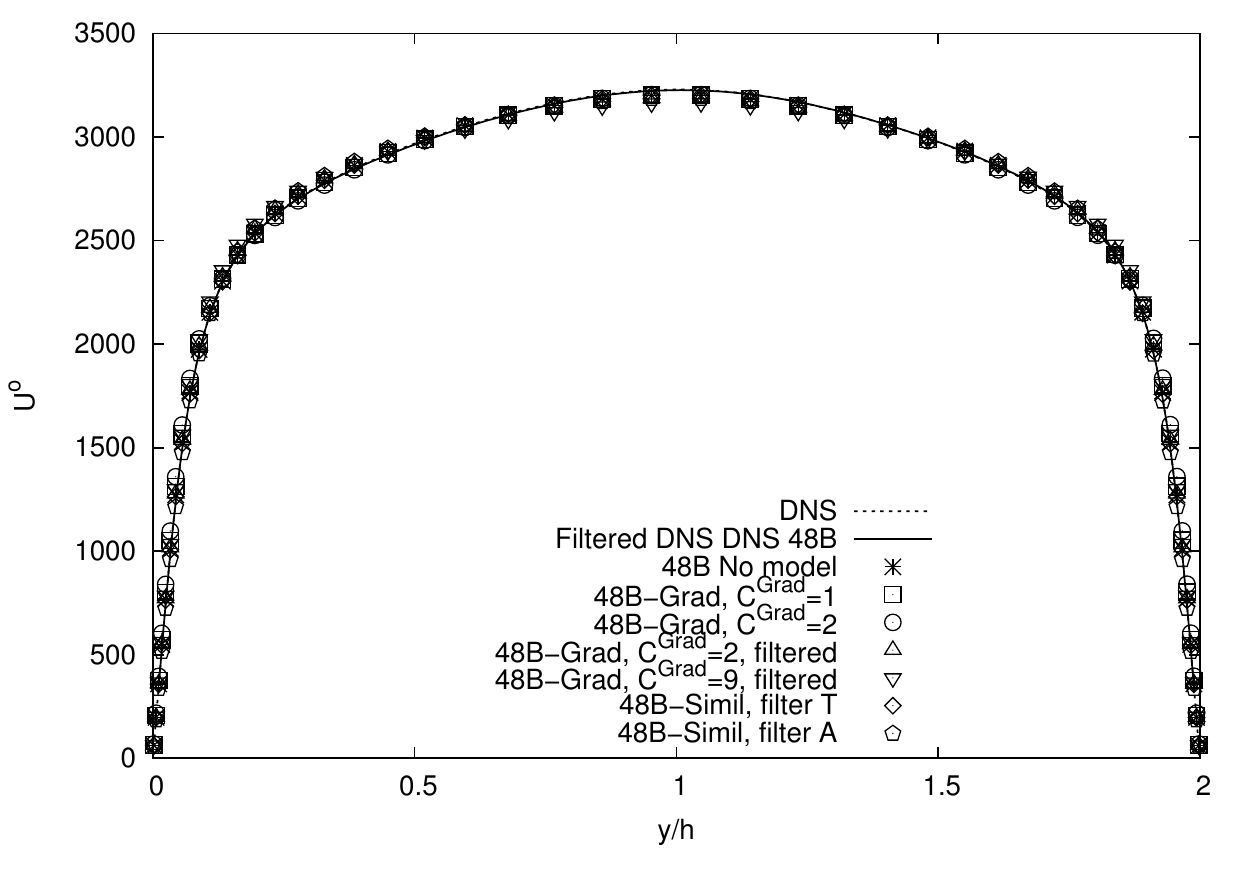}}
\subfigtopleft{\includegraphics[width=0.44\textwidth, trim={0 5 5 5}, clip]{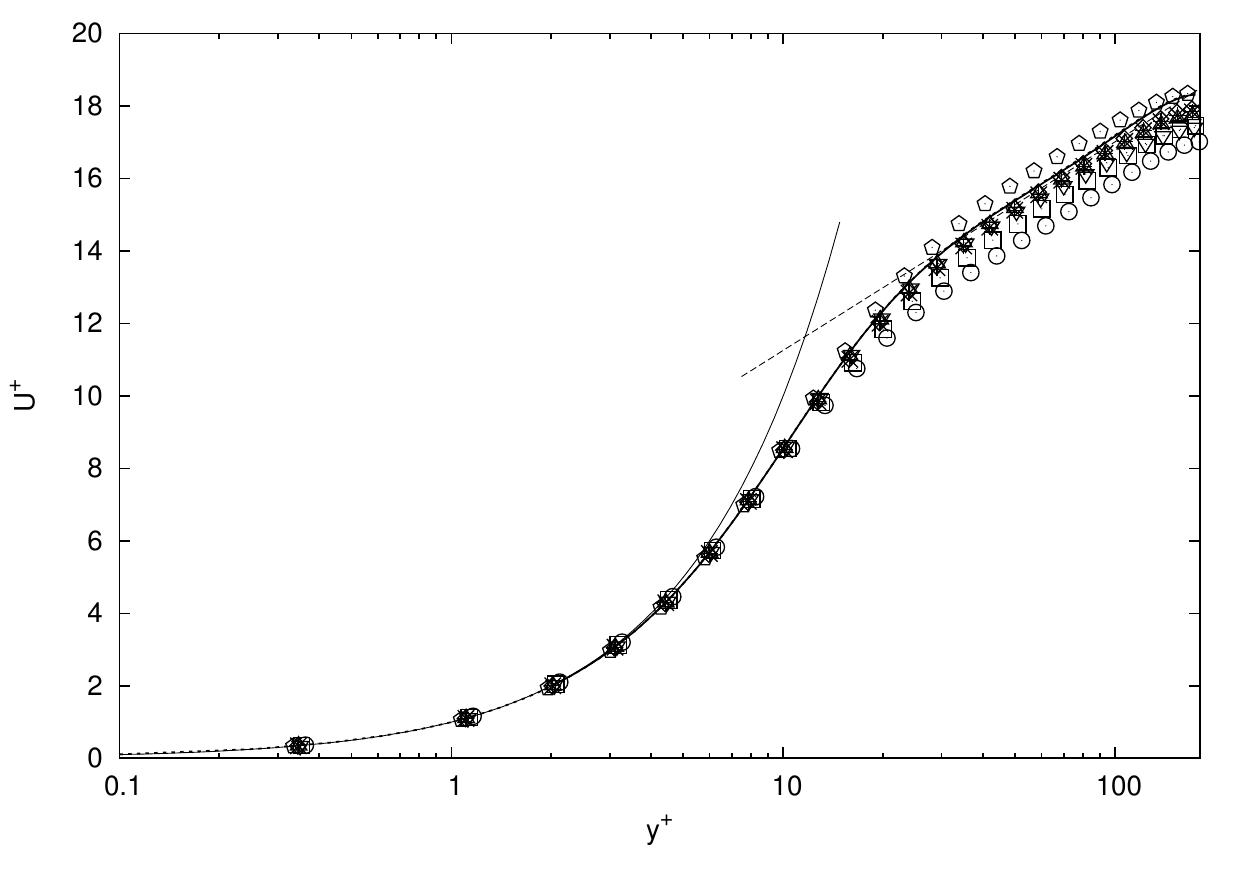}}
}\centerline{
\subfigtopleft{\includegraphics[width=0.44\textwidth, trim={0 5 5 5}, clip]{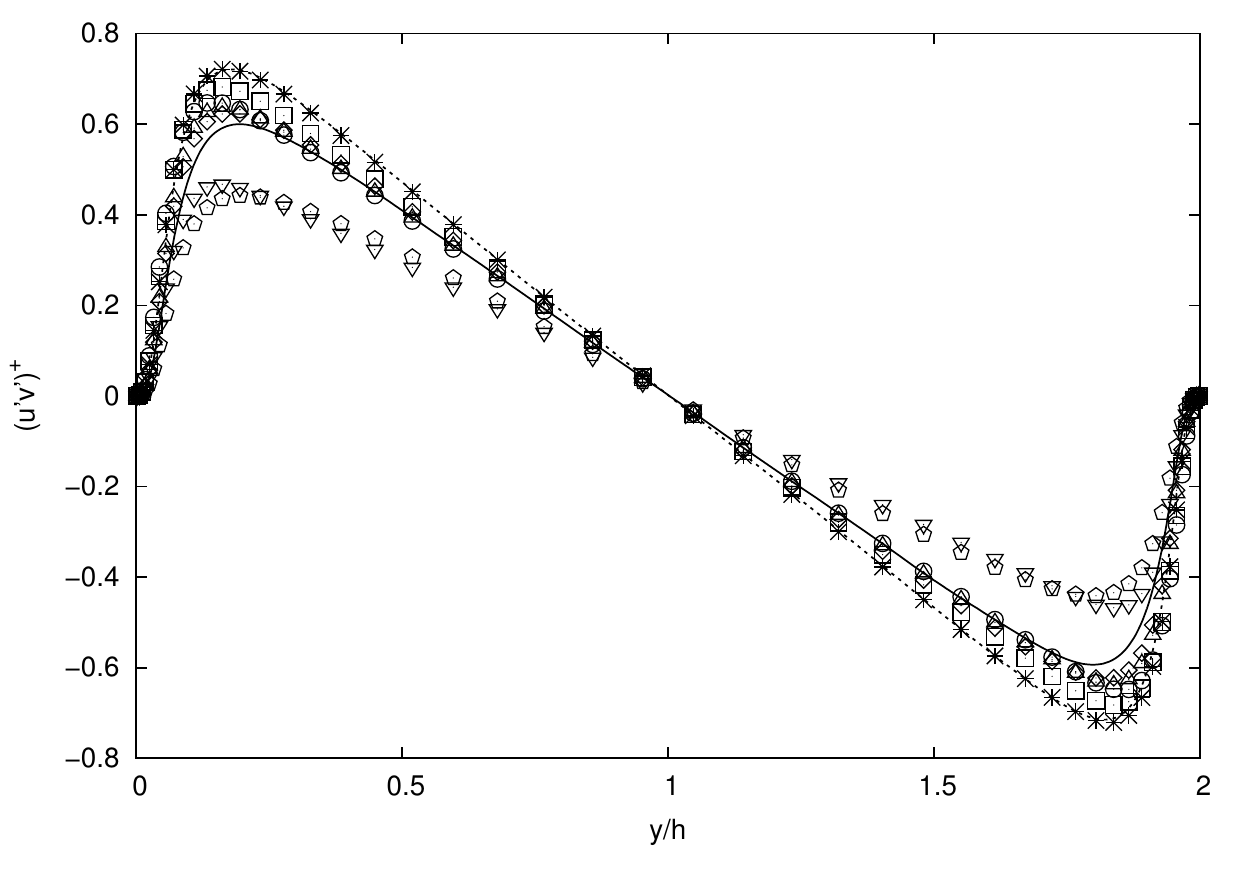}}
\subfigtopleft{\includegraphics[width=0.44\textwidth, trim={0 5 5 5}, clip]{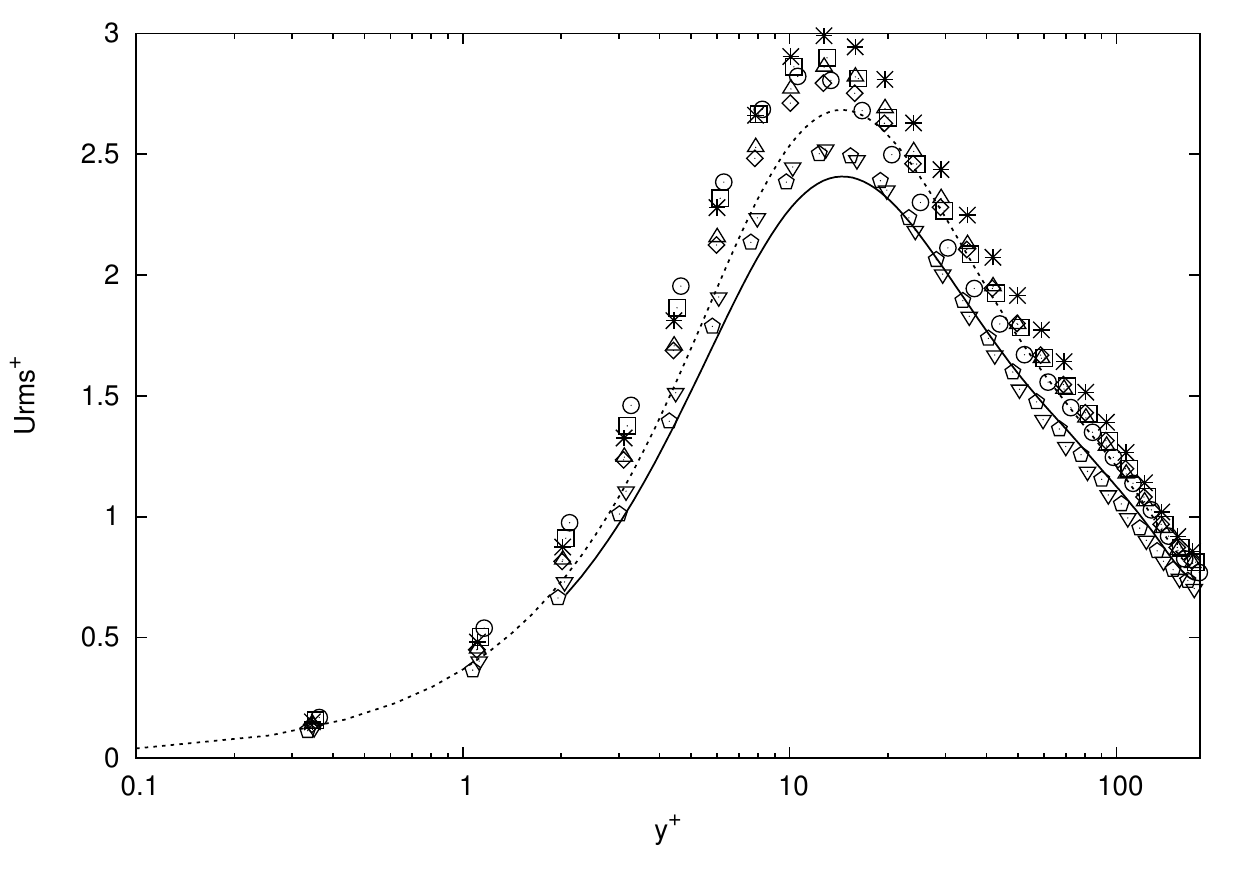}}
}\centerline{
\subfigtopleft{\includegraphics[width=0.44\textwidth, trim={0 5 5 5}, clip]{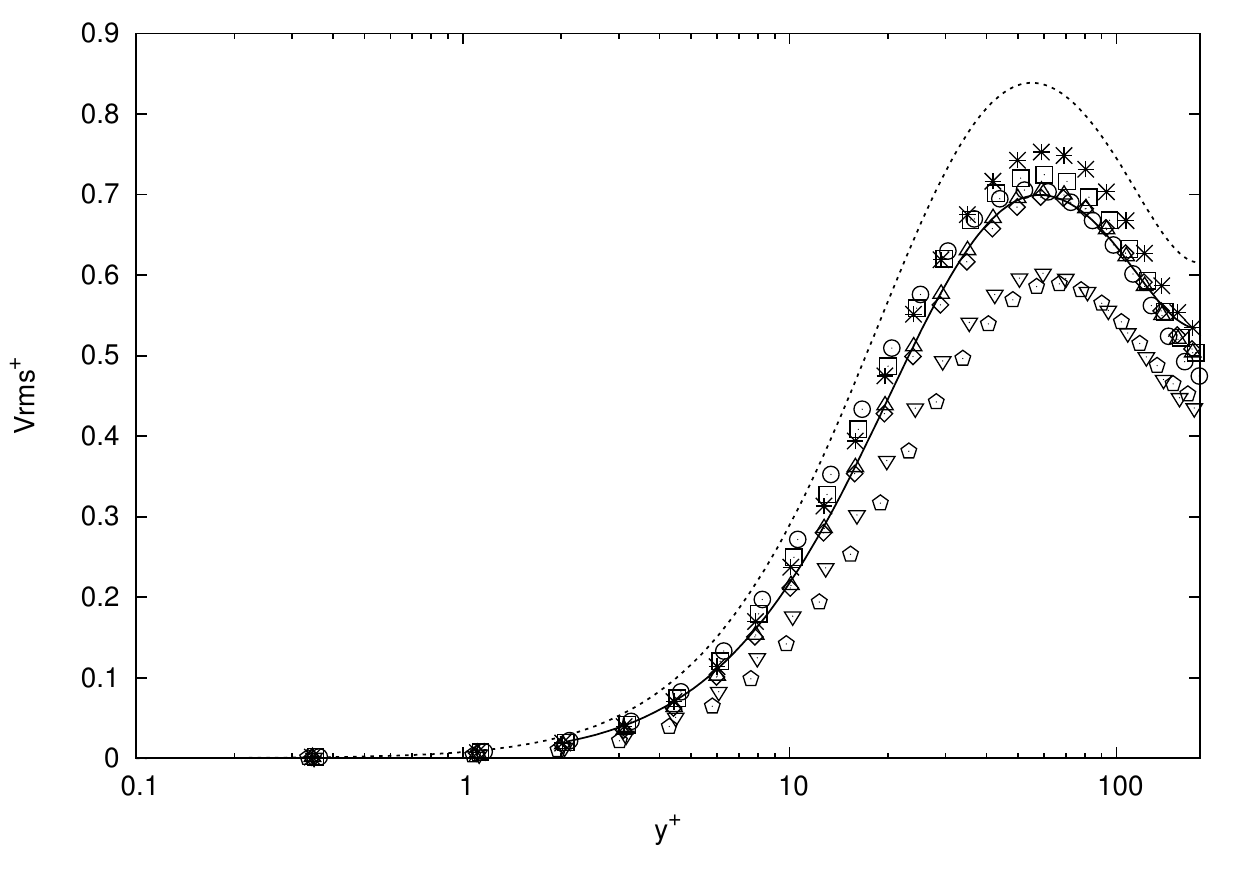}}
\subfigtopleft{\includegraphics[width=0.44\textwidth, trim={0 5 5 5}, clip]{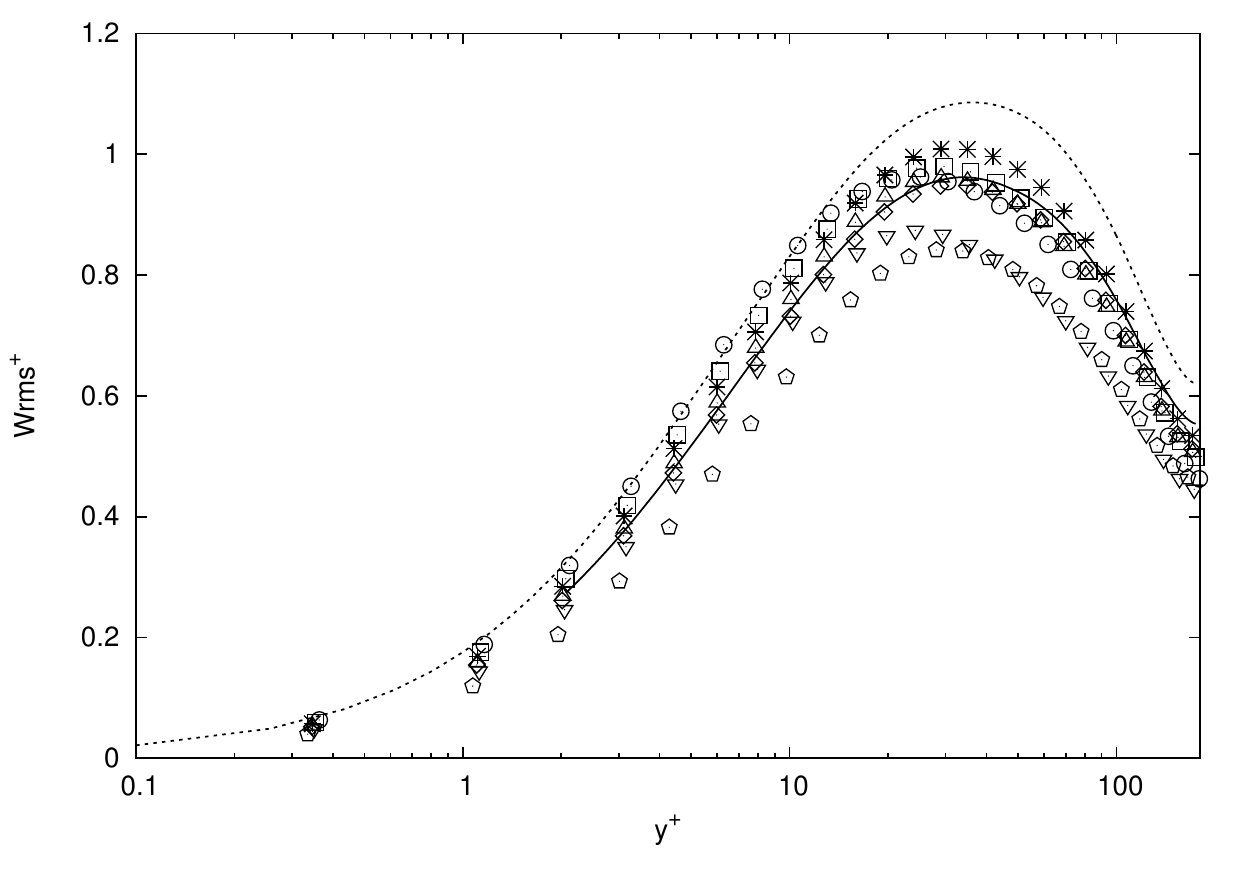}}
}
\caption[Comparison of large-eddy simulations with the gradient model using  $C^{\mathrm{Grad.}} = 1$ and $C^{\mathrm{Grad.}} = 2$ and the filtered gradient model using $C^{\mathrm{Grad.}} = 2$ and $C^{\mathrm{Grad.}} = 9$ and with the scale-similarity model using filter T and filter A.]{
Comparison of large-eddy simulations
with the gradient model using  $C^{\mathrm{Grad.}} = 1$ and $C^{\mathrm{Grad.}} = 2$ and the filtered gradient model using $C^{\mathrm{Grad.}} = 2$ and $C^{\mathrm{Grad.}} = 9$ and with the scale-similarity model using filter T and filter A
for the profiles of the mean streamwise velocity $\left\langle U_x \right\rangle$ (a, b), the covariance of streamwise and wall-normal velocity $\left\langle u_{\smash[b]{x}}' u_{\smash[b]{y}}' \right\rangle$ (c), the standard deviation of streamwise velocity $\smash[t]{\sqrt{\left\langle u_{\smash[b]{x}}'^2 \right\rangle}}$ (d), wall-normal velocity $\smash[t]{\sqrt{\left\langle u_{\smash[b]{y}}'^2 \right\rangle}}$ (e) and spanwise velocity $\smash[t]{\sqrt{\left\langle u_{\smash[b]{z}}'^2 \right\rangle}}$ (f)
with the mesh 48B.
The filtered gradient model uses the filter A.
\label{label18}}
\end{figure*}

\begin{figure*}
\setcounter{subfigcounter}{0}
\centerline{
\subfigtopleft{\includegraphics[width=0.44\textwidth, trim={0 5 5 5}, clip]{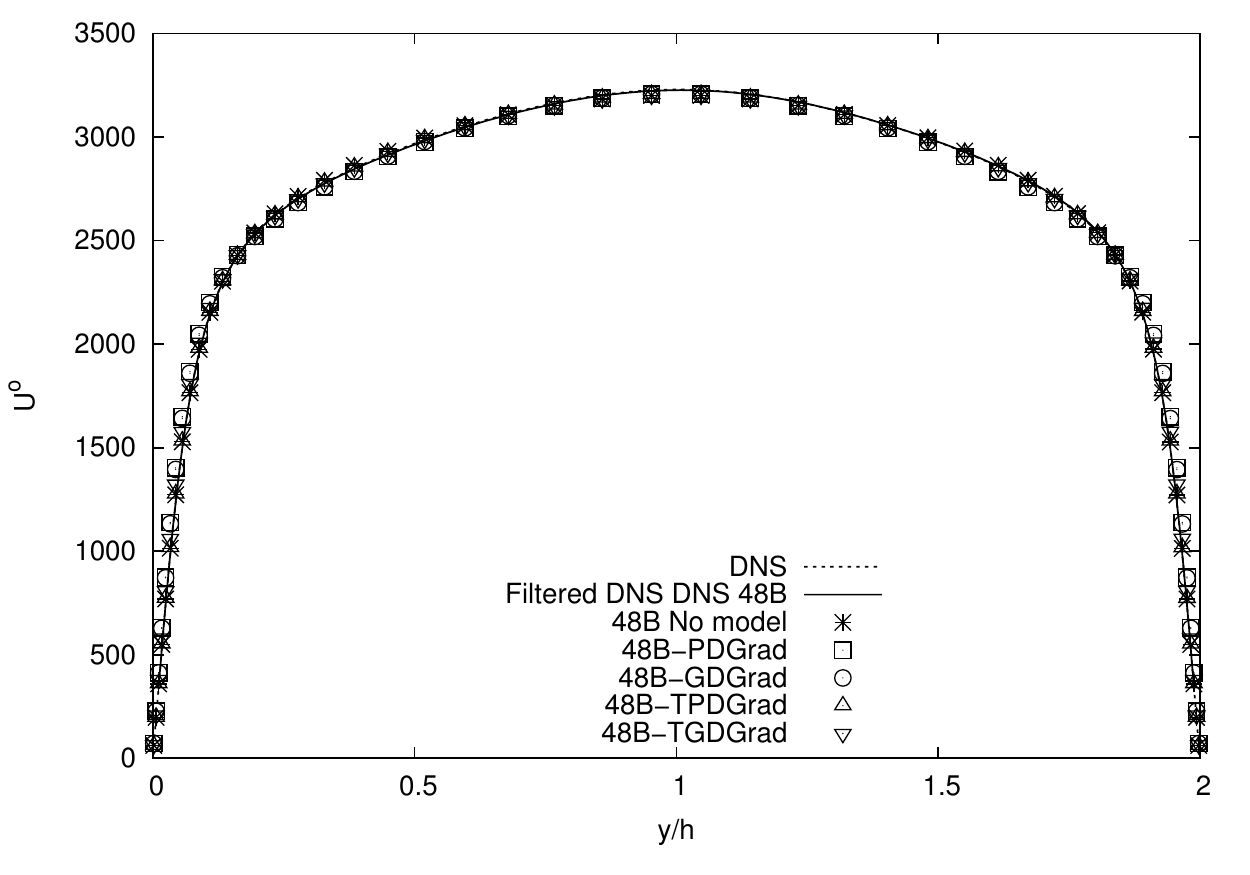}}
\subfigtopleft{\includegraphics[width=0.44\textwidth, trim={0 5 5 5}, clip]{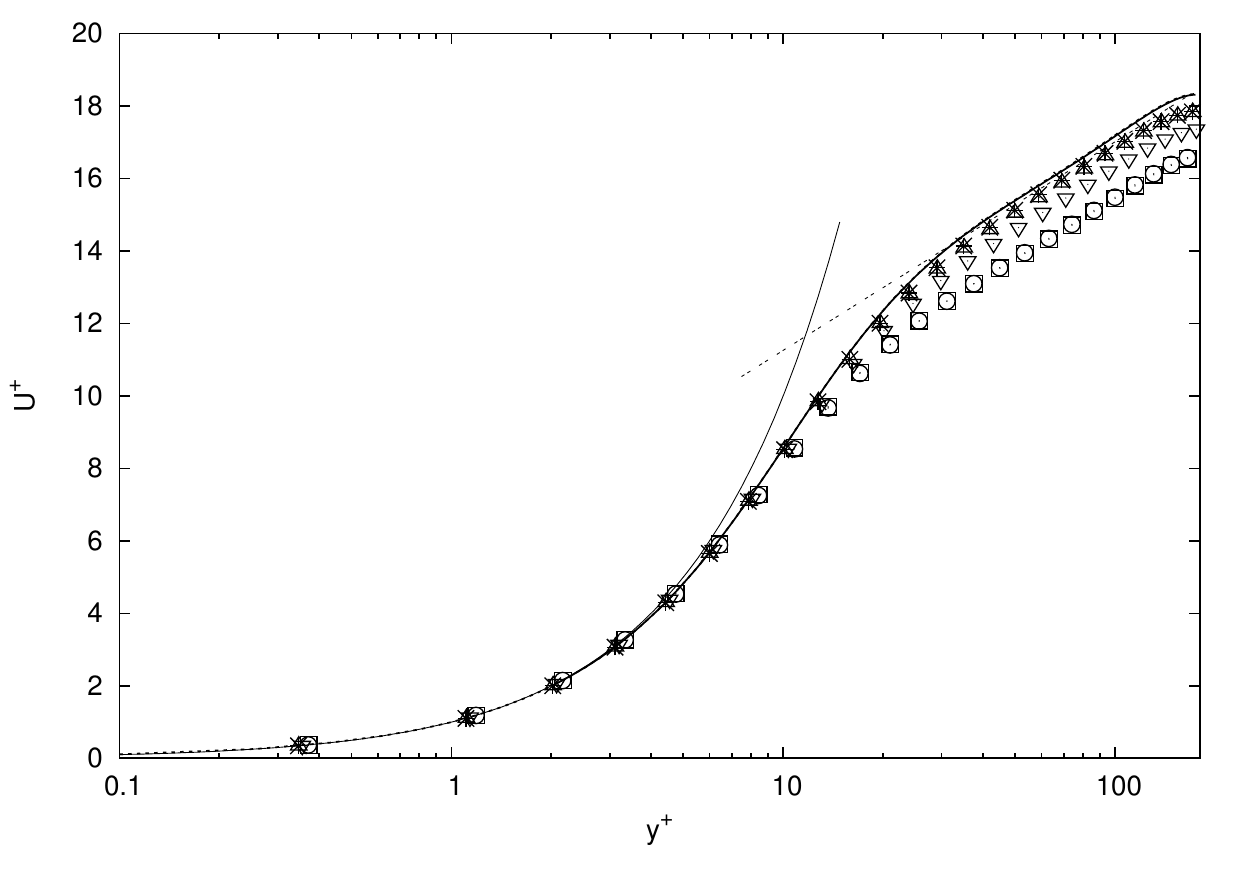}}
}\centerline{
\subfigtopleft{\includegraphics[width=0.44\textwidth, trim={0 5 5 5}, clip]{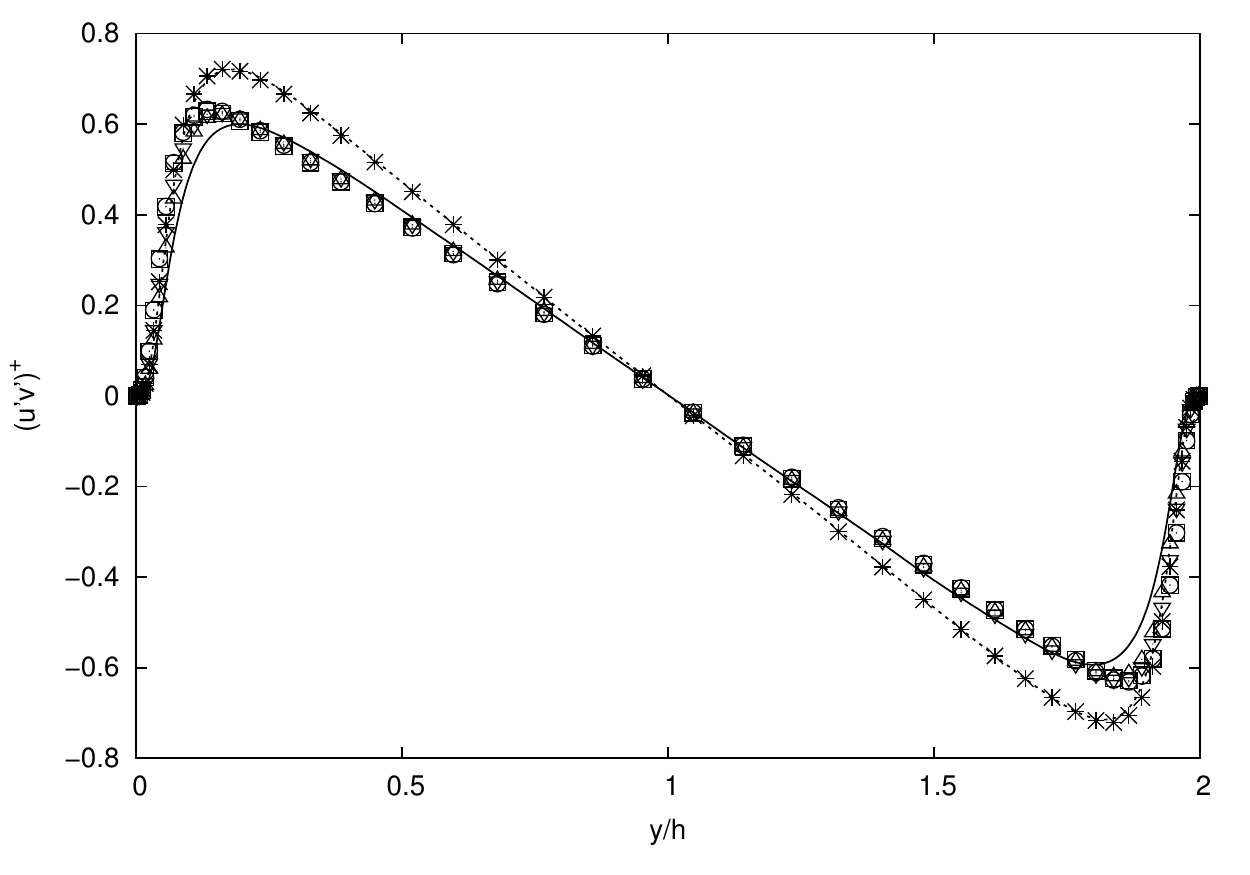}}
\subfigtopleft{\includegraphics[width=0.44\textwidth, trim={0 5 5 5}, clip]{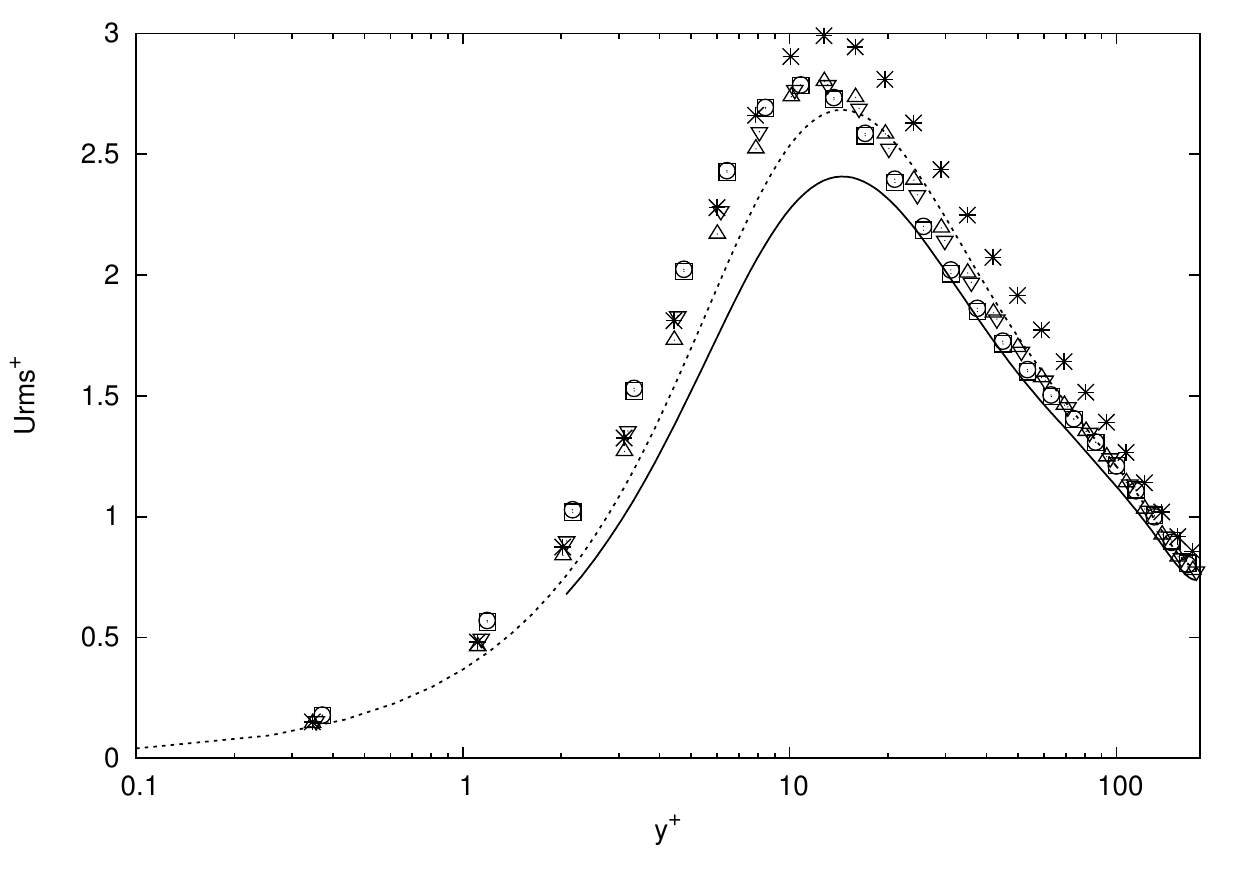}}
}\centerline{
\subfigtopleft{\includegraphics[width=0.44\textwidth, trim={0 5 5 5}, clip]{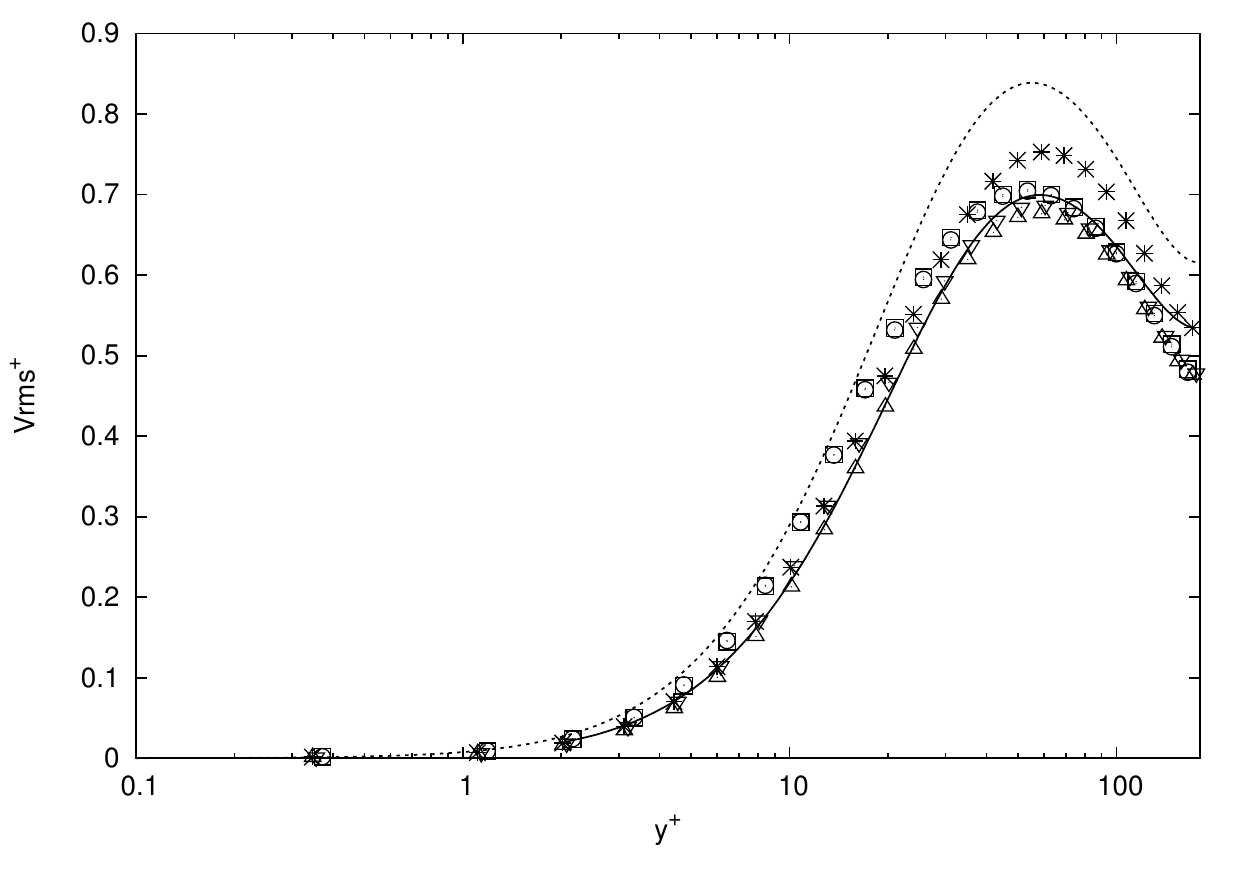}}
\subfigtopleft{\includegraphics[width=0.44\textwidth, trim={0 5 5 5}, clip]{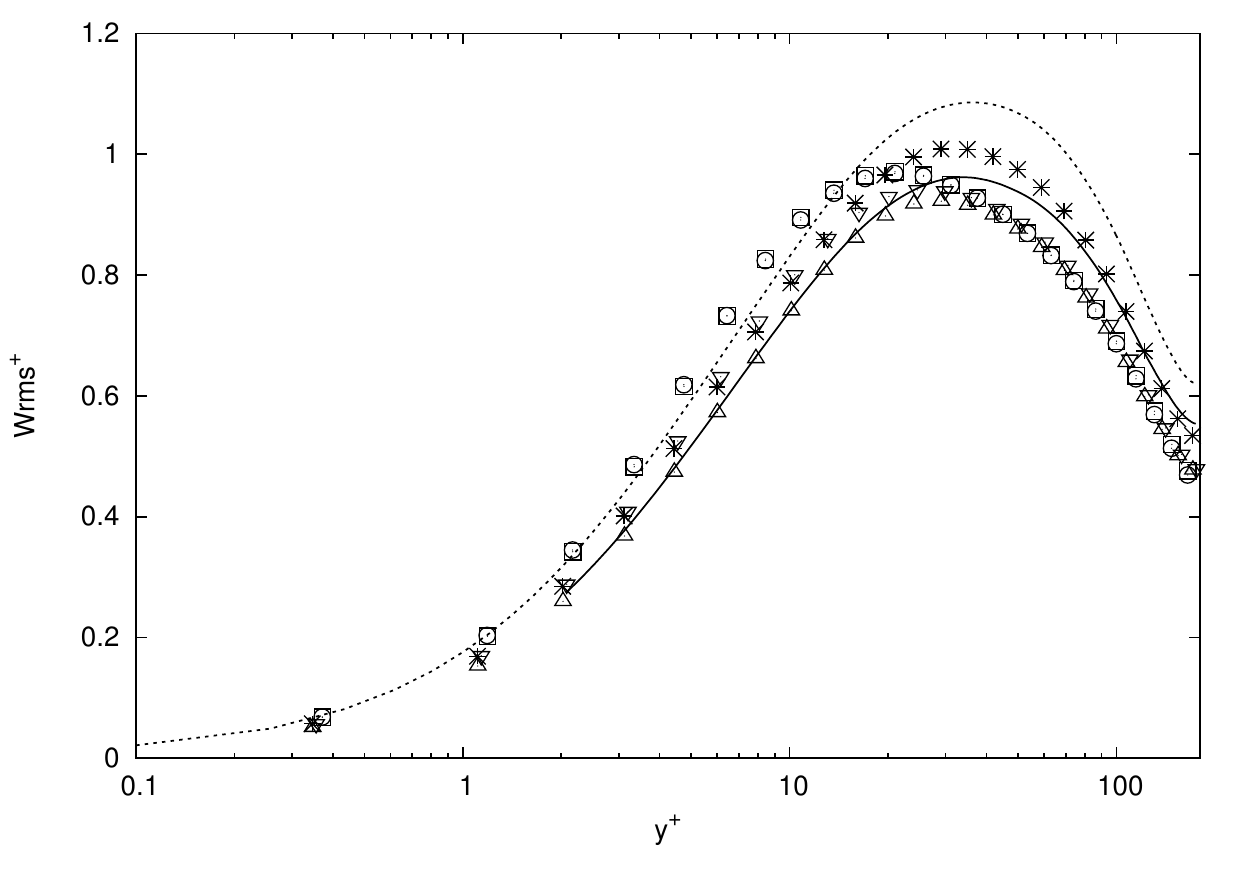}}
}
\caption[Comparison of large-eddy simulations with the plane-average, global-average, tensorial plane-average and tensorial global-average dynamic gradient models.]{
Comparison of large-eddy simulations
with the plane-average, global-average, tensorial plane-average and tensorial global-average dynamic gradient models
for the profiles of the mean streamwise velocity $\left\langle U_x \right\rangle$ (a, b), the covariance of streamwise and wall-normal velocity $\left\langle u_{\smash[b]{x}}' u_{\smash[b]{y}}' \right\rangle$ (c), the standard deviation of streamwise velocity $\smash[t]{\sqrt{\left\langle u_{\smash[b]{x}}'^2 \right\rangle}}$ (d), wall-normal velocity $\smash[t]{\sqrt{\left\langle u_{\smash[b]{y}}'^2 \right\rangle}}$ (e) and spanwise velocity $\smash[t]{\sqrt{\left\langle u_{\smash[b]{z}}'^2 \right\rangle}}$ (f)
with the mesh 48B.
\label{label20}}
\end{figure*}

\begin{figure*}
\setcounter{subfigcounter}{0}
\centerline{
\subfigtopleft{\includegraphics[width=0.44\textwidth, trim={0 5 5 5}, clip]{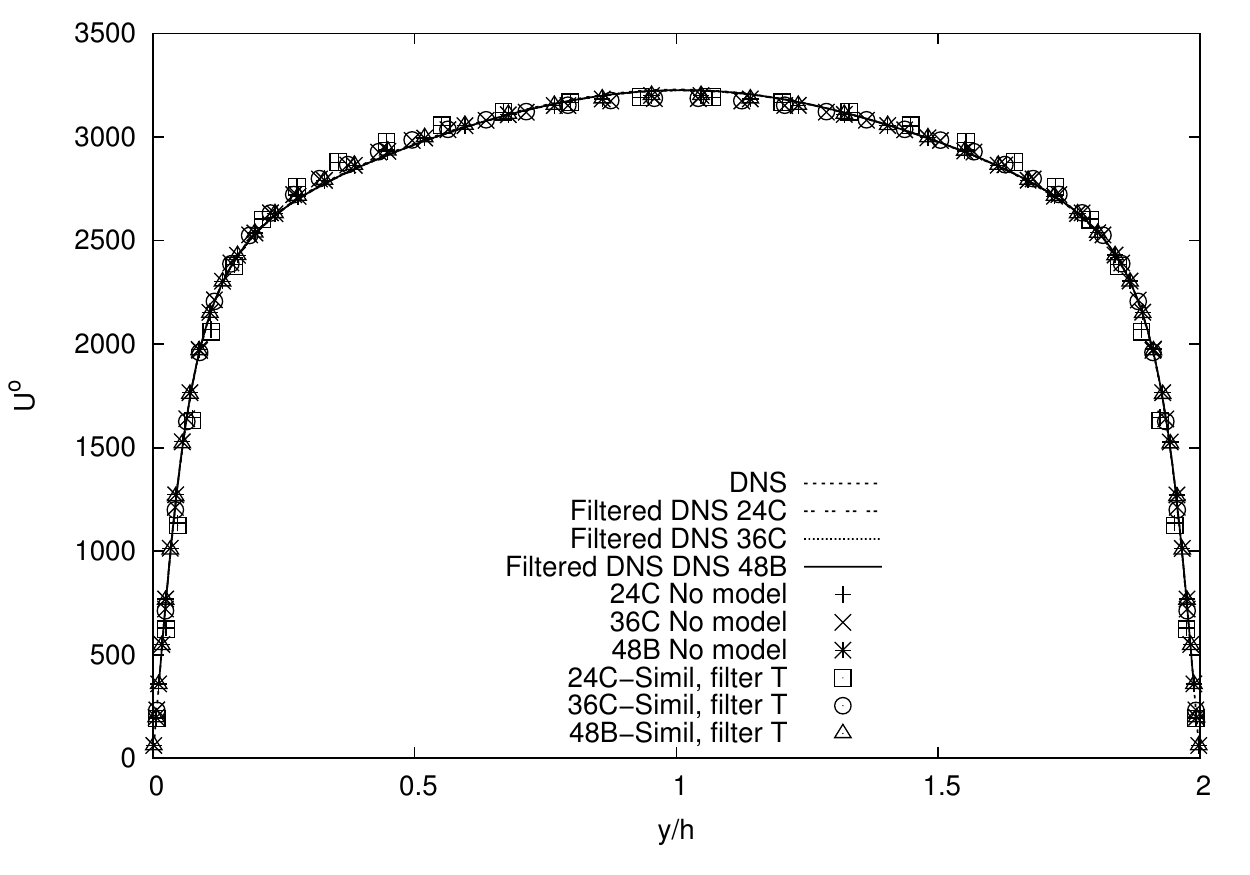}}
\subfigtopleft{\includegraphics[width=0.44\textwidth, trim={0 5 5 5}, clip]{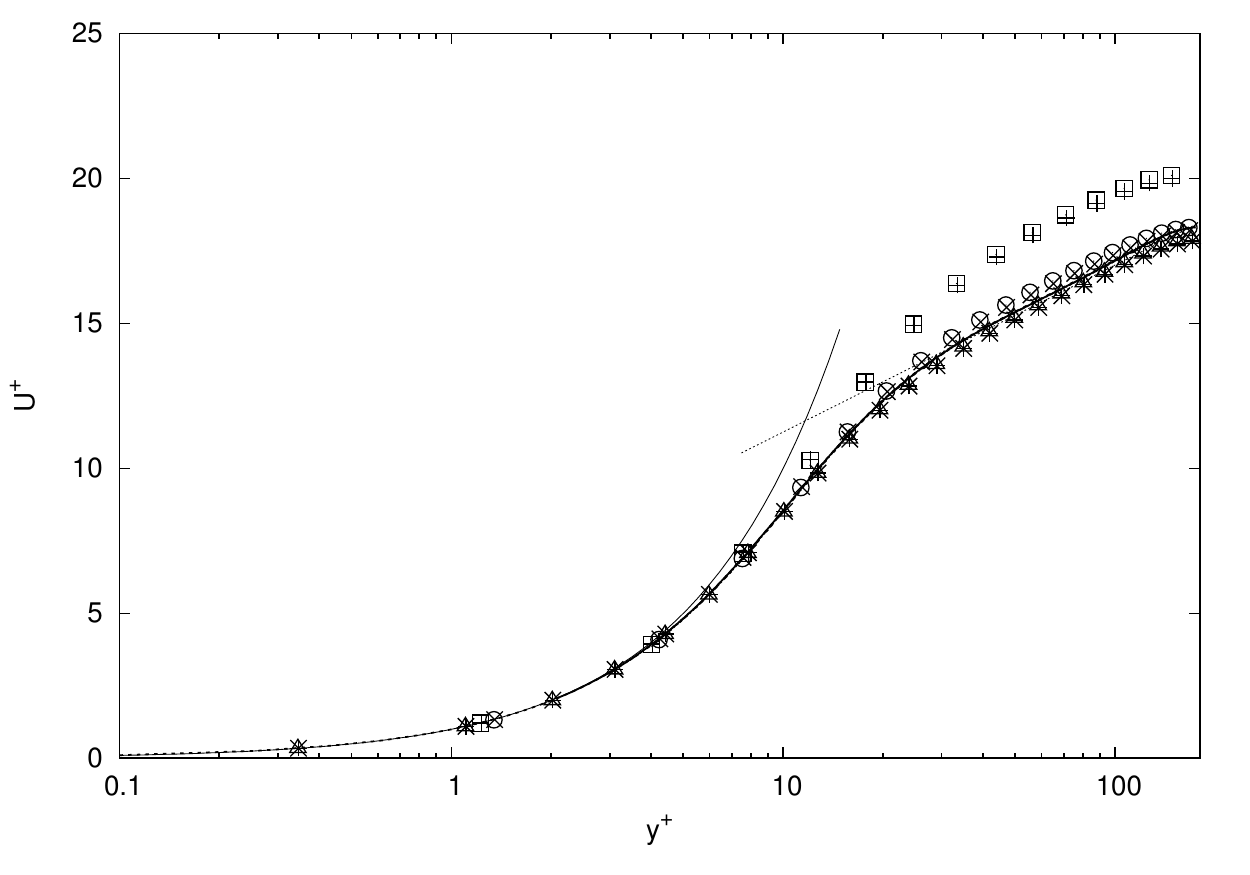}}
}\centerline{
\subfigtopleft{\includegraphics[width=0.44\textwidth, trim={0 5 5 5}, clip]{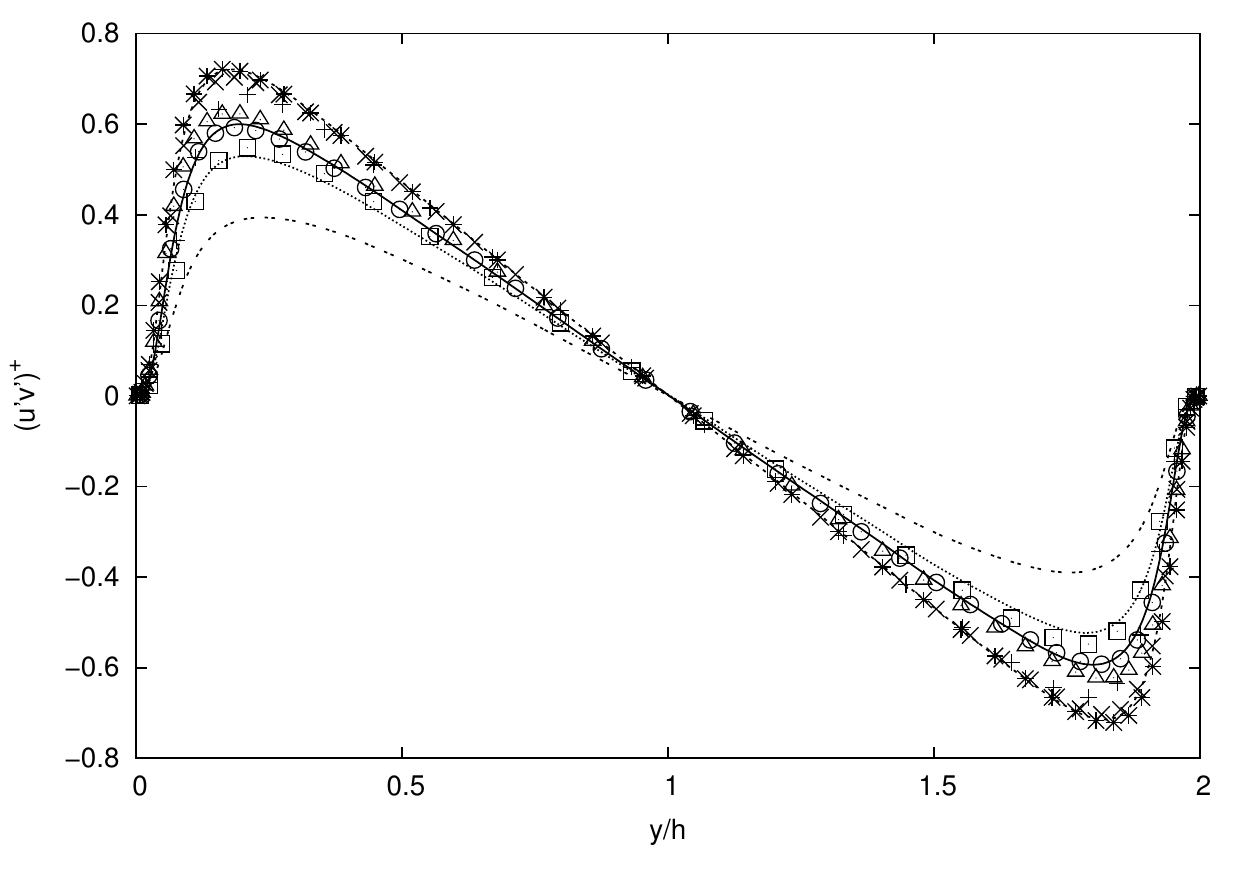}}
\subfigtopleft{\includegraphics[width=0.44\textwidth, trim={0 5 5 5}, clip]{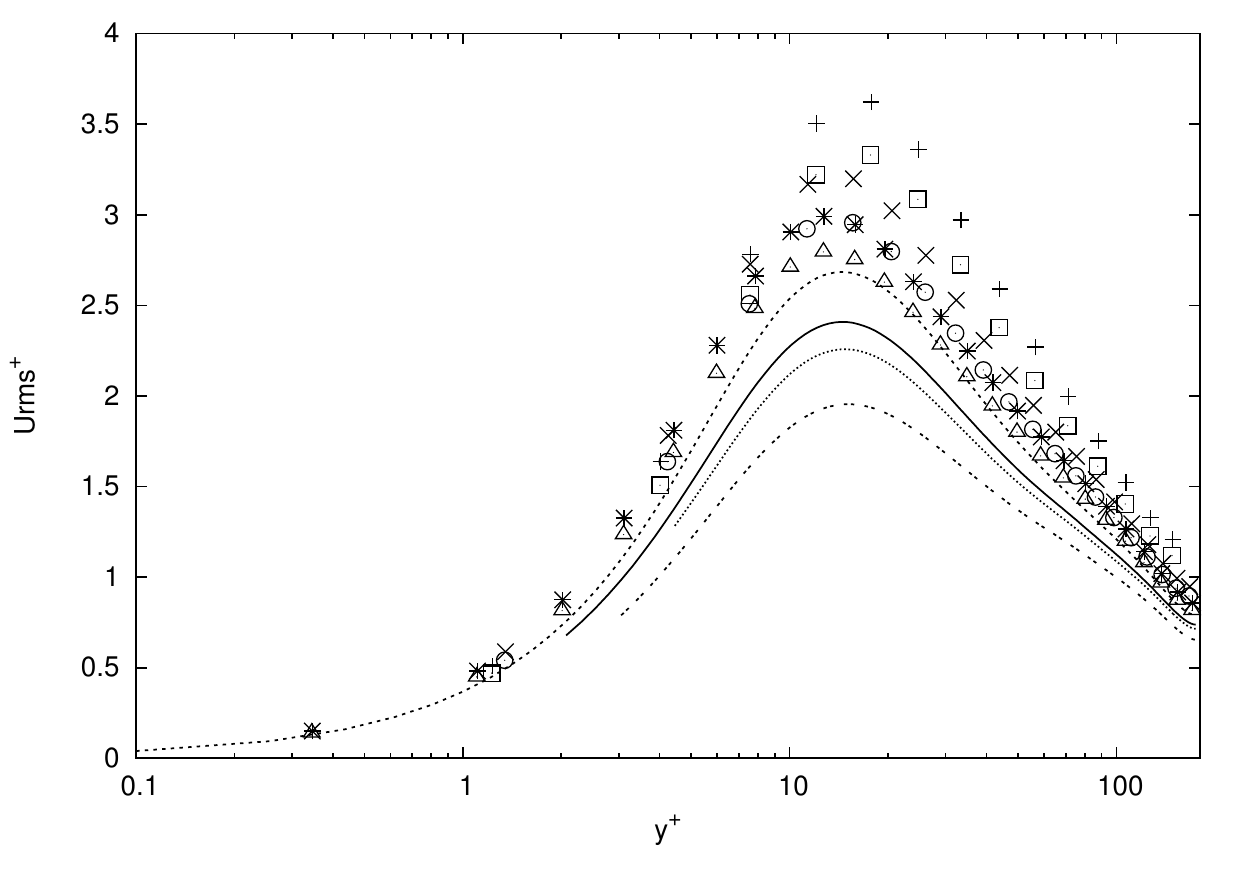}}
}\centerline{
\subfigtopleft{\includegraphics[width=0.44\textwidth, trim={0 5 5 5}, clip]{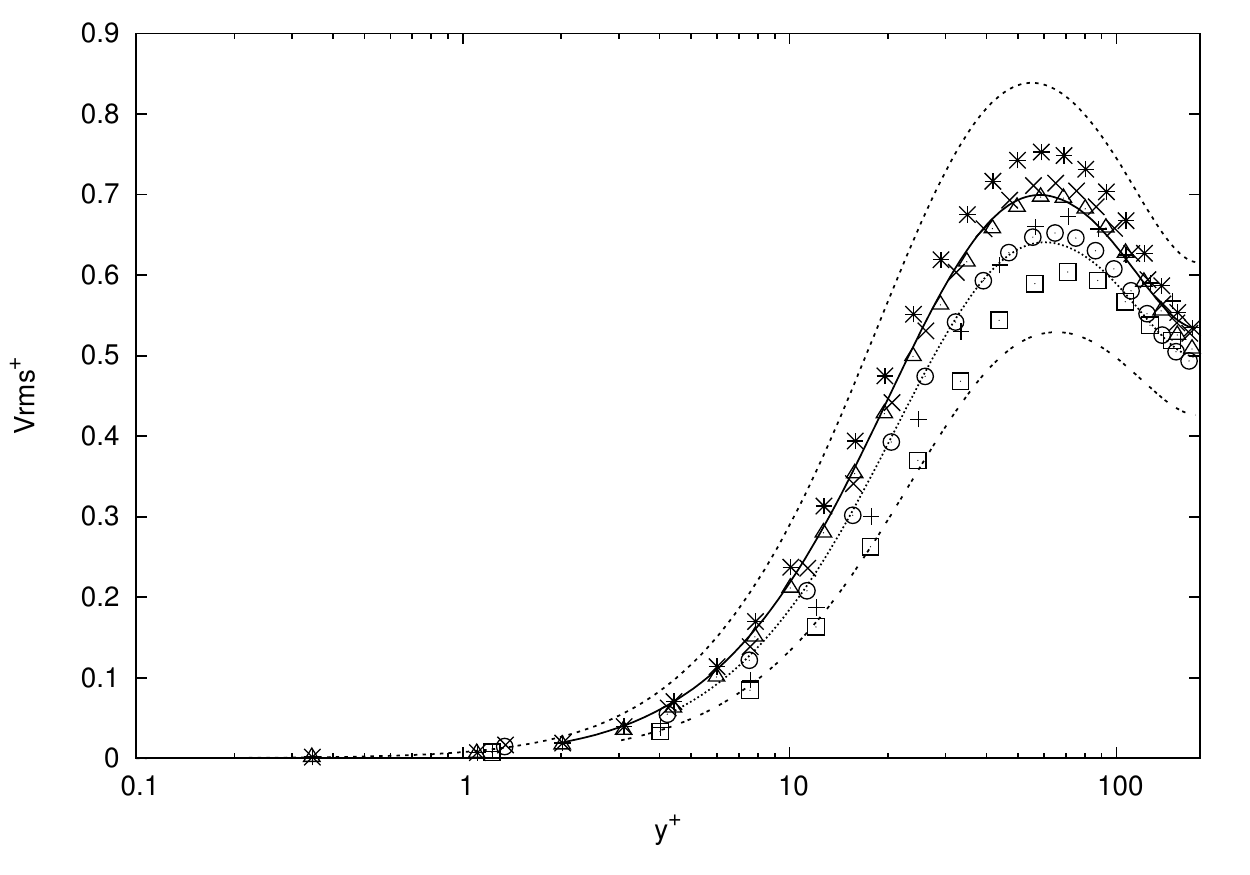}}
\subfigtopleft{\includegraphics[width=0.44\textwidth, trim={0 5 5 5}, clip]{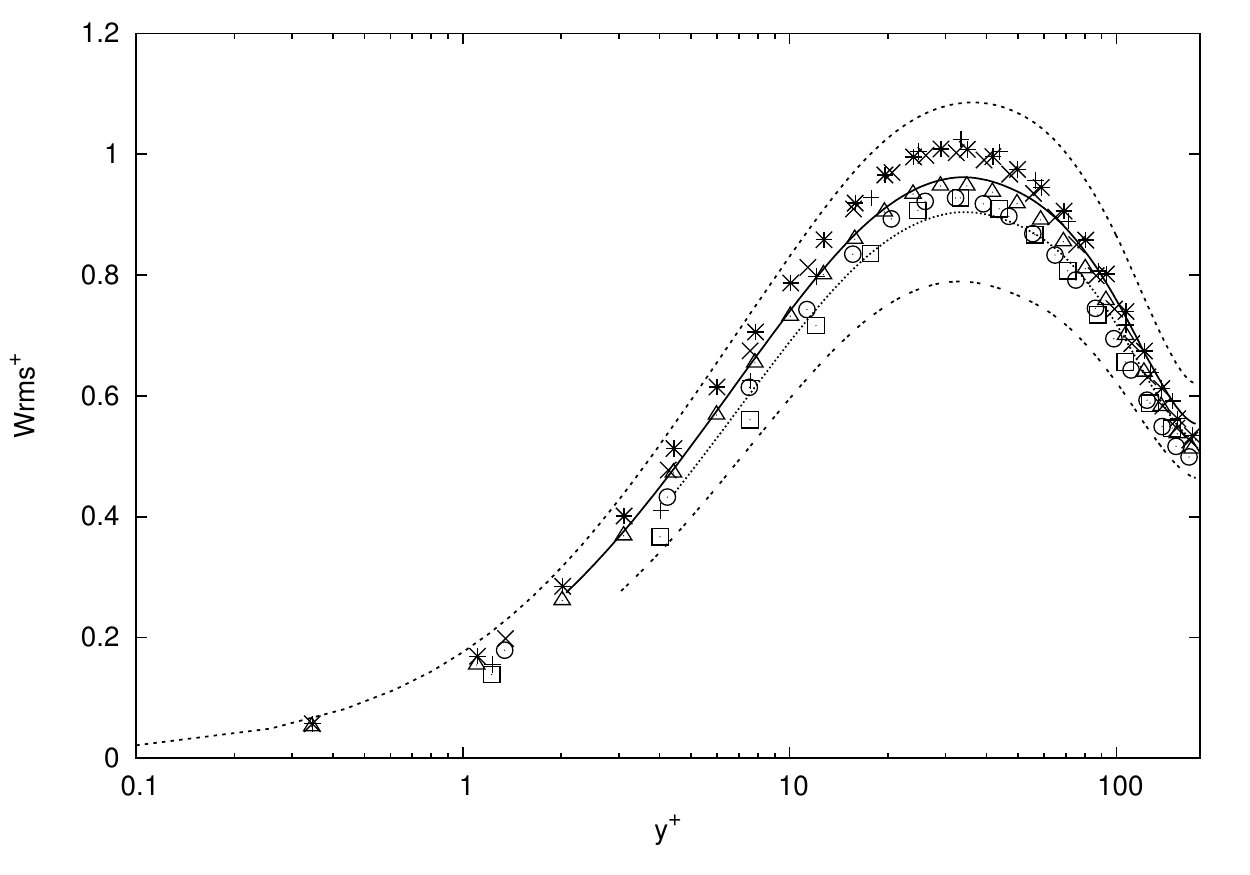}}
}
\caption[Comparison of large-eddy simulations with the scale-similarity model using filter T with the meshes 24C, 36C and 48B.]{
Comparison of large-eddy simulations
with the scale-similarity model using filter T
with the meshes 24C, 36C and 48B
for the profiles of the mean streamwise velocity $\left\langle U_x \right\rangle$ (a, b), the covariance of streamwise and wall-normal velocity $\left\langle u_{\smash[b]{x}}' u_{\smash[b]{y}}' \right\rangle$ (c), the standard deviation of streamwise velocity $\smash[t]{\sqrt{\left\langle u_{\smash[b]{x}}'^2 \right\rangle}}$ (d), wall-normal velocity $\smash[t]{\sqrt{\left\langle u_{\smash[b]{y}}'^2 \right\rangle}}$ (e) and spanwise velocity $\smash[t]{\sqrt{\left\langle u_{\smash[b]{z}}'^2 \right\rangle}}$ (f).
\label{label14}}
\end{figure*}

In this section, we investigate the structural modelling of the subgrid-scale tensor.
The structural models investigated are the gradient and scale-similarity
models, as well as the dynamic versions of the gradient model.
We give the results of large-eddy simulations
with the gradient and scale-similarity models
in figure \ref{label18}
with the mesh 48B.
The classical gradient model ($C^{\mathrm{Grad.}} = 1$) improves slightly the
standard deviation of streamwise velocity compared to the no-model
simulation, but deteriorates the profiles of the standard deviation of
wall-normal and spanwise velocity near the wall and the
prediction of the wall shear stress.
Nonetheless, the effects of the gradient model on the flow are rather small. 
To amplify the effects, we investigate gradient models
with a parameter $C^{\mathrm{Grad.}}$ larger than one.
The simulations are not stable using large multiplicative parameters.
The filtering of the gradient model improves the stability of the
simulation.
The resulting filtered gradient model may be seen as an
alternative formulation of the rational model
proposed by \citet{galdi2000approximation}\citep[see also][]{berselli2002mathematical, iliescu2003large, iliescu2004backscatter, berselli2005mathematics},
\begin{equation}
\tau_{ij}^{\mathrm{Grad, filtered}}(\vv{U}, \vv{\f{\Delta}}) = \FT{\tau_{ij}^{\mathrm{Grad.}}(\vv{U}, \vv{\f{\Delta}})} = \FT{\tfrac{1}{12} C^{\mathrm{Grad.}} \f{\Delta}_k^2 g_{ik} g_{jk}}.
\end{equation}
The test filter $\ft{\,\,\,\cdot\,\,}$ is computed using filter A.
The filtering alters the results of the simulation since
with $C^{\mathrm{Grad.}} = 2$, the predicted
wall shear stress is significantly different
for the nonfiltered and filtered gradient models (figure \ref{label18}).
With $C^{\mathrm{Grad.}} = 9$, the filtered gradient model
leads to a standard deviation of streamwise velocity
at the level of the filtered direct numerical simulation.
However, the covariance of streamwise and wall-normal
velocity and the standard deviation of wall-normal and spanwise
velocity are underestimated. Hence, there is no Pareto improvement compared
to the classical gradient model.

The plane-average and global-average dynamic gradient models give nearly
identical results because the plane-average dynamic parameter does not
strongly depend on the wall-normal coordinate (figure \ref{label20}).
They amplify on average the gradient model (table \ref{label7})
and provide similar results to the constant-parameter simulations.
The tensorial plane-average or global-average dynamic gradient model
amplify each component of the gradient model but increase in particular 
the relative amplitude of the ``$xx$'', ``$xy$'' and ``$yy$'' components (table \ref{label8}).
The tensorial dynamic gradient models, and the plane-average dynamic gradient
model in particular, provide a more accurate prediction of the wall shear
stress and the near-wall profile of the standard deviation of
wall-normal and spanwise velocity (figure \ref{label20}).

The results of the large-eddy simulations with the scale-similarity model
depend on the filter used (figure \ref{label18}).
Using filter A, the scale-similarity model is tied to the
gradient model with $C^{\mathrm{Grad.}} = 9$
according to the Taylor series expansion (\ref{label6}) of the test filter
in the scale-similarity model with $\ft{\Delta}_k^2 \approx 3 \f{\Delta}_k$.
Using filter T, the scale-similarity model is tied to the
gradient model with $C^{\mathrm{Grad.}} = 1$ since $\ft{\Delta}_k^2 \approx \f{\Delta}_k$.
However, the predictions with the scale-similarity and gradient model are not
the same, suggesting that the higher-order terms are relevant.
With the filter A, the scale-similarity model has with the mesh 48B
an excessive impact on the flow and deteriorates the profiles of the
turbulence statistics.
With the filter T, the model is more similar to the original model of \citet{bardina1980improved}.
The prediction of all turbulence statistics is with the mesh 48B improved
compared to the no-model simulation. In particular, the covariance of
streamwise and wall-normal velocity and the standard deviation of
wall-normal and spanwise velocity are in agreement with the filtered direct
numerical simulation. The standard deviation of wall-normal and
spanwise velocity remains overestimated but is decreased compared to the
no-model simulation.
These satisfactory results do not generalise very well to the 24C and 36C
meshes. Indeed, the effects of the scale-similarity model on the turbulence
statistics is similar for the three meshes and does not seem to correctly take
into account the variations of filter width. As a result, the covariance of
streamwise and wall-normal velocity and the standard deviation of
velocity components are overestimated with the meshes 24C and 36C (figure \ref{label14}).

\subsection{Tensorial models and tensorial mixed models}

\begin{figure*}
\setcounter{subfigcounter}{0}
\centerline{
\subfigtopleft{\includegraphics[width=0.44\textwidth, trim={0 5 5 5}, clip]{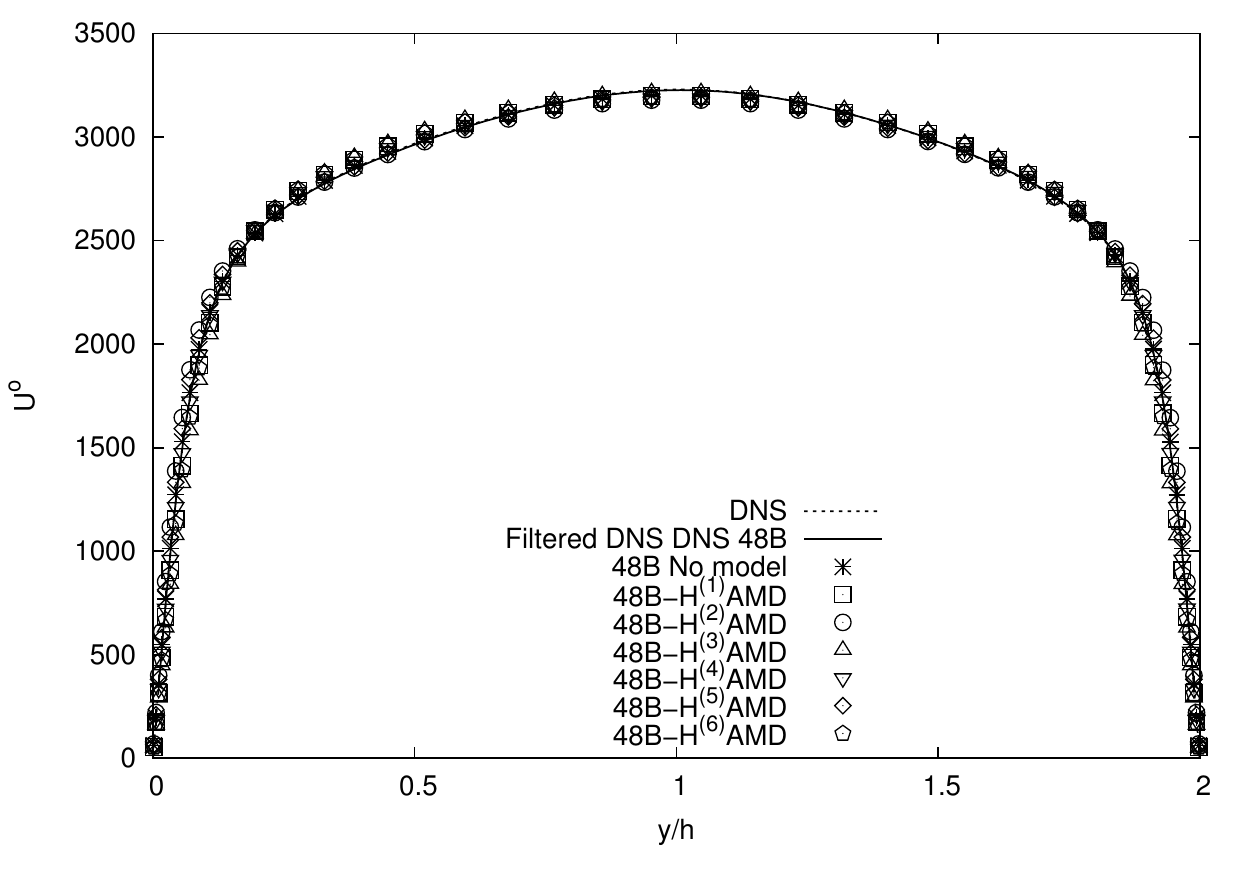}}
\subfigtopleft{\includegraphics[width=0.44\textwidth, trim={0 5 5 5}, clip]{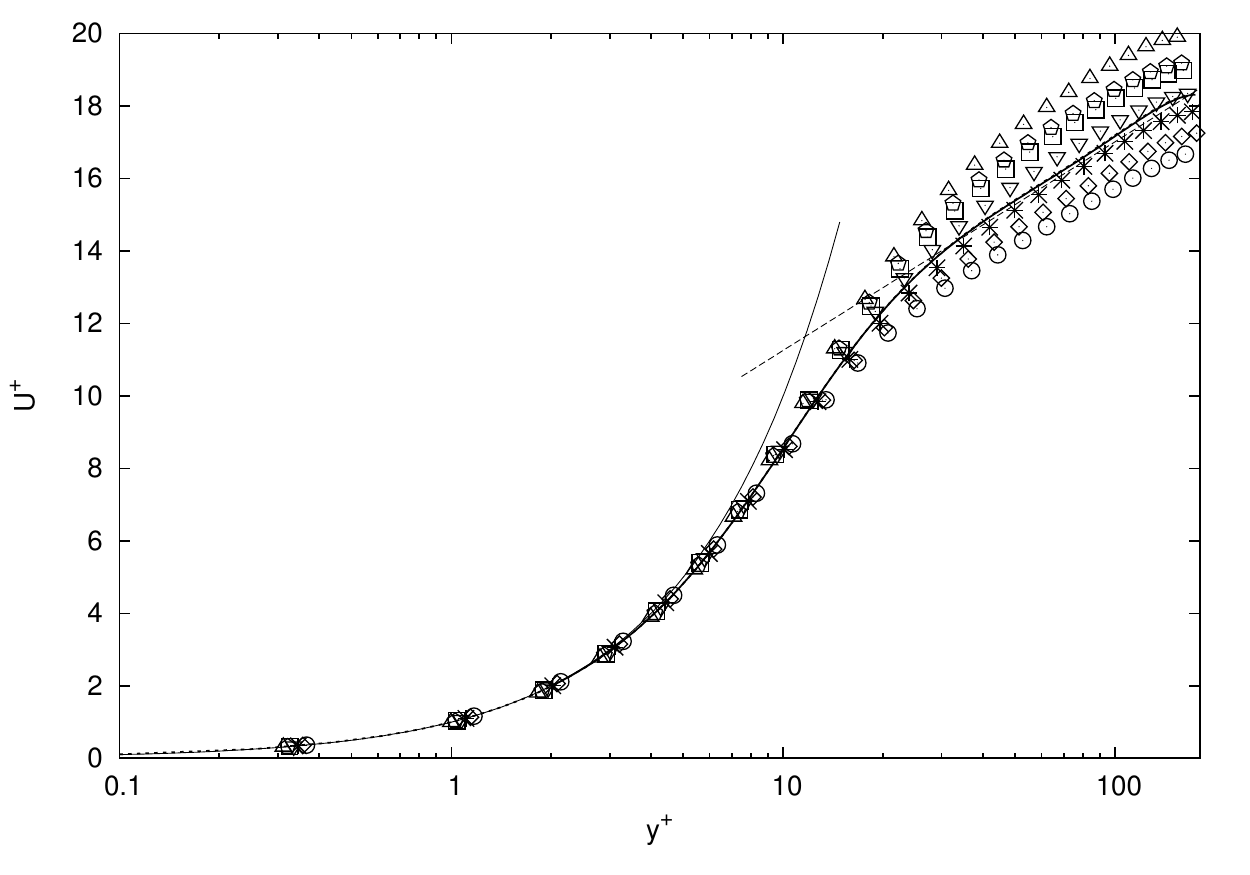}}
}\centerline{
\subfigtopleft{\includegraphics[width=0.44\textwidth, trim={0 5 5 5}, clip]{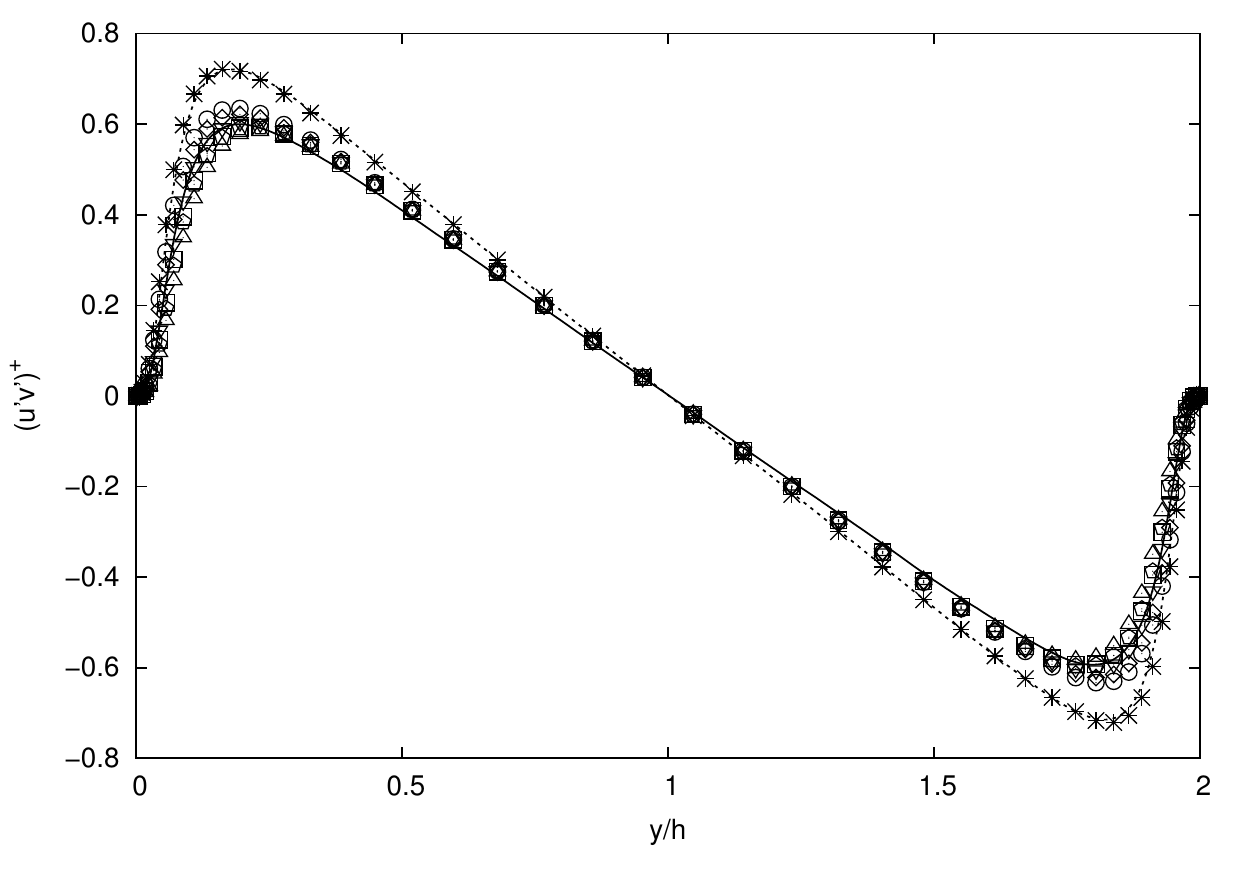}}
\subfigtopleft{\includegraphics[width=0.44\textwidth, trim={0 5 5 5}, clip]{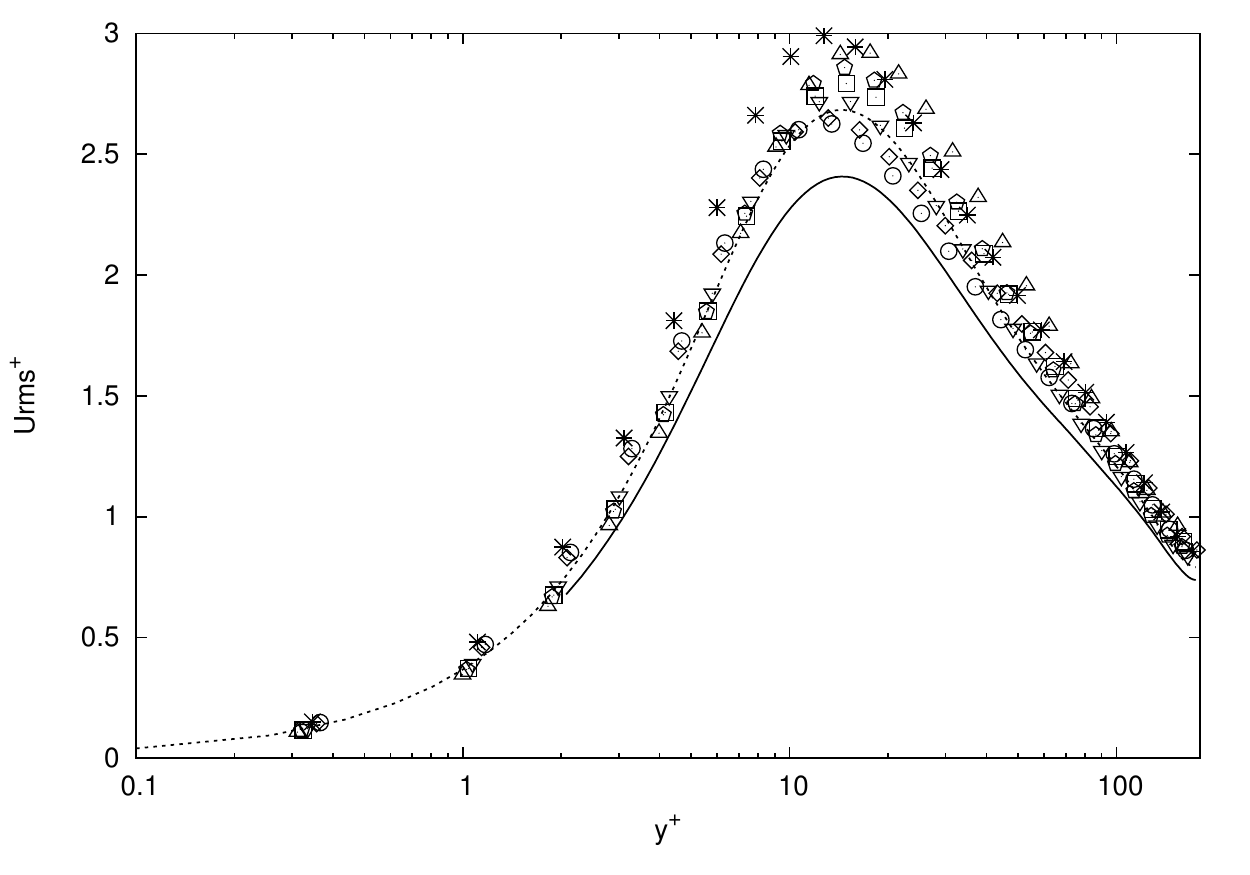}}
}\centerline{
\subfigtopleft{\includegraphics[width=0.44\textwidth, trim={0 5 5 5}, clip]{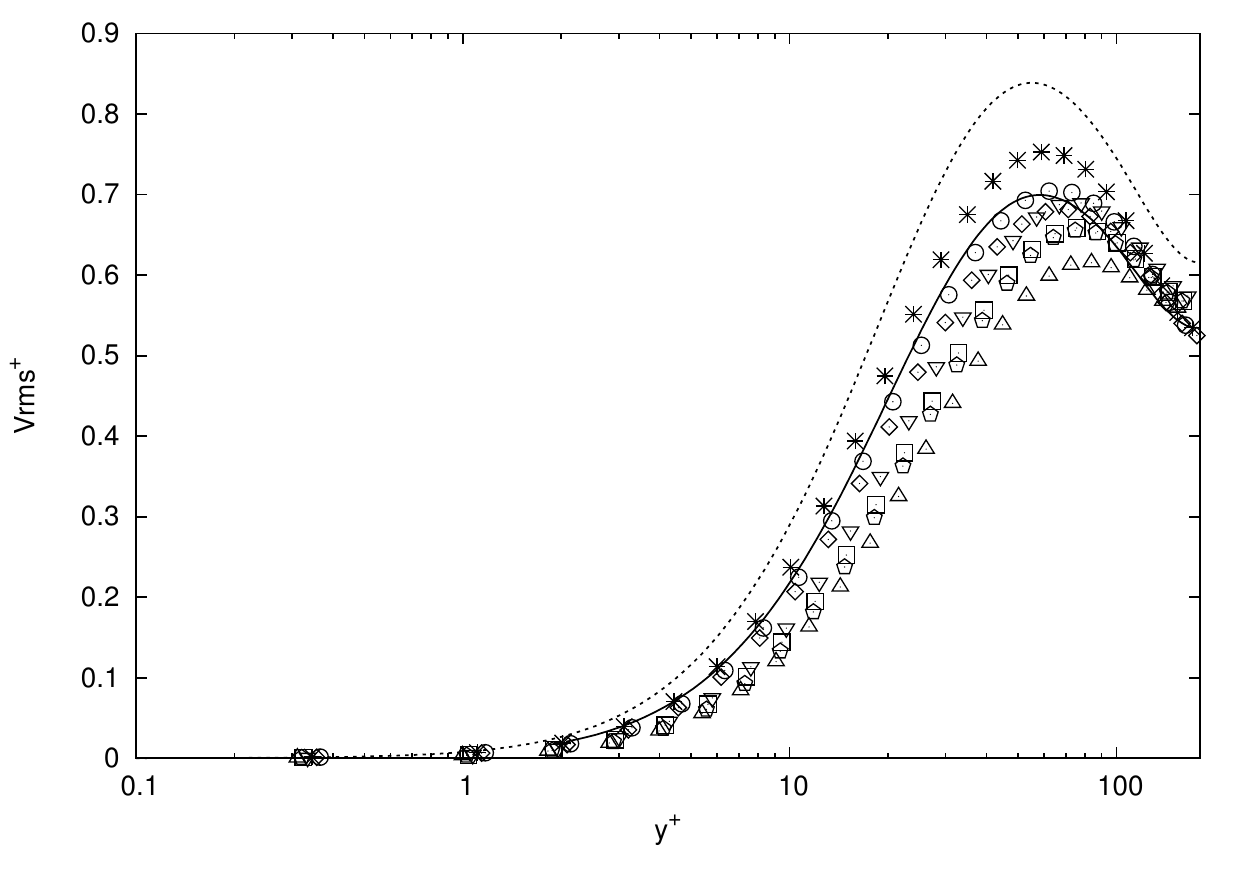}}
\subfigtopleft{\includegraphics[width=0.44\textwidth, trim={0 5 5 5}, clip]{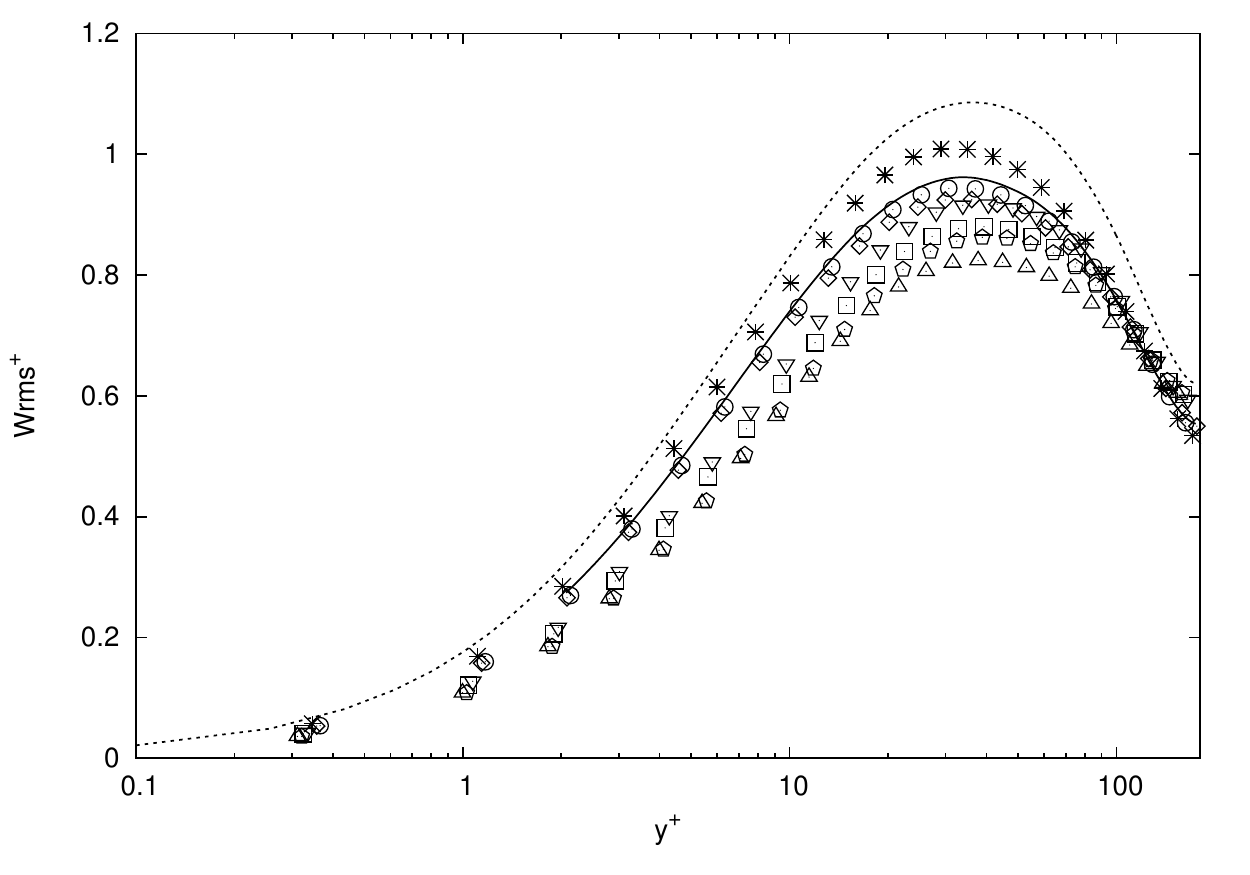}}
}
\caption[Comparison of large-eddy simulations with tensorial AMD models.]{
Comparison of large-eddy simulations
with tensorial AMD models
for the profiles of the mean streamwise velocity $\left\langle U_x \right\rangle$ (a, b), the covariance of streamwise and wall-normal velocity $\left\langle u_{\smash[b]{x}}' u_{\smash[b]{y}}' \right\rangle$ (c), the standard deviation of streamwise velocity $\smash[t]{\sqrt{\left\langle u_{\smash[b]{x}}'^2 \right\rangle}}$ (d), wall-normal velocity $\smash[t]{\sqrt{\left\langle u_{\smash[b]{y}}'^2 \right\rangle}}$ (e) and spanwise velocity $\smash[t]{\sqrt{\left\langle u_{\smash[b]{z}}'^2 \right\rangle}}$ (f)
with the mesh 48B.
\label{label27}}
\end{figure*}

\begin{figure*}
\setcounter{subfigcounter}{0}
\centerline{
\subfigtopleft{\includegraphics[width=0.44\textwidth, trim={0 5 5 5}, clip]{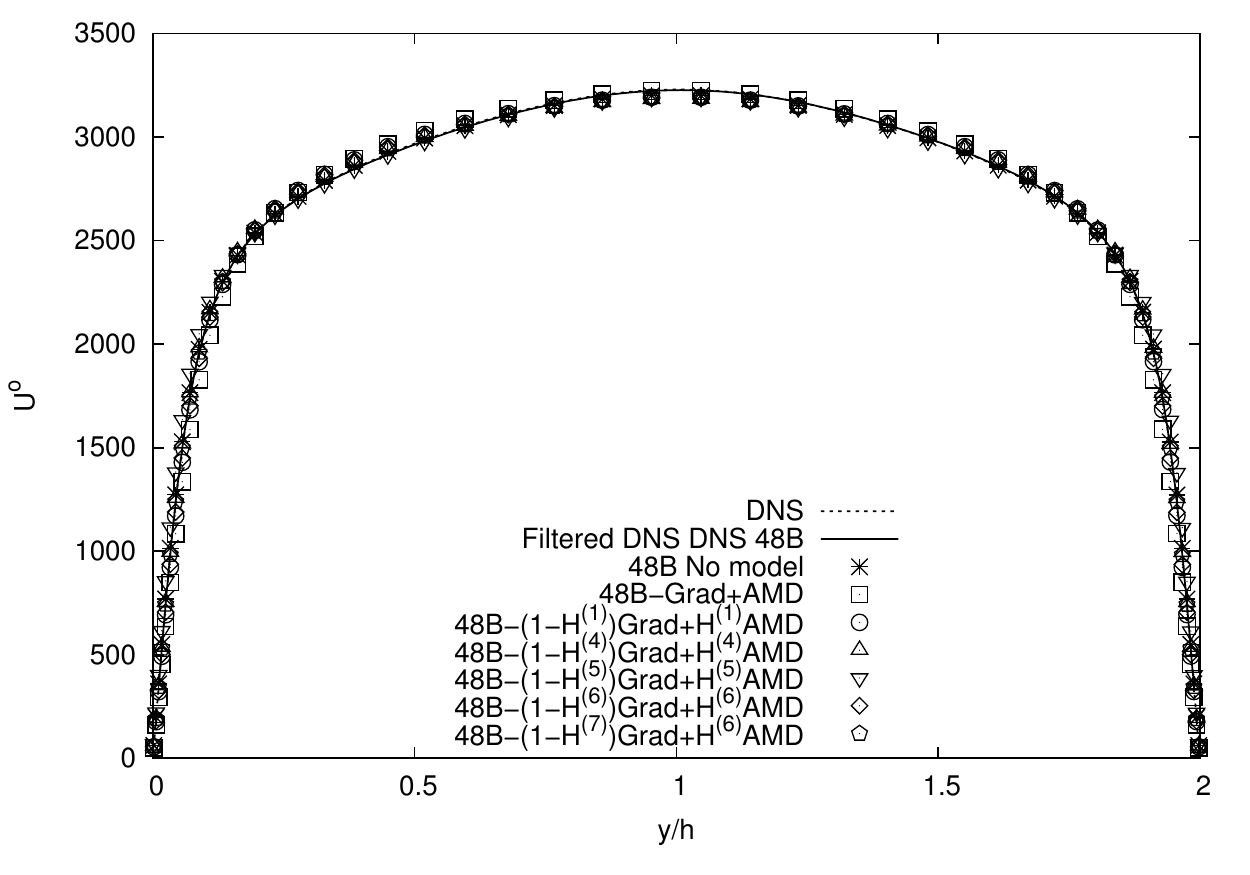}}
\subfigtopleft{\includegraphics[width=0.44\textwidth, trim={0 5 5 5}, clip]{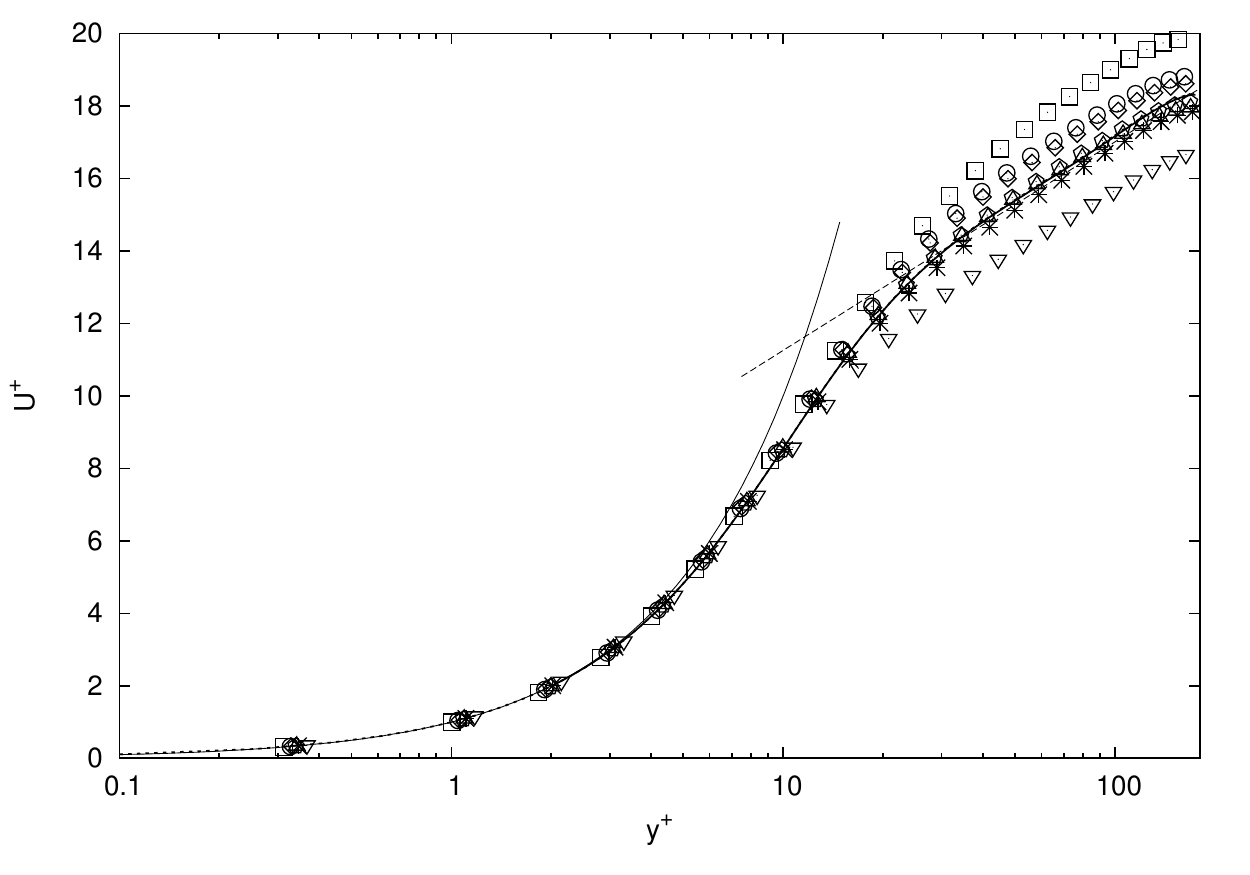}}
}\centerline{
\subfigtopleft{\includegraphics[width=0.44\textwidth, trim={0 5 5 5}, clip]{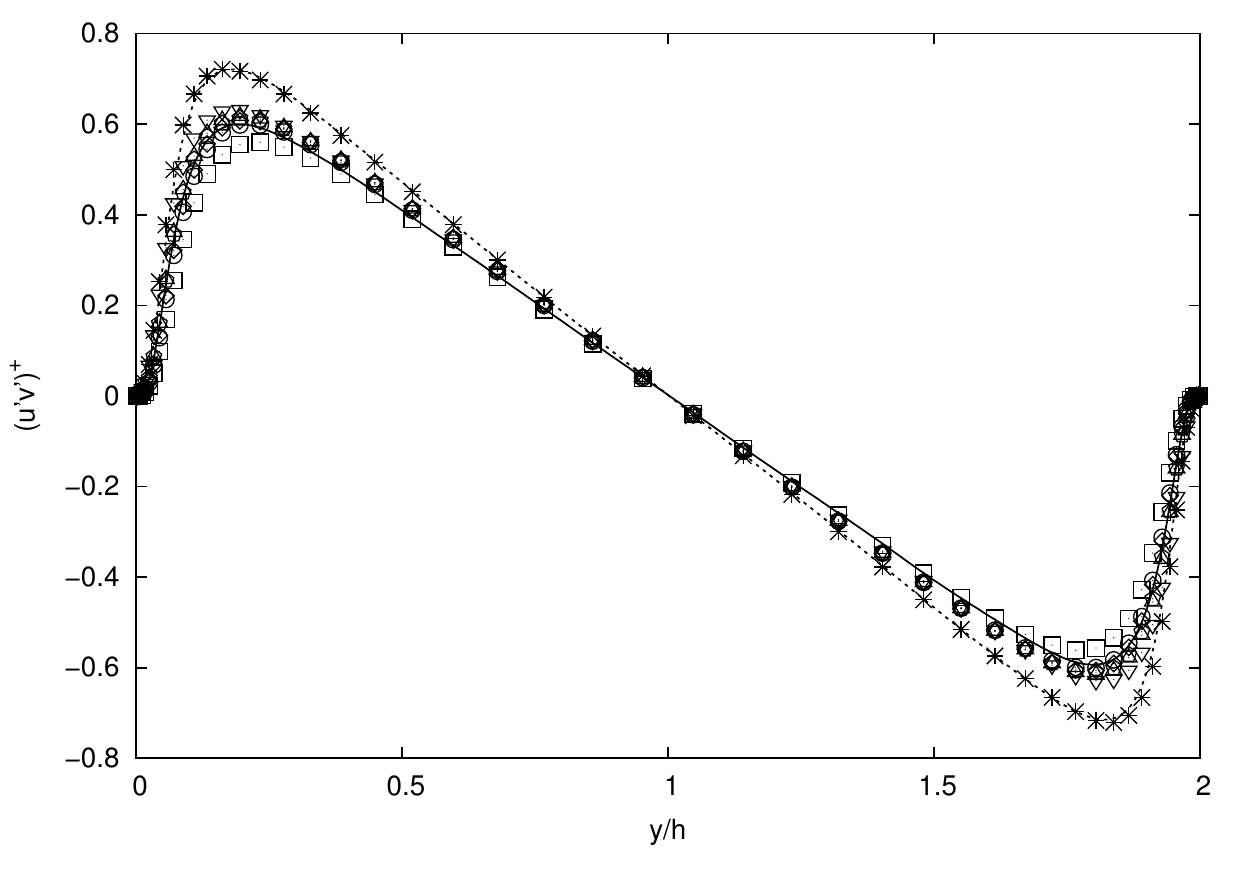}}
\subfigtopleft{\includegraphics[width=0.44\textwidth, trim={0 5 5 5}, clip]{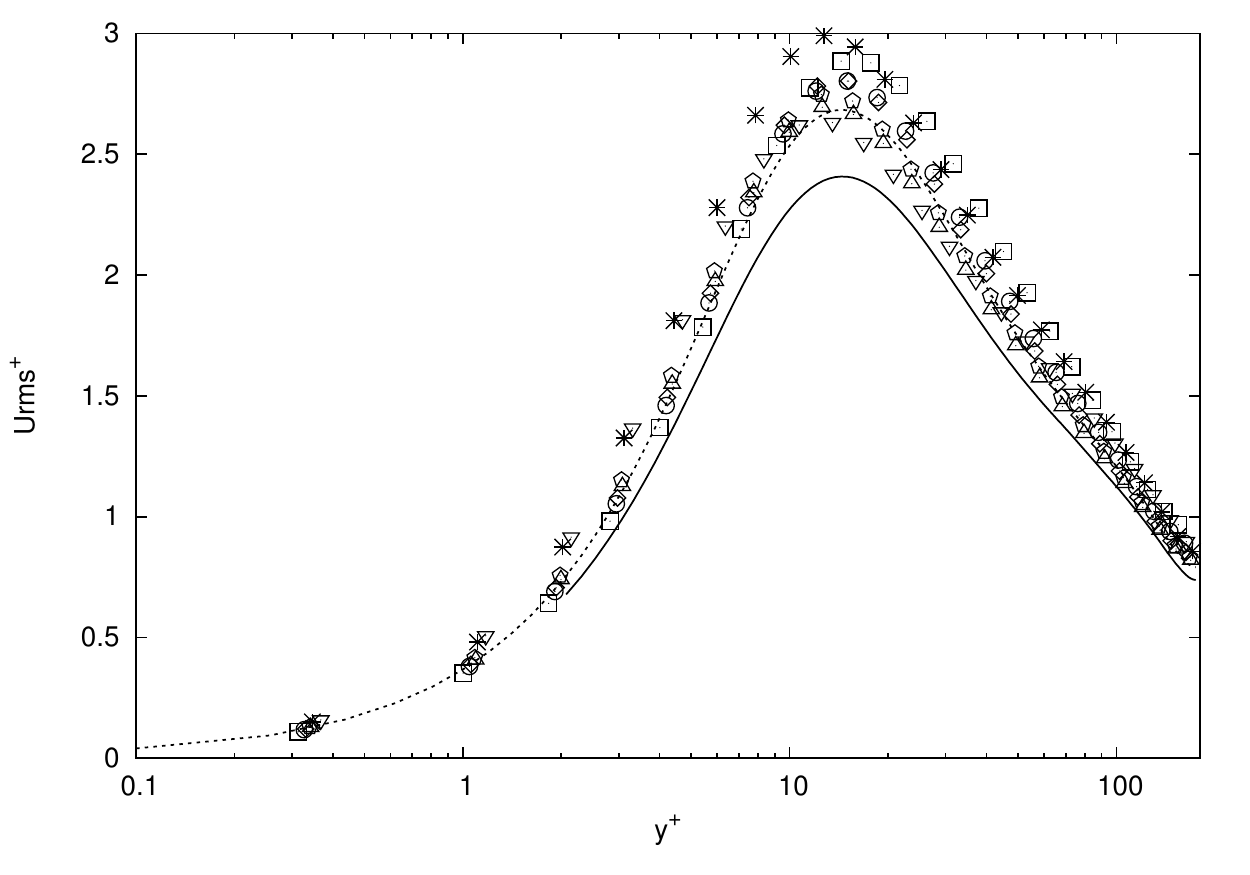}}
}\centerline{
\subfigtopleft{\includegraphics[width=0.44\textwidth, trim={0 5 5 5}, clip]{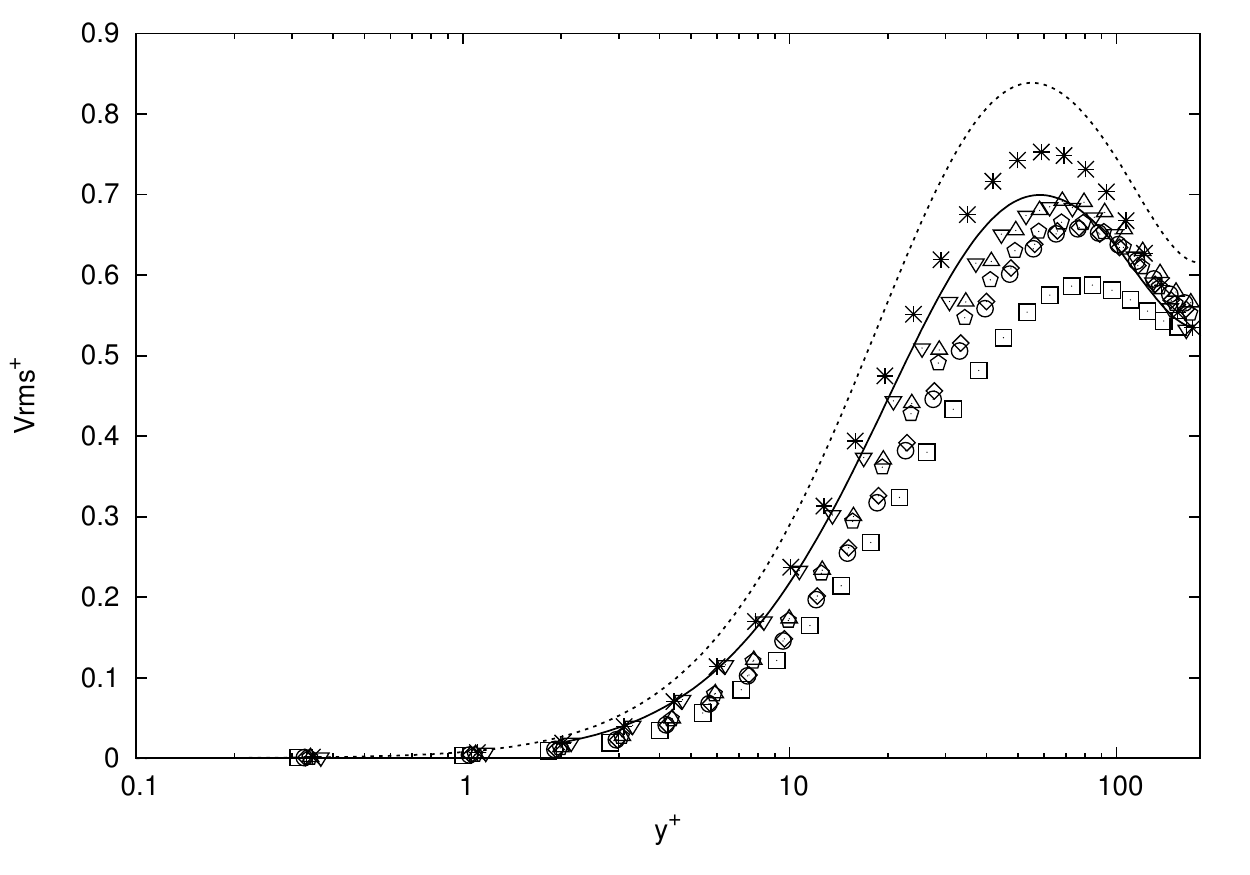}}
\subfigtopleft{\includegraphics[width=0.44\textwidth, trim={0 5 5 5}, clip]{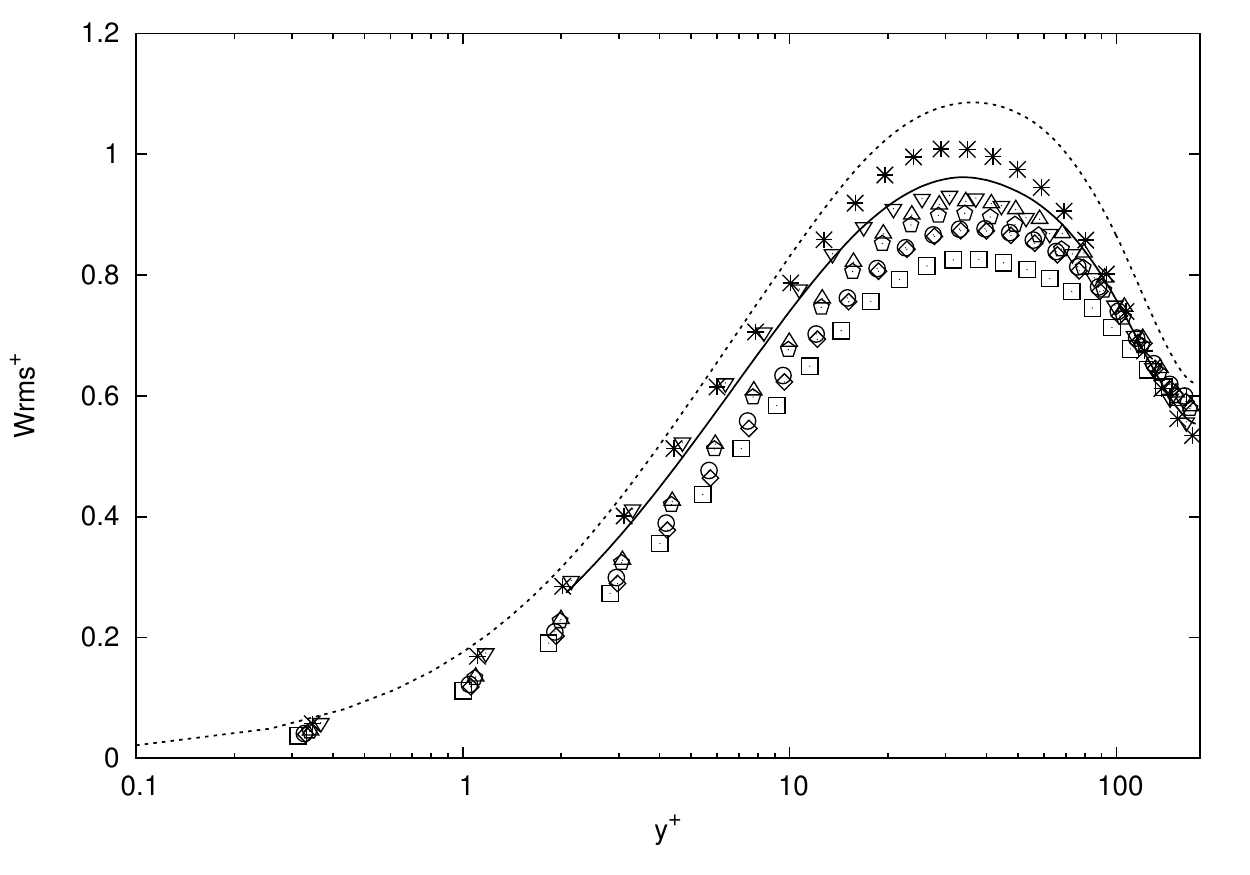}}
}
\caption[Comparison of large-eddy simulations with the gradient-AMD mixed model and tensorial gradient-AMD mixed models.]{
Comparison of large-eddy simulations
with the gradient-AMD mixed model and tensorial gradient-AMD mixed models
for the profiles of the mean streamwise velocity $\left\langle U_x \right\rangle$ (a, b), the covariance of streamwise and wall-normal velocity $\left\langle u_{\smash[b]{x}}' u_{\smash[b]{y}}' \right\rangle$ (c), the standard deviation of streamwise velocity $\smash[t]{\sqrt{\left\langle u_{\smash[b]{x}}'^2 \right\rangle}}$ (d), wall-normal velocity $\smash[t]{\sqrt{\left\langle u_{\smash[b]{y}}'^2 \right\rangle}}$ (e) and spanwise velocity $\smash[t]{\sqrt{\left\langle u_{\smash[b]{z}}'^2 \right\rangle}}$ (f)
with the mesh 48B.
\label{label17}}
\end{figure*}

\begin{figure*}
\setcounter{subfigcounter}{0}
\centerline{
\subfigtopleft{\includegraphics[width=0.44\textwidth, trim={0 5 5 5}, clip]{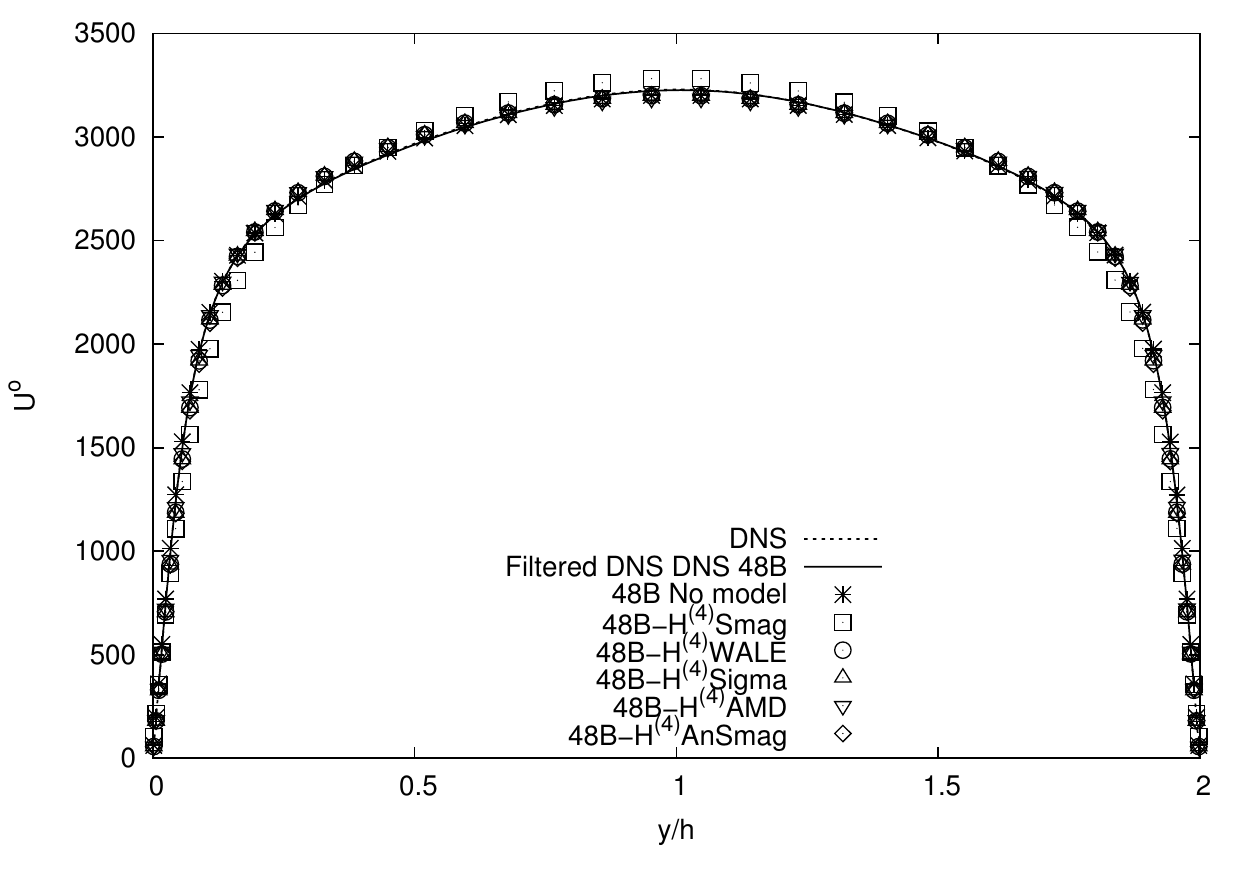}}
\subfigtopleft{\includegraphics[width=0.44\textwidth, trim={0 5 5 5}, clip]{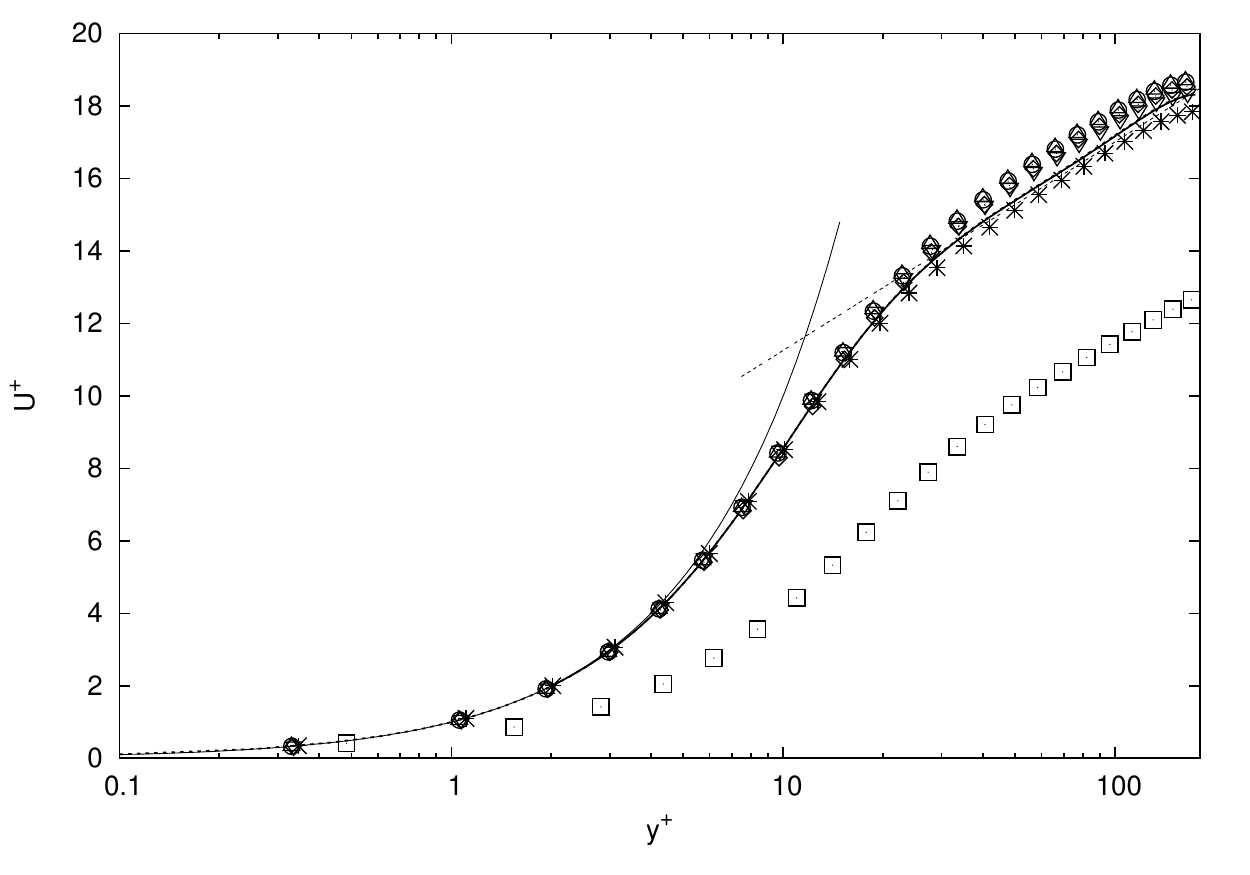}}
}\centerline{
\subfigtopleft{\includegraphics[width=0.44\textwidth, trim={0 5 5 5}, clip]{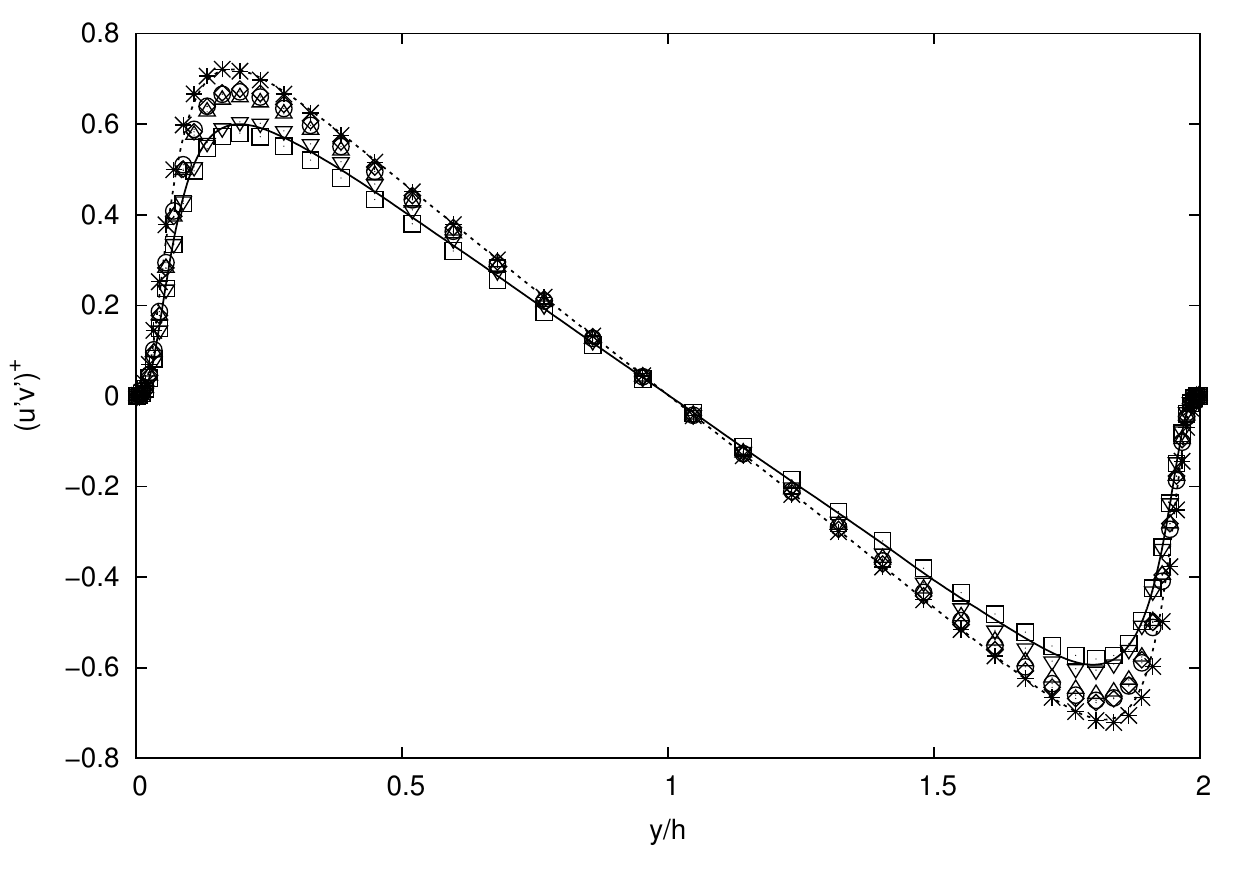}}
\subfigtopleft{\includegraphics[width=0.44\textwidth, trim={0 5 5 5}, clip]{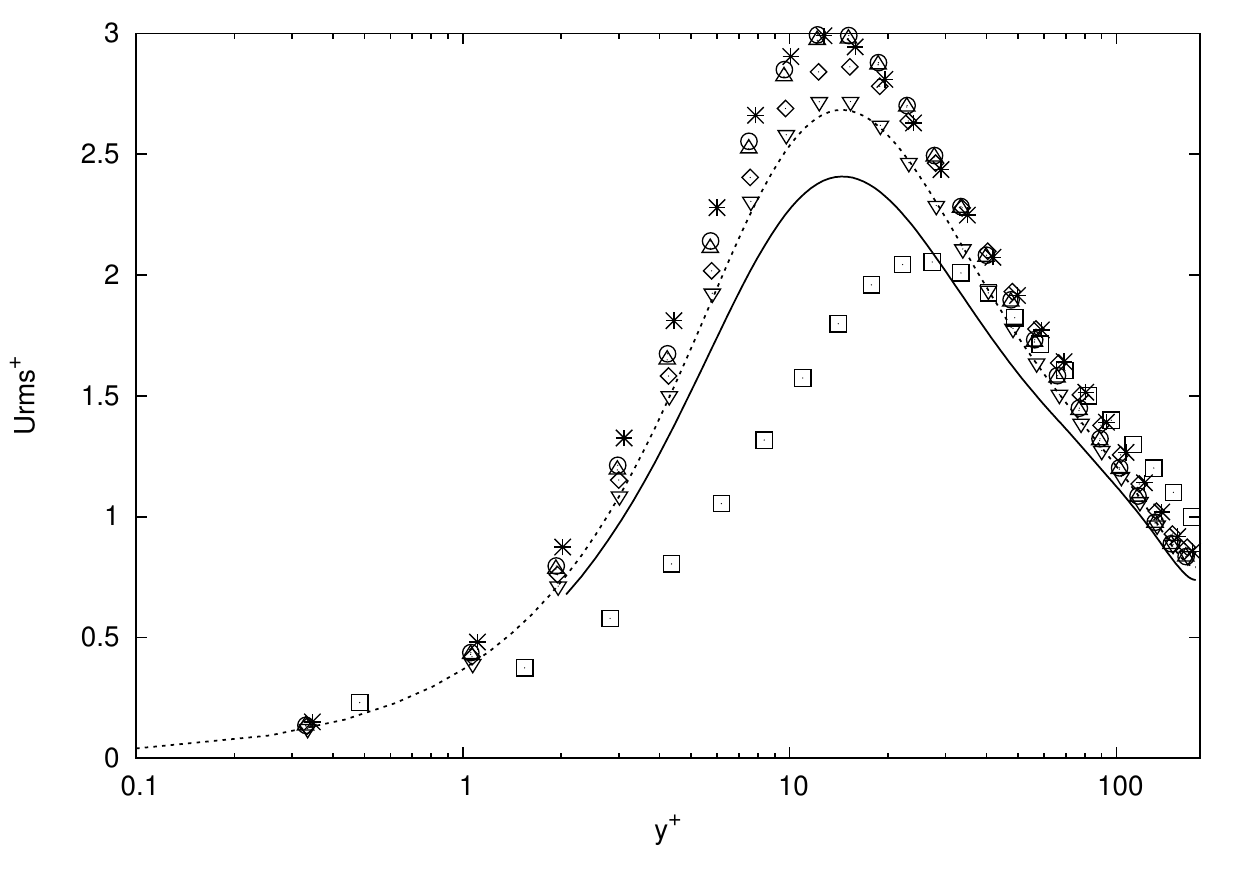}}
}\centerline{
\subfigtopleft{\includegraphics[width=0.44\textwidth, trim={0 5 5 5}, clip]{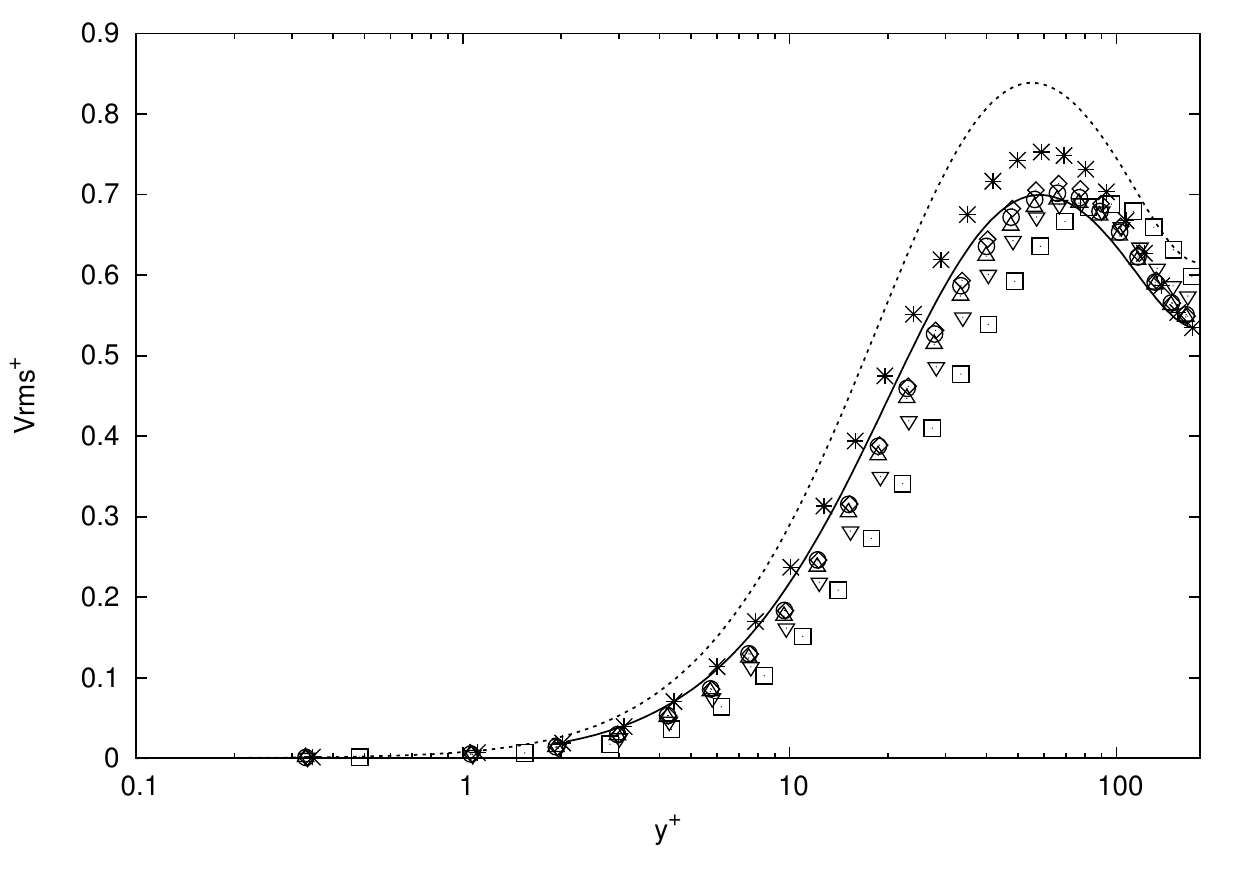}}
\subfigtopleft{\includegraphics[width=0.44\textwidth, trim={0 5 5 5}, clip]{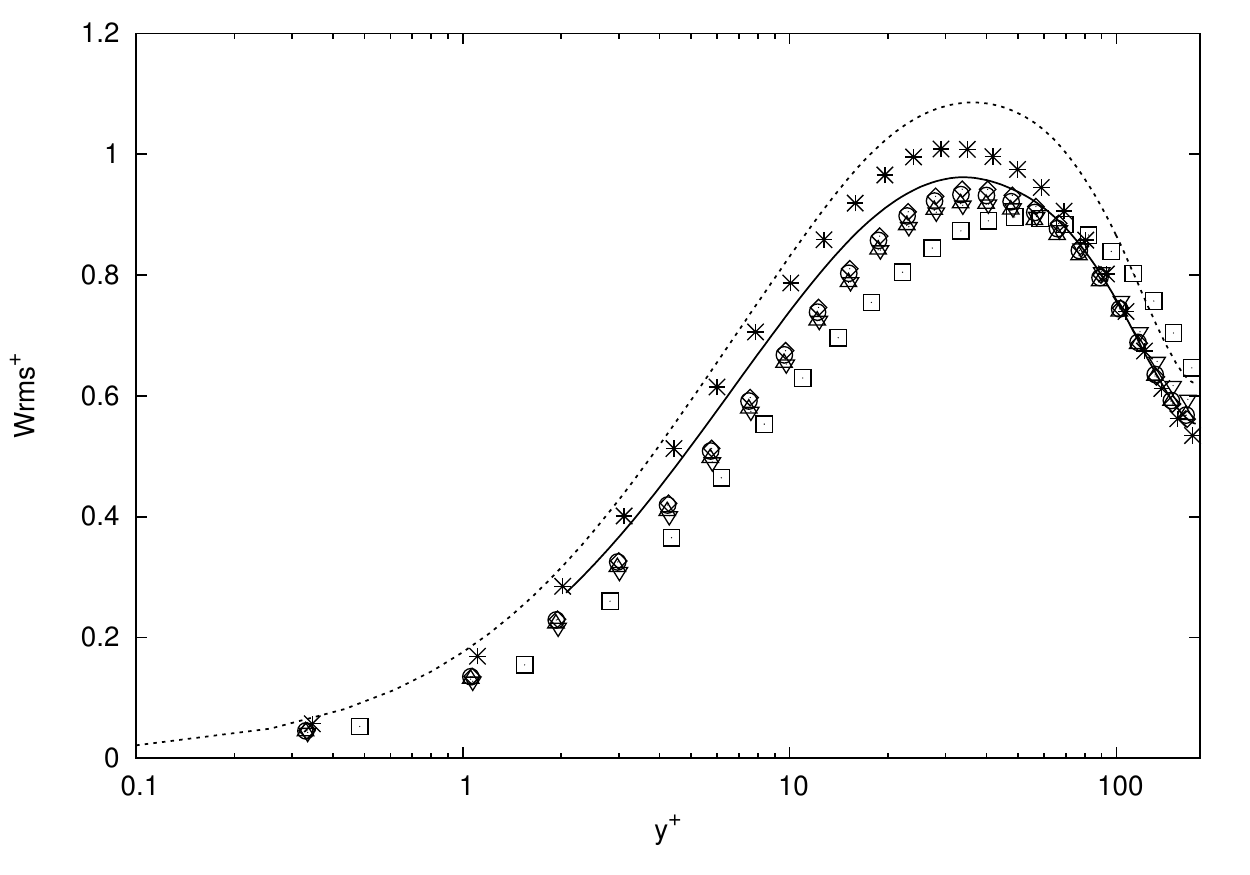}}
}
\caption[Comparison of large-eddy simulations with tensorial gradient-Smagorinsky, gradient-WALE, gradient-Sigma, gradient-AMD and gradient-Anisotropic-Smagorinsky mixed models.]{
Comparison of large-eddy simulations
with tensorial gradient-Smagorinsky, gradient-WALE, gradient-Sigma, gradient-AMD and gradient-Anisotropic-Smagorinsky mixed models
for the profiles of the mean streamwise velocity $\left\langle U_x \right\rangle$ (a, b), the covariance of streamwise and wall-normal velocity $\left\langle u_{\smash[b]{x}}' u_{\smash[b]{y}}' \right\rangle$ (c), the standard deviation of streamwise velocity $\smash[t]{\sqrt{\left\langle u_{\smash[b]{x}}'^2 \right\rangle}}$ (d), wall-normal velocity $\smash[t]{\sqrt{\left\langle u_{\smash[b]{y}}'^2 \right\rangle}}$ (e) and spanwise velocity $\smash[t]{\sqrt{\left\langle u_{\smash[b]{z}}'^2 \right\rangle}}$ (f)
with the mesh 48B.
\label{label21}}
\end{figure*}

\begin{figure*}
\setcounter{subfigcounter}{0}
\centerline{
\subfigtopleft{\includegraphics[width=0.44\textwidth, trim={0 5 5 5}, clip]{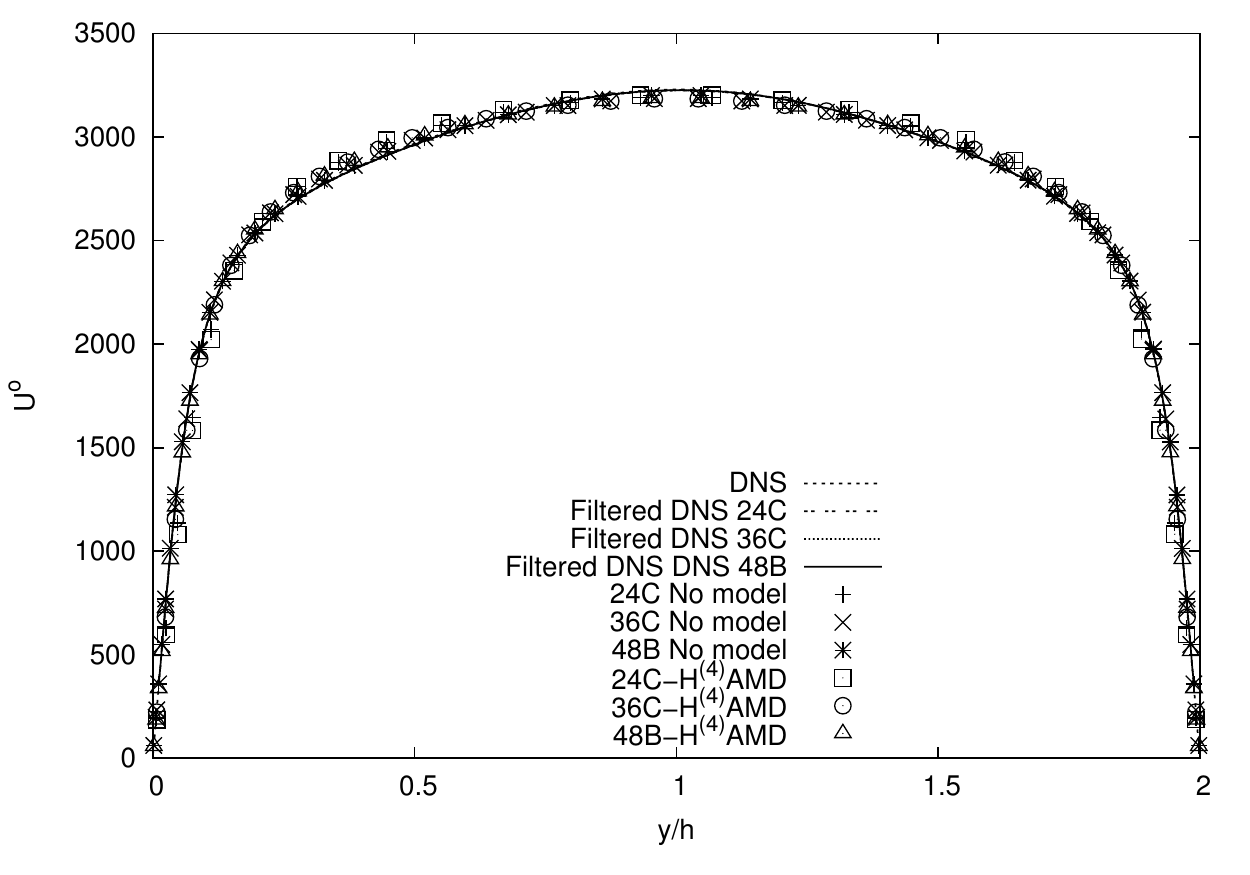}}
\subfigtopleft{\includegraphics[width=0.44\textwidth, trim={0 5 5 5}, clip]{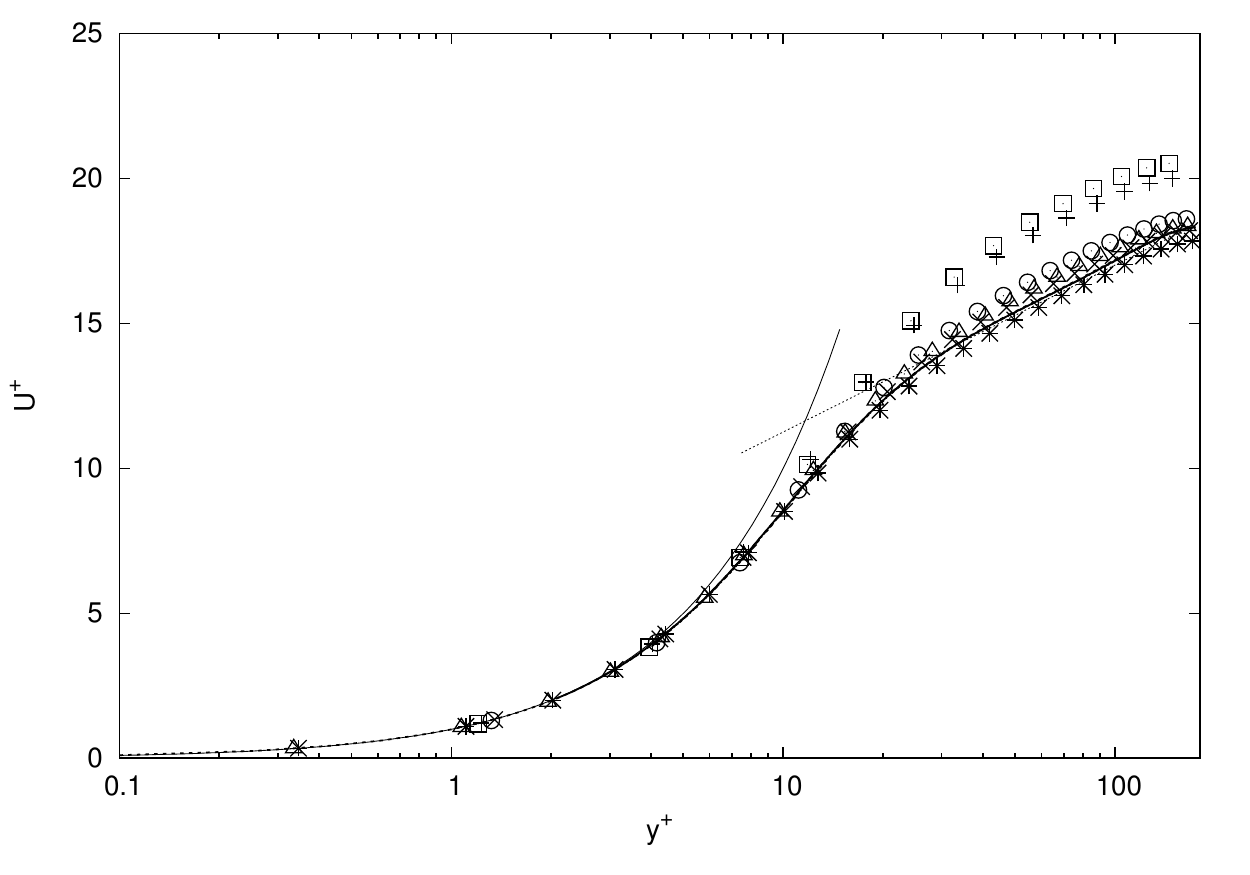}}
}\centerline{
\subfigtopleft{\includegraphics[width=0.44\textwidth, trim={0 5 5 5}, clip]{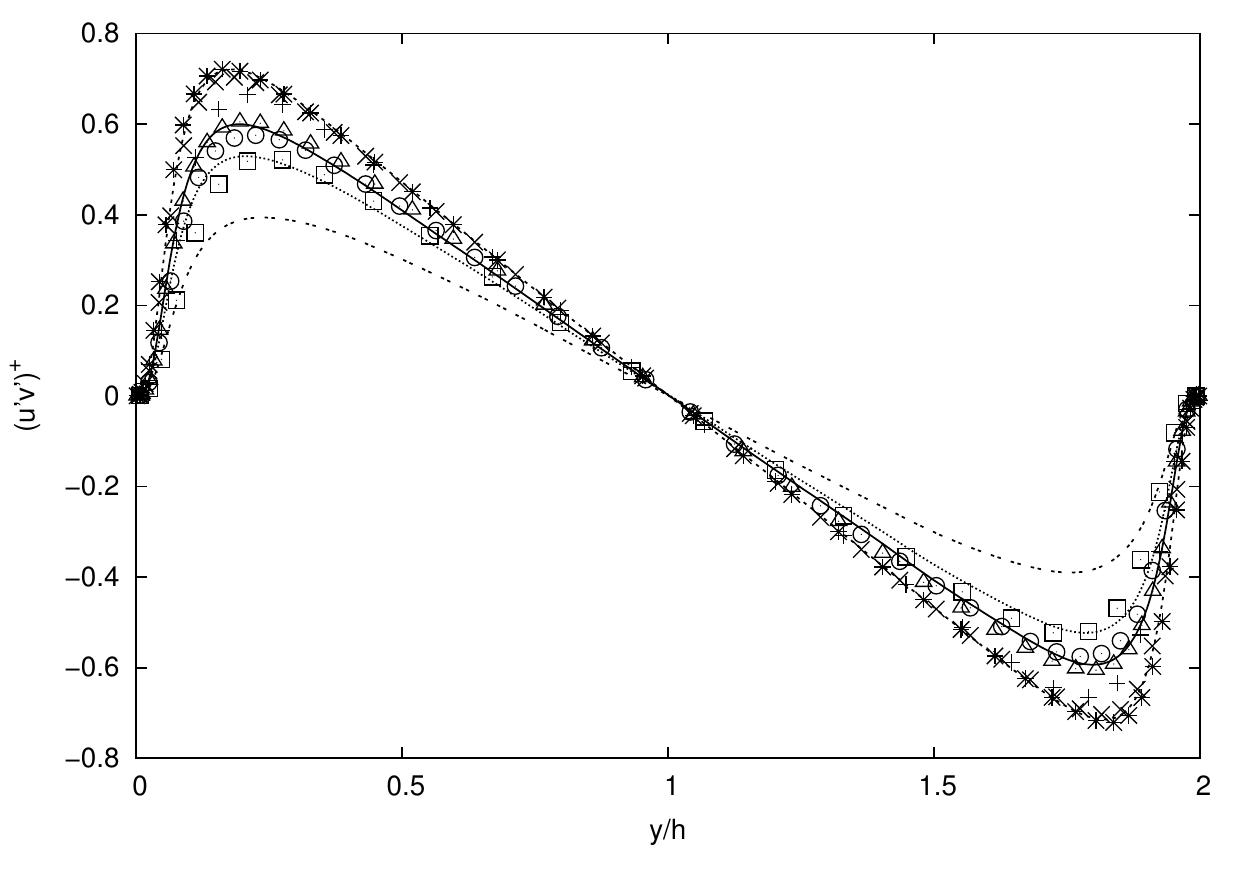}}
\subfigtopleft{\includegraphics[width=0.44\textwidth, trim={0 5 5 5}, clip]{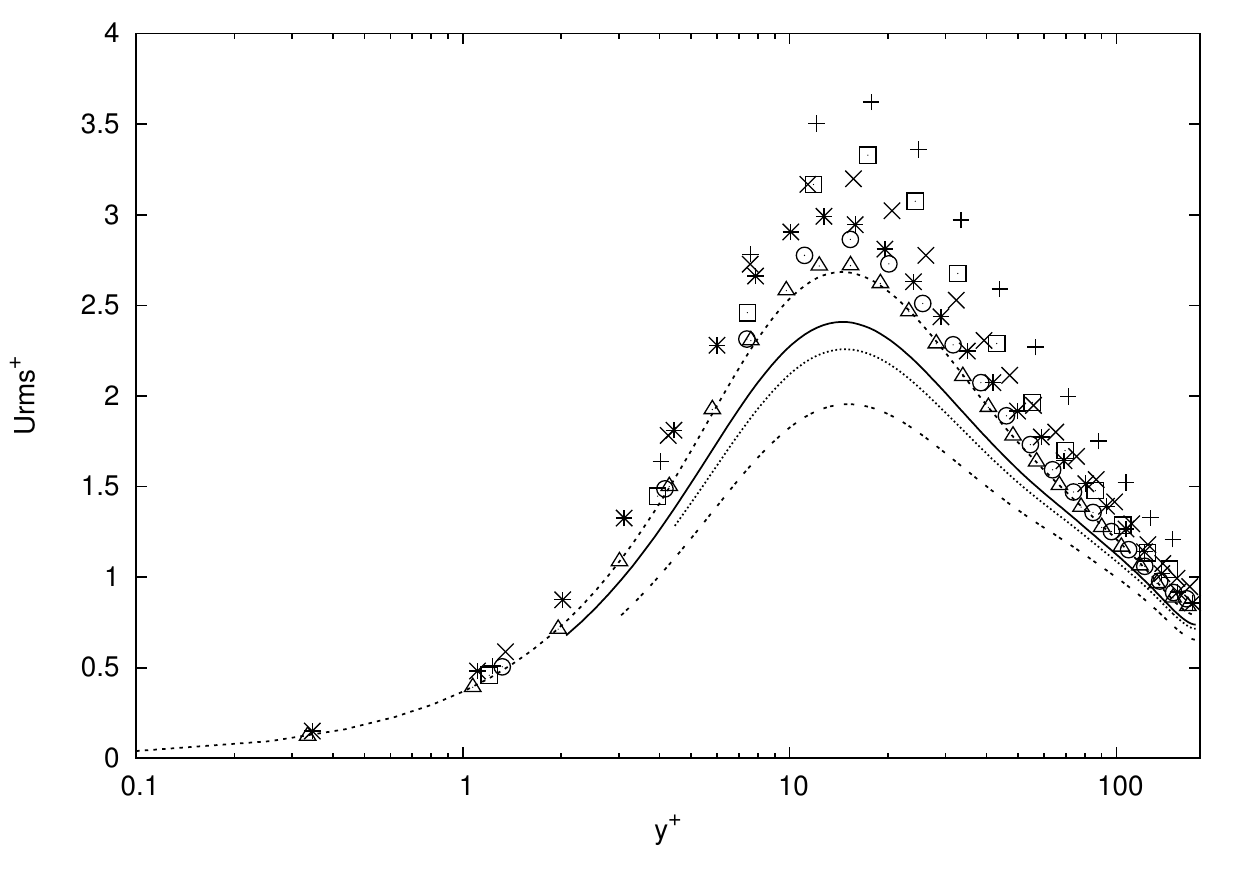}}
}\centerline{
\subfigtopleft{\includegraphics[width=0.44\textwidth, trim={0 5 5 5}, clip]{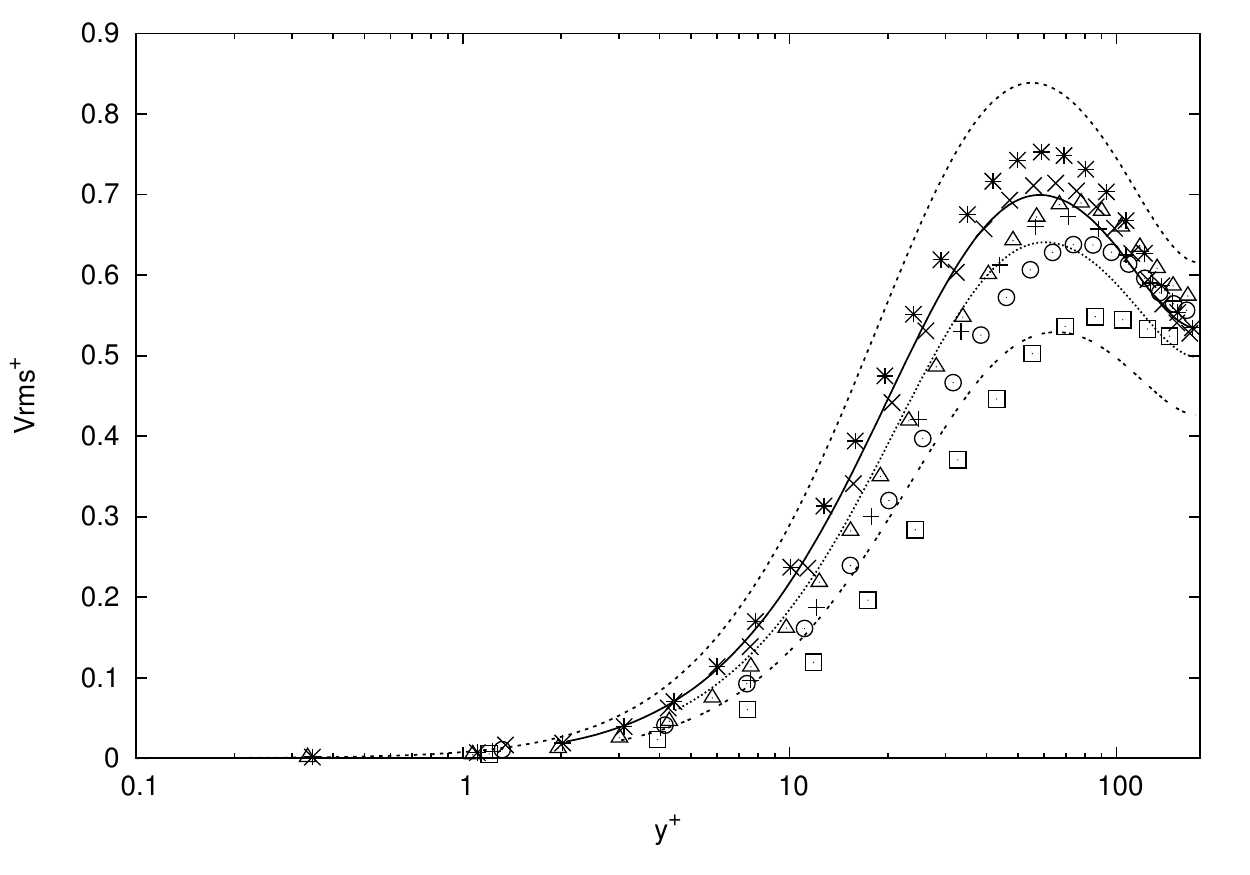}}
\subfigtopleft{\includegraphics[width=0.44\textwidth, trim={0 5 5 5}, clip]{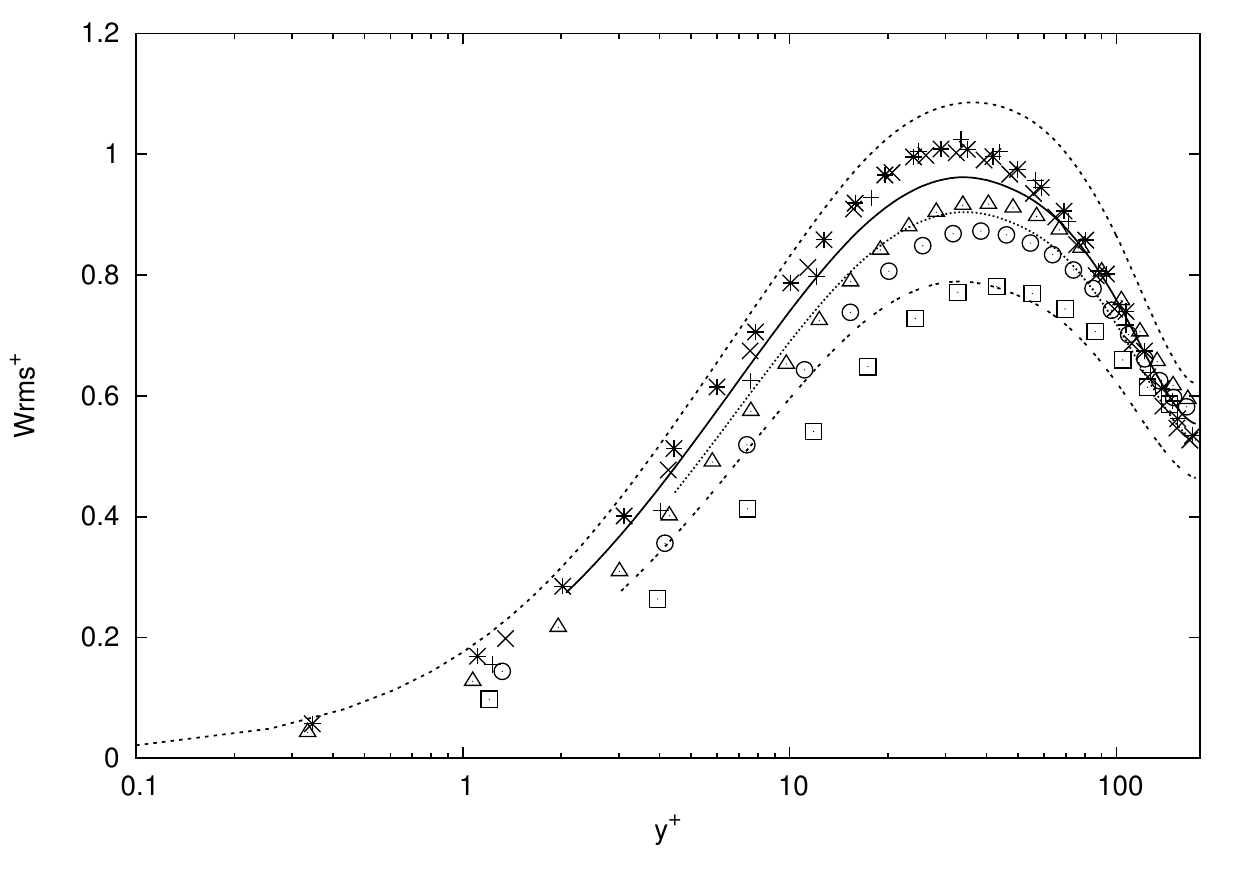}}
}
\caption[Comparison of large-eddy simulations with the tensorial gradient-AMD mixed model based on the $H^{(4)}$ tensor with the meshes 24C, 36C and 48B.]{
Comparison of large-eddy simulations
with the tensorial gradient-AMD mixed model based on the $H^{(4)}$ (equation \ref{label2}) tensor
with the meshes 24C, 36C and 48B
for the profiles of the mean streamwise velocity $\left\langle U_x \right\rangle$ (a, b), the covariance of streamwise and wall-normal velocity $\left\langle u_{\smash[b]{x}}' u_{\smash[b]{y}}' \right\rangle$ (c), the standard deviation of streamwise velocity $\smash[t]{\sqrt{\left\langle u_{\smash[b]{x}}'^2 \right\rangle}}$ (d), wall-normal velocity $\smash[t]{\sqrt{\left\langle u_{\smash[b]{y}}'^2 \right\rangle}}$ (e) and spanwise velocity $\smash[t]{\sqrt{\left\langle u_{\smash[b]{z}}'^2 \right\rangle}}$ (f).
\label{label13}}
\end{figure*}

In this section, we investigate the modelling of the subgrid-scale tensor
with tensorial, mixed and tensorial mixed models,
as well as the dynamic versions of these models.
We focus in particular on models based on the AMD model.
The results of large-eddy simulations with various tensorial AMD models is
compared in figure \ref{label27}
with the mesh 48B.
The models based on the $H^{(4)}$ (equation \ref{label2}) tensor leads to the best prediction of the
wall shear stress while those based on the $H^{(2)}$ (equation \ref{eqdefh2}), $H^{(3)}$ (equation \ref{eqdefh3}) and $H^{(6)}$ (equation \ref{label4})
tensors heavily underestimate or overestimate the wall shear stress.
Compared to the classical AMD model, the tensorial AMD models based on the
$H^{(2)}$ (equation \ref{eqdefh2}), $H^{(4)}$ (equation \ref{label2}) and $H^{(5)}$ (equation \ref{label3}) tensors give better predictions of the
covariance of streamwise and wall-normal velocity and of the standard
deviation of velocity components.
Besides, while functional models were found unable to decrease the
maximum value of the standard deviation of streamwise velocity
compared to the no-model simulation (figure \ref{label22}),
all tensorial AMD models investigated verify this property.
The behaviour of the tensorial AMD models upon mesh derefinement is as the
scale-similarity model not satisfactory for the covariance of streamwise
and wall-normal velocity and the standard deviation of streamwise
velocity. Indeed, the reduction of the maximum amplitude is not sufficiently enhanced
with the coarser meshes (figure \ref{label13}).
It is however more acceptable for the standard deviation of
wall-normal and spanwise velocity in the sense that the profiles undergo the
expected decrease of maximum amplitude with increased filter width.

Contrary to tensorial AMD models, the tensorial models based on
the Smagorinsky, WALE or Sigma model do not decrease the maximum value standard
deviation of streamwise velocity with the mesh 48B (figure \ref{label21}).
A decrease is also obtained using a tensorial Anistropic Smagorinsky model, but
the effect is smaller than with the AMD model.
While the underlying explication is not known, this is to some extent consistent
with the results of tensorial global-average dynamic models (figure \ref{label26}),
in which the AMD model led to a stronger decrease of the standard deviation of
streamwise velocity.

Tensorial gradient-AMD mixed models complement tensorial AMD models
using the gradient model to close the components of the subgrid term
not modelled by the AMD model.
The addition of the gradient model to the AMD model has only a small effect on
the turbulence statistics (figure \ref{label17}).
It decreases the estimated wall shear stress, providing an improvement
for the classical AMD model and the tensorial AMD models
based on the $H^{(1)}$ (equation \ref{eqdefh1}), $H^{(3)}$ (equation \ref{eqdefh3}), $H^{(4)}$ (equation \ref{label2}) and $H^{(6)}$ (equation \ref{label4}) tensors,
which overestimate the wall shear stress,
and a degradation for the tensorial AMD models
based on the $H^{(2)}$ (equation \ref{eqdefh2}) and $H^{(5)}$ (equation \ref{label3}) tensors,
which underestimate the wall shear stress.

\begin{figure*}
 \setcounter{subfigcounter}{0}
 \centerline{
 \subfigtopleft{\includegraphics[width=0.44\textwidth, trim={0 5 5 5}, clip]{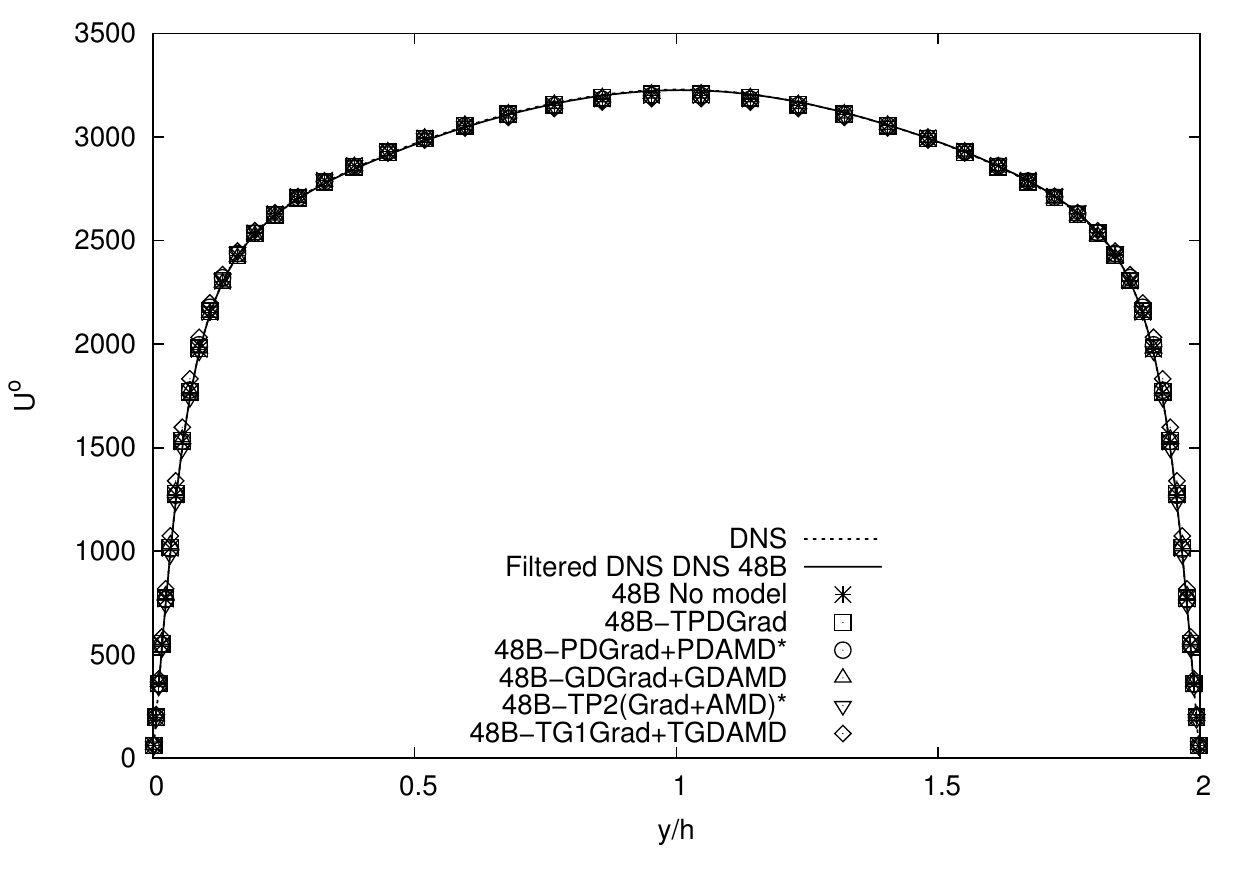}}
 \subfigtopleft{\includegraphics[width=0.44\textwidth, trim={0 5 5 5}, clip]{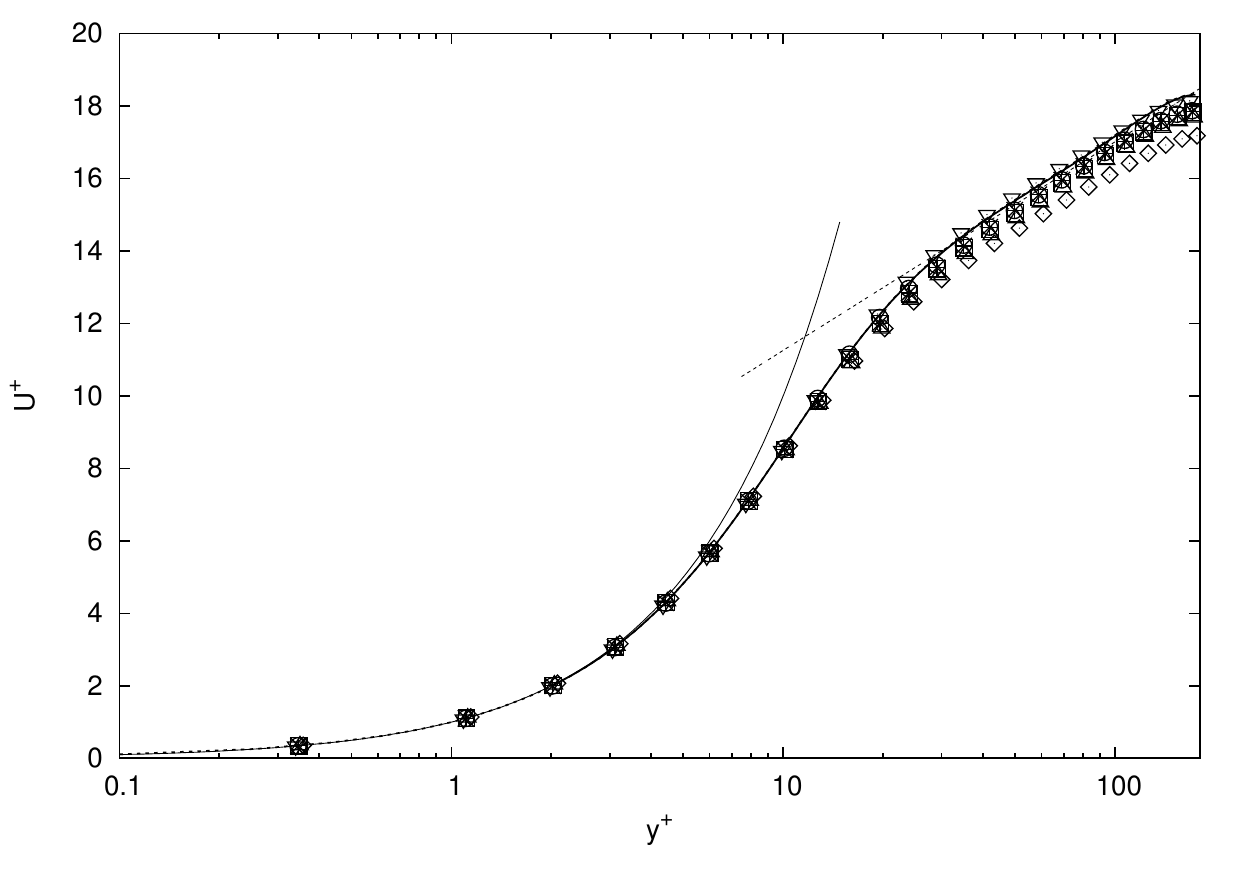}}
 }\centerline{
 \subfigtopleft{\includegraphics[width=0.44\textwidth, trim={0 5 5 5}, clip]{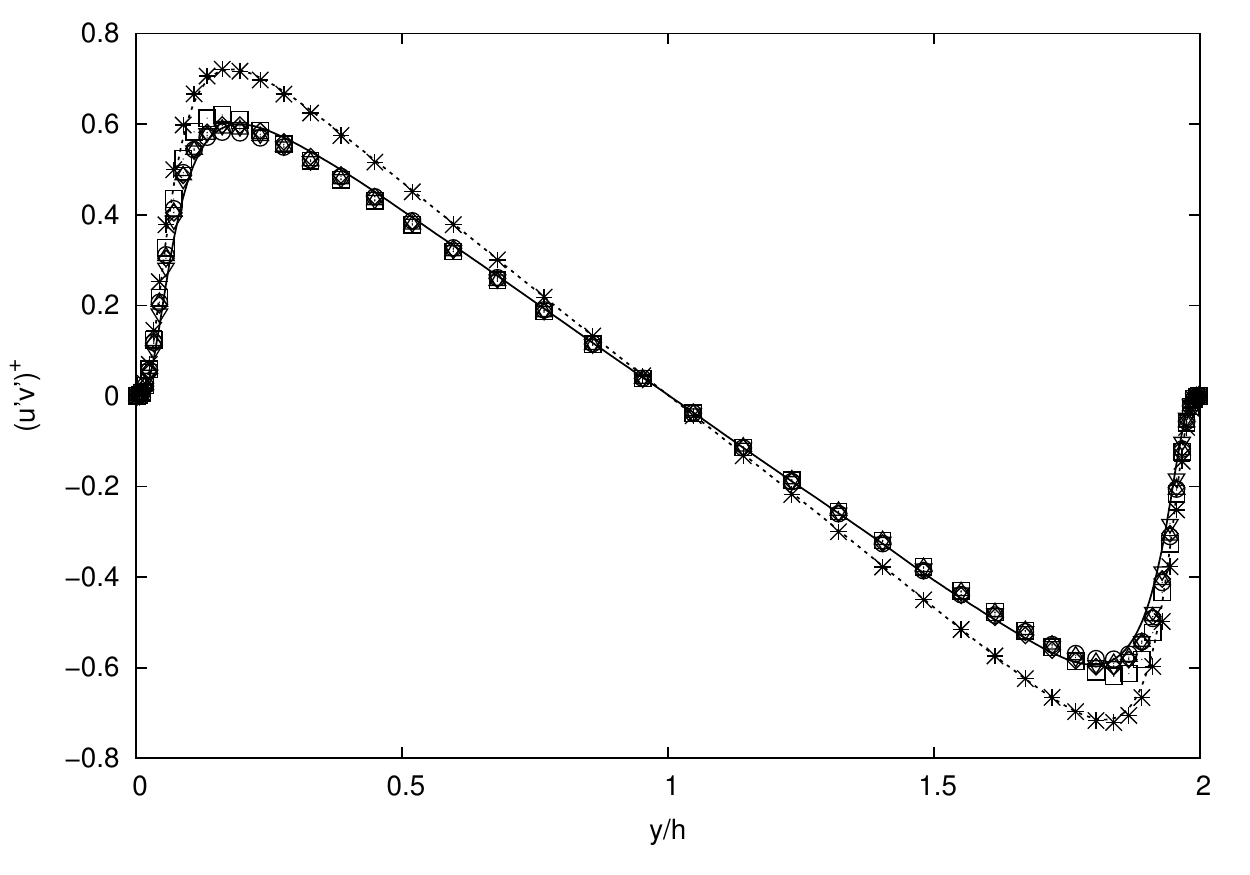}}
 \subfigtopleft{\includegraphics[width=0.44\textwidth, trim={0 5 5 5}, clip]{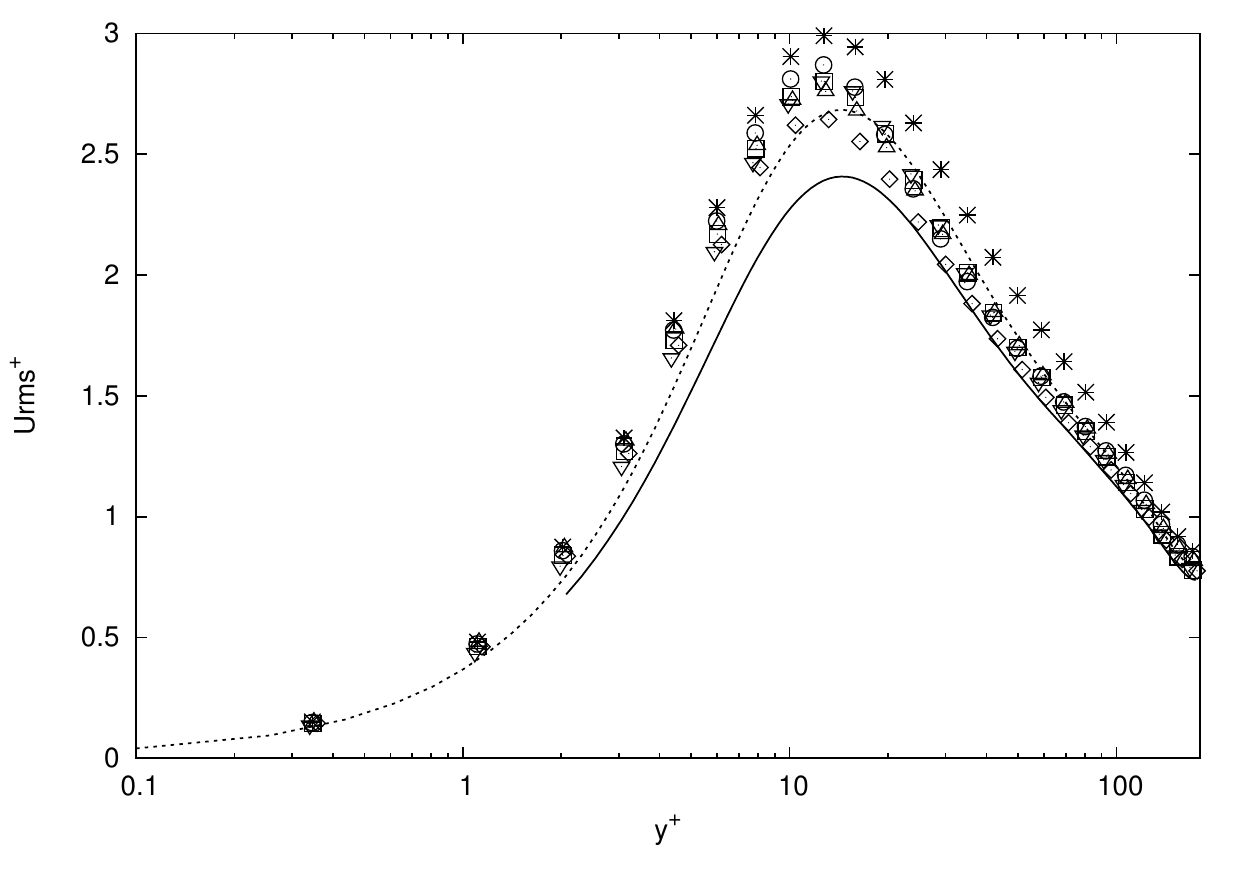}}
 }\centerline{
 \subfigtopleft{\includegraphics[width=0.44\textwidth, trim={0 5 5 5}, clip]{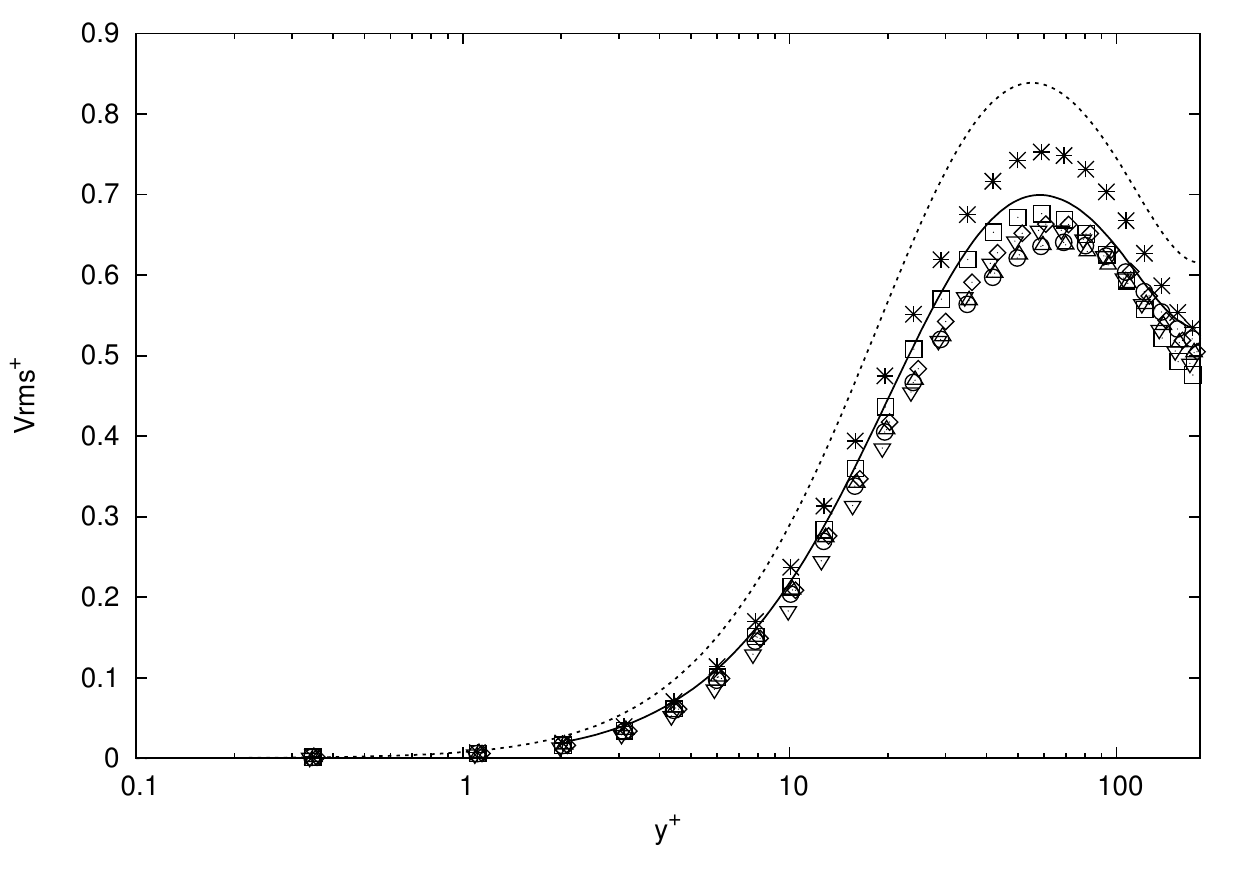}}
 \subfigtopleft{\includegraphics[width=0.44\textwidth, trim={0 5 5 5}, clip]{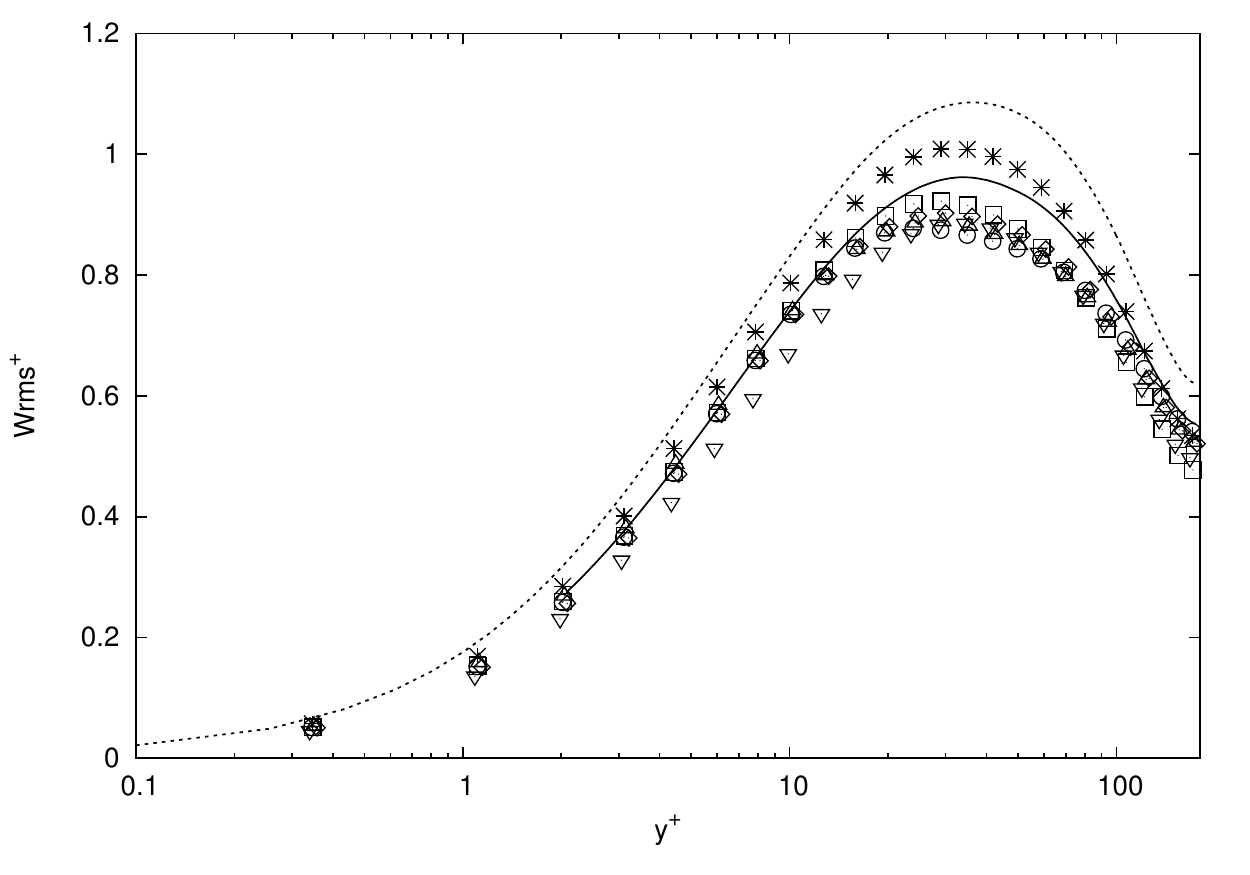}}
 }
 \caption[Comparison of large-eddy simulations with dynamic gradient-AMD mixed models .]{
 Comparison of large-eddy simulations
 with dynamic gradient-AMD mixed models
 for the profiles of the mean streamwise velocity $\left\langle U_x \right\rangle$ (a, b), the covariance of streamwise and wall-normal velocity $\left\langle u_{\smash[b]{x}}' u_{\smash[b]{y}}' \right\rangle$ (c), the standard deviation of streamwise velocity $\smash[t]{\sqrt{\left\langle u_{\smash[b]{x}}'^2 \right\rangle}}$ (d), wall-normal velocity $\smash[t]{\sqrt{\left\langle u_{\smash[b]{y}}'^2 \right\rangle}}$ (e) and spanwise velocity $\smash[t]{\sqrt{\left\langle u_{\smash[b]{z}}'^2 \right\rangle}}$ (f) 
 with the mesh 48B.
 An asterisk (*) indicates the clipping of the AMD-related part.
 \label{label19}}
 \end{figure*}

We investigated various dynamic versions of gradient-AMD mixed models.
Dynamic gradient-AMD mixed models may be based on a plane average, a
global average, a tensorial plane average or a tensorial global average.
Some dynamic procedures are not stable. The stability of the dynamic
procedures investigated is reported in table \ref{label10}.
Plane-average dynamic methods are only stable if the AMD-related part of the
model is not dynamic. Global-average dynamic methods are more stable.
All dynamic procedures investigated are stable if the negative values of the
dynamic parameters of the AMD model are clipped.
However, this makes the AMD model negligible using a global average
(table \ref{leslsdgfunparamb}) or a tensorial global average (table \ref{label9}).
If the dynamic procedure is not tensorial, the one-parameter dynamic method
based on the prior computation of the AMD model with the classical dynamic
procedure (P1Grad+PDAMD or G1Grad+GDAMD) and the two-parameter dynamic method
(P2(Grad+AMD) or G2(Grad+AMD)) give similar results.
The one-parameter dynamic method based on the prior computation of the
gradient model with the classical dynamic method (PDGrad+P1AMD or GDGrad+G1AMD)
and the use of the classical dynamic method for the gradient and AMD models
(PDGrad+PDAMD or GDGrad+GDAMD) also give similar results.
We compare in figure \ref{label19}
a selection of the best-performing models for each type of dynamic procedure.
The dynamic gradient-AMD mixed models do not provide significant improvements
over the constant-parameter tensorial gradient-AMD mixed models.
The best results are achieved with tensorial dynamic procedures.

All in all, while none of the models investigated
is able to properly reproduce the effect of the subgrid-scale tensor
on the flow, some models improves the predictions of
the simulation compared to the no-model case.
We recommend the use of the scale-similarity model and the constant-parameter
or dynamic tensorial AMD model, which provide the most promising results.

 \begin{table*}
 \centerline{\begin{tabular}{lcccc}
           &   Type of          & \multicolumn{2}{c}{Dynamic method of each model} &         \\
           &     averaging      & Gradient               & AMD                               &             Stability\\[.5em]
 \hline\\[-.5em]
     P1Grad+AMD          &                Plane-average                & One-parameter      & Not dynamic                         &      Stable                  \\        
     PDGrad+PDAMD        &                Plane-average                & Classical          & Classical                           &  Not stable                 \\        
     PDGrad+P1AMD        &                Plane-average                & Classical          & One-parameter                       &  Not stable                 \\        
     P1Grad+PDAMD        &                Plane-average                & One-parameter      & Classical                           &  Not stable                 \\        
     P2(Grad+AMD)        &                Plane-average                & Two-parameter      & Two-parameter                       &  Not stable                 \\        
 TP1Grad+AMD             &       Tensorial plane-average                & One-parameter      & Not dynamic                         &      Stable                  \\        
 TPDGrad+TPDAMD          &       Tensorial plane-average                & Classical          & Classical                           &  Not stable                 \\        
 TPDGrad+TP1AMD          &       Tensorial plane-average                & Classical          & One-parameter                       &  Not stable                 \\        
 TP1Grad+TPDAMD          &       Tensorial plane-average                & One-parameter      & Classical                           &  Not stable                 \\        
 TP2(Grad+AMD)           &       Tensorial plane-average                & Two-parameter      & Two-parameter                       &  Not stable\\[.5em]  
 \hline\\[-.5em]                                                             
     G1Grad+AMD          &               Global-average                & One-parameter      & Not dynamic                         &     Stable                   \\        
     GDGrad+GDAMD        &               Global-average                & Classical          & Classical                           &     Stable                   \\        
     GDGrad+G1AMD        &               Global-average                & Classical          & One-parameter                       &     Stable                   \\        
     G1Grad+GDAMD        &               Global-average                & One-parameter      & Classical                           & Not stable                  \\        
     G2(Grad+AMD)        &               Global-average                & Two-parameter      & Two-parameter                       & Not stable                  \\        
 TG1Grad+AMD             &       Tensorial global-average                & One-parameter      & Not dynamic                         &     Stable                   \\        
 TGDGrad+TGDAMD          &       Tensorial global-average                & Classical          & Classical                           &     Stable                   \\        
 TGDGrad+TG1AMD          &       Tensorial global-average                & Classical          & One-parameter                       & Not stable                  \\        
 TG1Grad+TGDAMD          &       Tensorial global-average                & One-parameter      & Classical                           &     Stable                   \\        
 TG2(Grad+AMD)           &       Tensorial global-average                & Two-parameter      & Two-parameter                       & Not stable                  \\        
 \end{tabular}}
 \caption[Stability of the large-eddy simulations
 with a
 dynamic method.]{Stability of the large-eddy simulations
 with plane-average, global-average, tensorial plane-average and tensorial global-average
 dynamic methods for gradient-AMD mixed models 
 with the mesh 48B.
 The clipping only concerns the negative dynamic parameters of the AMD model.
 \label{label10}}
 \end{table*}

\begin{table}
\centerline{\begin{tabular}{lcc}
                             & \multicolumn{2}{l}{ \rlap{\hspace{-9em}          Average of the dynamic parameter (standard deviation), } } \\
                             & \multicolumn{2}{c}{           $\left\langle C^{\mathrm{mod}} \right\rangle$ ($\sqrt{\left\langle (C^{\mathrm{mod}})^2 \right\rangle - \left\langle C^{\mathrm{mod}} \right\rangle^2}$)}\\
                             & AMD-related                             & Gradient-related\\[.5em]
\hline\\[-.5em]
 G1Grad+AMD                  & ---                                     & $\numprint{1.760}$ ($\numprint{0.047}$)\\
 GDGrad+GDAMD                & $\numprint{0.424}$ ($\numprint{0.017}$) & $\numprint{2.245}$ ($\numprint{0.054}$)\\
 GDGrad+G1AMD                & $\numprint{0.424}$ ($\numprint{0.017}$) & $\numprint{2.208}$ ($\numprint{0.053}$)\\
 G1Grad+GDAMD*               & $\numprint{0.003}$ ($\numprint{0.007}$) & $\numprint{2.587}$ ($\numprint{0.053}$)\\
 G2(Grad+AMD)*               & $\numprint{0.003}$ ($\numprint{0.008}$) & $\numprint{2.589}$ ($\numprint{0.054}$)\\
\end{tabular}}
\caption[Dynamic parameters of large-eddy simulations with global-average dynamic gradient-AMD mixed models.]{Average and normalised standard deviation of the AMD-related and gradient-related dynamic parameters
of the large-eddy simulations
with global-average dynamic gradient-AMD mixed models
with the mesh 48B.
An asterisk (*) indicates the clipping of the AMD-related part.
\label{leslsdgfunparamb}}
\end{table}

\begin{table*}
\subtable[AMD-related dynamic parameters]{%
\centerline{\begin{tabular}{lcccccc}
                             & \multicolumn{6}{c}{           Average of the dynamic parameter (standard deviation), } \\
                             & \multicolumn{6}{c}{           $\left\langle C^{\mathrm{mod}} \right\rangle$ ($\sqrt{\left\langle (C^{\mathrm{mod}})^2 \right\rangle - \left\langle C^{\mathrm{mod}} \right\rangle^2}$)} \\
                             & $xx$                                               & $xy$                                     & $xz$                                    & $yy$                                              & $zy$                                               & $zz$\\[.5em]
\hline\\[-.5em]                                                                                                                                                                                                                                                                 
 TG1Grad+AMD                 &            ---                                     & ---                                      & ---                                     &            ---                                    & ---                                                & ---   \\
 TGDGrad+TGDAMD              & $          \numprint{ 0.634}$ ($\numprint{0.080}$) & $\numprint{0.761} $ ($\numprint{0.031}$) & $\numprint{0.255}$ ($\numprint{0.037}$) & $\!\!\!\!\!\numprint{-0.020}$ ($\numprint{0.033}$)& $\!\!\!\!\!\numprint{ 0.049}$ ($\numprint{0.021}$) & $\!\!\!\!\!\numprint{-0.076}$ ($\numprint{0.025}$)\\
 TGDGrad+TG1AMD*             & $          \numprint{ 0.000}$ ($\numprint{0.000}$) & $\numprint{0.213} $ ($\numprint{0.019}$) & $\numprint{0.125}$ ($\numprint{0.022}$) & $          \numprint{ 0.100}$ ($\numprint{0.022}$)& $\!\!\!\!\!\numprint{ 0.003}$ ($\numprint{0.005}$) & $          \numprint{ 0.000}$ ($\numprint{0.000}$)\\
 TG1Grad+TGDAMD              & $          \numprint{ 0.584}$ ($\numprint{0.076}$) & $\numprint{0.799} $ ($\numprint{0.033}$) & $\numprint{0.258}$ ($\numprint{0.038}$) & $\!\!\!\!\!\numprint{-0.008}$ ($\numprint{0.034}$)& $\!\!\!\!\!\numprint{ 0.050}$ ($\numprint{0.021}$) & $\!\!\!\!\!\numprint{-0.073}$ ($\numprint{0.025}$)\\
 TG2(Grad+AMD)*              & $          \numprint{ 0.000}$ ($\numprint{0.000}$) & $\numprint{0.327} $ ($\numprint{0.030}$) & $\numprint{0.121}$ ($\numprint{0.022}$) & $          \numprint{ 0.101}$ ($\numprint{0.023}$)& $\!\!\!\!\!\numprint{ 0.003}$ ($\numprint{0.005}$) & $          \numprint{ 0.000}$ ($\numprint{0.000}$)\\[.5em]
\end{tabular}}
}%
\\%
\subtable[Gradient-related dynamic parameters]{%
\centerline{\begin{tabular}{lcccccc}
                             & \multicolumn{6}{c}{           Average of the dynamic parameter (standard deviation), } \\
                             & \multicolumn{6}{c}{           $\left\langle C^{\mathrm{mod}} \right\rangle$ ($\sqrt{\left\langle (C^{\mathrm{mod}})^2 \right\rangle - \left\langle C^{\mathrm{mod}} \right\rangle^2}$)} \\
                             & $xx$                                               & $xy$                                     & $xz$                                    & $yy$                                              & $zy$                                               & $zz$\\[.5em]
\hline\\[-.5em]                                                                                                                                                                                                                                                                 
 TG1Grad+AMD                 & $          \numprint{ 1.818}$ ($\numprint{0.051}$) & $\numprint{0.465} $ ($\numprint{0.070}$) & $\numprint{1.244}$ ($\numprint{0.053}$) & $          \numprint{ 2.326}$ ($\numprint{0.077}$)& $\!\!\!\!\!\numprint{ 1.466}$ ($\numprint{0.060}$) & $          \numprint{ 1.668}$ ($\numprint{0.049}$)\\
 TGDGrad+TGDAMD              & $          \numprint{ 2.280}$ ($\numprint{0.053}$) & $\numprint{2.271} $ ($\numprint{0.060}$) & $\numprint{1.422}$ ($\numprint{0.034}$) & $          \numprint{ 2.681}$ ($\numprint{0.067}$)& $\!\!\!\!\!\numprint{ 1.598}$ ($\numprint{0.034}$) & $          \numprint{ 1.853}$ ($\numprint{0.030}$)\\
 TGDGrad+TG1AMD*             & $          \numprint{ 2.473}$ ($\numprint{0.056}$) & $\numprint{2.359} $ ($\numprint{0.060}$) & $\numprint{1.408}$ ($\numprint{0.037}$) & $          \numprint{ 2.840}$ ($\numprint{0.066}$)& $\!\!\!\!\!\numprint{ 1.599}$ ($\numprint{0.038}$) & $          \numprint{ 1.933}$ ($\numprint{0.030}$)\\
 TG1Grad+TGDAMD              & $          \numprint{ 2.311}$ ($\numprint{0.055}$) & $\numprint{1.302} $ ($\numprint{0.048}$) & $\numprint{1.388}$ ($\numprint{0.035}$) & $          \numprint{ 2.727}$ ($\numprint{0.068}$)& $\!\!\!\!\!\numprint{ 1.587}$ ($\numprint{0.035}$) & $          \numprint{ 1.875}$ ($\numprint{0.029}$)\\
 TG2(Grad+AMD)*              & $          \numprint{ 2.486}$ ($\numprint{0.057}$) & $\numprint{1.962} $ ($\numprint{0.061}$) & $\numprint{1.399}$ ($\numprint{0.037}$) & $          \numprint{ 2.863}$ ($\numprint{0.067}$)& $\!\!\!\!\!\numprint{ 1.601}$ ($\numprint{0.039}$) & $          \numprint{ 1.940}$ ($\numprint{0.031}$)\\[.5em]
\end{tabular}}
}%
\caption[Dynamic parameters of large-eddy simulations with tensorial global-average dynamic gradient-AMD mixed models.]{Average and normalised standard deviation of the AMD-related and gradient-related dynamic parameters
of the large-eddy simulations
with tensorial global-average dynamic gradient-AMD mixed models
with the mesh 48B.
An asterisk (*) indicates the clipping of the AMD-related part.
\label{label9}}
\end{table*}

\section{Conclusion}

Subgrid-scale models are investigated using large-eddy simulations of a fully developed
turbulent channel flow.
The large-eddy simulations implementing the models are carried out using a
finite method in a staggered grid system with a third-order Runge--Kutta time
scheme.
To examine the influence of the modelling, the large-eddy simulations are
compared to a filtered direct numerical simulation.
The modelling of the subgrid-scale tensor governs the wall shear stress and the
turbulence anisotropy.
The gradient model is not sufficiently
impactful in a large-eddy simulation and must be filtered and amplified
to alters significantly the flow.
Functional eddy-viscosity models do not accurately
represent the turbulence anisotropy as
the standard deviation of streamwise velocity is insufficiently decreased compared
to the wall-normal and spanwise components.
Scalar dynamic procedures avoid the need for an arbitrary model parameter but
does not improve significantly the prediction of models with a proper asymptotic near-wall behaviour.
On the other hand, tensorial eddy-viscosity models can provide more accurate predictions of the
wall shear stress and the turbulence anisotropy
than scalar eddy-viscosity models using either a constant tensorial coefficient or
a tensorial dynamic procedure.
The effects are particularly salient with the AMD model.

\section*{Acknowledgments}
This work was funded by the French Investments for the future (``Investissements d'Avenir'')
programme managed by the National Agency for Research (ANR) under contract ANR-10-LABX-22-01 (labex SOLSTICE).
The authors gratefully acknowledge the CEA for the development of the TRUST
platform. This work was granted access to the HPC resources of CINES under the
allocations 2017-A0022A05099 and 2018-A0042A05099 made by GENCI.